\newcommand\papertitle{Variance of dust temperature and spectral index in \Planck\ polarization data using spin-moment expansion }

\documentclass[twocolumn]{aa}

\makeatletter
\renewcommand*\aa@pageof{, page \thepage{} of \pageref*{LastPage}}
\makeatother
\usepackage{bbm}
\usepackage{graphicx}
\usepackage{lastpage}
\usepackage{enumitem}

\usepackage{placeins}

\usepackage{txfonts}
\usepackage{url}
\usepackage{natbib}
\usepackage{color}
\usepackage{colortbl} 
\usepackage[dvipsnames]{xcolor}
\usepackage{enumitem}
\usepackage{tipa}
\usepackage{cuted}
\usepackage[normalem]{ulem}
\usepackage[switch]{lineno}
\usepackage{diagbox}

\usepackage[breaklinks, colorlinks, citecolor=blue]{hyperref}
\usepackage{subfigure}
\usepackage{epsfig}
\usepackage{multirow}
\usepackage{array}

\def\i{\mathbbm{i}}
\usepackage{url}
\usepackage[T1]{fontenc}
\usepackage{rotating}
\usepackage{soul}
\usepackage{wrapfig}
\usepackage{tikz}

\def\Nside{\ensuremath{N_{\mathrm{side}}}}
\newcommand{\srolltwo}{{\tt SRoll2}}
\newcommand{\npipe}{{\tt NPIPE}}
\newcommand{\sr}{{\tt SR}}

\makeatletter
\def\env@cases{
  \let\@ifnextchar\new@ifnextchar
  \left\lbrace
  \def\arraystretch{1.2}
  \array{l@{}l@{}}
}
\makeatother

\definecolor{mygreen}{RGB}{104,198,107}
\definecolor{myred}{RGB}{252,137,125}
\definecolor{myyellow}{RGB}{252,225,126}
\definecolor{mygrey}{RGB}{215,215,215}

\newcommand\LEt[1]{}

\def\pysmthree{{\sc PySM}~3}

\def\dmod{{\tt d1}}

\def\third{0.33\textwidth}
\def\half{0.48\textwidth}

\def\tlab{\emph{(top)}}

\def\blab{\emph{(bottom)}}
\def\clab{\emph{(center)}}
\def\rlab{\emph{(right)}}
\def\llab{\emph{(left)}}

\def\GHz{\,GHz}
\def\deg{^\circ}

\def\hmo{{^{\tt HM1}}}
\def\hmt{{^{\tt HM2}}}
\def\fullo{{^{\tt MX1}}}
\def\fullt{{^{\tt MX2}}}

\def\COMMANDER{\textit{Commander}}
\def\Nsims{{N_{\rm sims}}}

\def\d{{\rm d}}

\def\i{\mathbbm{i}}
\def\Re{{\operatorname{Re}}}
\def\Im{{\operatorname{Im}}}
\def\Cov{{\rm Cov}}
\def\fsky{f_{\rm sky}}

\def\ms{{\overline{s}}}
\def\mT{{\overline{T}}}
\def\mbeta{{\overline{\beta}}}
\def\meps{{\overline{\varepsilon}}}

\def\ain{a_{i,n}}
\def\T{T}

\def\asin{\ain^s}
\def\ajn{a_{j,n}}

\def\asjn{\ajn^s}

\def\aim{a_{i,m}}

\def\asim{\aim^s}

\def\ajm{a_{j,m}}

\def\asjm{\ajm^s}

\def\sumn{\sum\nolimits_{n=1}^\infty}
\def\sumallnm{\sum\nolimits_{n,m=1}^\infty}

\def\aio{a_{i,1}}
\def\ajo{a_{j,1}}
\def\ait{a_{i,2}}

\def\aid{a_{i,3}}

\def\asio{a_{i,1}^s}
\def\asjo{a_{j,1}^s}

\def\aiu{a_{i,1}}
\def\aid{a_{i,2}}
\def\ait{a_{i,3}}

\def\atn{a^\T_{\nu,n}}
\def\abn{a^\beta_{\nu,n}}
\def\asn{a^s_{\nu,n}}

\def\abun{a^\beta_{\nu,1}}

\def\atun{a^\T_{\nu,1}}

\def\nuref{{\nu_0}}
\def\unitij{\mu K_{\rm CMB}^2 \times \frac{\meps_0^2}{\meps_i\meps_j}}

\def\R{\mathbf{R}}
\def\Rt{R^\psi}
\def\RP{R^P}

\def\Rcxi{\R_i}
\def\Rcxj{\R_j}
\def\Rti{{\Rt_i}}
\def\Rtj{{\Rt_j}}
\def\RPi{{\RP_i}}
\def\RPj{{\RP_j}}

\def\Reps{R^\varepsilon}
\def\Repsi{{\Reps_i}}
\def\Repsj{{\Reps_j}}

\def\cvar{c}
\def\varPcx{\boldsymbol{\rm c}^\P}
\def\varP{\cvar^P}
\def\varpsi{\cvar^\psi}
\def\varpsiP{\cvar^{\psi P}}
\def\varPpsi{\cvar^{P \psi}}
\def\vareps{\cvar^{\varepsilon}}
\def\hvareps{\hat{\cvar}^{\varepsilon}}

\def\varspsi{\sigma_s^\psi}
\def\varseps{\sigma_s^\varepsilon}
\def\varTpsi{\sigma_T^\psi}
\def\varTeps{\sigma_T^\varepsilon}
\def\varbpsi{\sigma_\beta^\psi}
\def\varbeps{\sigma_\beta^\varepsilon}
\def\varTpsiP{\Sigma_T^{\psi P}}

\def\varij{\cvar_{ij}}
\def\varii{\cvar_{ii}}
\def\varjj{\cvar_{jj}}

\def\varPcxij{\boldsymbol{\rm c}^\P_{ij}}
\def\varPij{\cvar^P_{ij}}
\def\varpsiij{\cvar^\psi_{ij}}
\def\varpsiPij{\cvar^{\psi P}_{ij}}
\def\varPpsiij{\cvar^{P \psi}_{ij}}
\def\varpsiPji{\varPpsiij}
\def\varepsij{\cvar^\varepsilon_{ij}}

\def\cdeb{\hat{\cvar}}

\def\pref{p_0}
\def\Iref{I_0}
\def\Pref{P_0}

\def\P{\mathbf{P}}

\def\Ims{\left\langle \Iref^2\right\rangle}

\def\Pms{\left\langle P_0^2\right\rangle}

\def\mrat{{\overline{\eta}}}
\def\p{\mathbf{p}}
\def\deltap{\delta\hat{\p}}
\def\pmax{p_{\rm max}}

\def\Mosym{\Delta}

\def\Mosn{{\vec{\Mosym}^s_n}}
\def\Mosm{\vec{\Mosym}^s_m}

\def\Mon{\vec{\Mosym}^s_n}

\def\Mom{\vec{\Mosym}^s_m}

\def\W{{\vec{\mathcal{W}}}}
\def\IW{\Im\W^s}
\def\RW{\Re\W^s}
\def\Wone{{\vec{\mathcal{W}^s_1}}}

\def\Wto{{\vec{\mathcal{W}_1^\T}}}

\def\Wso{{\vec{\mathcal{W}_1^s}}}

\def\Wbn{{\vec{\mathcal{W}}_n^{\beta}}}
\def\Wtn{{\vec{\mathcal{W}}_n^\T}}
\def\Wn{\vec{\mathcal{W}}^s_n}

\def\Wsn{{\vec{\mathcal{W}}^s_n}}

\def\mWsn{{\overline{\W}^s_n}}

\def\mIsn{{\overline{\omega}^s_n}}

\def\mMosn{\overline{\vec{\Mosym}^s_n}}

\def\mWone{{\overline{\W}^s_1}}

\def\Ibn{{{\omega_n^\beta}}}
\def\Itn{{{\omega_n^\T}}}
\def\In{{{\omega^s_n}}}
\def\Iso{{{\omega_1^s}}}

\def\Isn{{{\omega_n^s}}}
\def\Ism{{{\omega_m^s}}}

\def\Wsm{{\vec{\mathcal{W}}_m^s}}

\def\vecB{\overrightarrow{B}}
\def\vecDB{\overrightarrow{\Delta B}}
\newcommand{\expf}[1]{{\rm e}^{#1}}

\def\lc{\mathcal{L}}

\def\anu{a_\nu}

\def\rP{\mathbf{r}}
\def\hrP{\hat{\mathbf{r}}}
\def\mRP{\mathbf{\overline{r}}}

\def\Planck{\textit{Planck}}

\begin{document}

\title{\papertitle}
\authorrunning{Guillet et al.}
\titlerunning{Variance of dust temperature and spectral index using spin-moment expansion
}

\author{Vincent Guillet\inst{\ref{IAS},\ref{LUPM}},
Léo Vacher \inst{\ref{SISSA},\ref{INFN},\ref{IFPU}},
Jonathan Aumont\inst{\ref{IRAP}},
François Boulanger\inst{\ref{ENS}},
Alessia Ritacco\inst{\ref{ENS},\ref{INP},\ref{INAF-OAC}},
Jean-Marc Delouis\inst{\ref{IFREMER}}, and
Andrea Bracco\inst{\ref{ENS},\ref{INAF}}.
}

\date{Accepted at A\&A, 19 october 2025}

\institute{
\label{IAS}Institut d'Astrophysique Spatiale, CNRS, Univ. Paris-Sud, Universit\'{e} Paris-Saclay, B\^{a}t. 121, 91405 Orsay cedex, France 
\and
\label{LUPM}Laboratoire Univers et Particules de Montpellier, Universit{\'e} de Montpellier, CNRS/IN2P2, CC 72, Place Eug{\`e}ne Bataillon, 34095 Montpellier Cedex 5, France 
\and
\label{SISSA}International School for Advanced Studies (SISSA), Via Bonomea 265, 34136, Trieste, Italy
\and
\label{INFN}Istituto Nazionale di Fisica Nucleare (INFN), Sezione di Trieste, via Valerio 2, 34127 Trieste, Italy 
\and
\label{IFPU}Institute for Fundamental Physics of the Universe (IFPU), Via Beirut, 2, 34151 Grignano, Trieste, Italy
\and
\label{IRAP}IRAP, Universit\'e de Toulouse, CNRS, CNES, UPS, (Toulouse), France 
\and
\label{ENS}Laboratoire de Physique de l'Ecole Normale Sup\'erieure, ENS, Universit\'e PSL, CNRS, Sorbonne Universit\'e, Universit\'e de Paris, F-75005 Paris, France 
\and
\label{INP}Univ. Grenoble Alpes, CNRS, Grenoble INP, LPSC-IN2P3, 53, avenue des Martyrs, 38000 Grenoble, France 
\and
\label{INAF-OAC}INAF - Osservatorio Astronomico di Cagliari, Via della Scienza 5, 09047 Selargius, Italy
\and
\label{IFREMER}Laboratoire d'Oc\'eanographie Physique et Spatiale (LOPS), Univ. Brest, CNRS, Ifremer, IRD, 29200 Brest, France
\and
\label{INAF}INAF - Osservatorio Astrofisico di Arcetri, Largo E. Fermi 5, 50125 Firenze, Italy 
}

\abstract{
Thermal dust is the major polarized foreground hindering the detection of primordial cosmic microwave background (CMB) $B$-modes. Its signal exhibits complex behavior in frequency space, arising from the combined variation in our Galaxy of the orientation of magnetic fields and the spectral properties of dust grains aligned with magnetic field lines. In this work, we present a new framework for analyzing the thermal dust signal using polarized microwave data. We introduce residual maps, represented as complex quantities, which capture deviations of the local polarized spectral energy distribution (SED) from the mean complex SED averaged over the sky mask. 
We present simple predictions that relate the values of the statistical correlation and covariances between the residual maps to the physical properties of the emitting aligned grains. Testing these predictions provides valuable information about the nature of the dust signal. We evaluated our predictions using \Planck\ data over a 97\% mask excluding the inner Galactic plane. Despite its simplicity, our model captures a significant part of the statistical properties of the data. For the $\srolltwo$ version of the data, the spectral dependence of the covariances between residual maps is  compatible with a dust model that includes only temperature variations rather than spectral index variations. In contrast, for the PR4 \Planck\ official release, it is incompatible with both models.  Our methodology
can be used to analyze future high-precision polarization data and to build more accurate dust models for use by the CMB community.}


\maketitle

\section{Introduction}

Observations from the visible to the submillimeter wavelengths (submm) provide extensive evidence on the variations in dust composition and optical properties through the Galaxy. The importance of dust processing in the interstellar medium (ISM) is demonstrated first by the large variations seen in the extinction curve and the depletions of refractory elements \citep{Draine03,Jenkins09,vardustdisk2}. Additional evidence comes from variations in the spectral energy distribution (SED) and the far-infrared and submm dust opacity of molecular clouds and the diffuse ISM, observed by the {\it Herschel} \citep{Ysard2013} and \Planck\ space missions \citep{Planck13_XI,Planck2014dust}. These variations of the dust SED include, but are not fully explained by, the dependence of dust temperature on the local intensity of the interstellar radiation field \citep{Fanciullo2015}. 

The observational signatures of dust evolution in the dust SED in polarization are harder to identify due to the reduced amplitude of the signal compared to the total intensity \citep{pelgrims2021,Ritacco2023}. The interpretation of polarization data considers the combined variations in the structure of the Galactic magnetic field and the dust SED, both in the plane of the sky and along the line of sight in the three-dimensional Galaxy \citep{Tassis2015,PlanckL,McBride2023,Vacher2025}. This is the specific problem that our work addresses. We present a modeling framework that incorporates variations in dust properties within a turbulent magnetic field, designed to quantify the variance in dust spectral properties from submm polarization data.

In submm polarization maps, the diffuse polarized emission from dust grains that are locally aligned with magnetic-field orientation is summed over the light cone, a small region of the sky observed by the instrumental beam containing both plane-of-the-sky and line-of-sight variations. Variations in the magnetic field orientation within the light cone lead to depolarization \citep{planck2016-XLIV,Clark19}, and their combined variations with changes in dust properties lead to the frequency-dependence of the polarization angle \citep{Tassis2015,PlanckL,Vacher2023a}. 
For studies of the cosmic microwave background (CMB) polarization, the variation over the sky of dust spectral parameters, and the non-linear distortion they induce in the SEDs, together with their impact on the frequency dependence of polarization angles, makes the separation of the dust and CMB signals a difficult task, especially given the high sensitivity of next-generation experiments \citep{Remazeilles2016,Ptep,Wolz2024}. 
The modeling of spatial variations in the dust polarization SED is therefore of prime importance for the search for primordial $B$-modes from inflation and for the putative $EB$ correlation from cosmic birefringence \citep{Diego2022BiRe,CosmicBirefringence,Vacher2023b,Jost2023,Sullivan2025}. 

\Planck\ observations show that the mean SED of dust polarized emission in the high-latitude diffuse ISM is remarkably well fitted by a modified black body (MBB) law from 353\GHz\ to microwave frequencies \citep{Planck18_XI}. This observational result leads us to model spatial variations of the dust polarization SED using the moment expansion method around an MBB, as introduced by \citet{Chluba}. Previous studies carried out in preparation for future experiments, such as the Simons Observatory and LiteBIRD, show that when spatial variations are present in the foreground SED, the moment-expansion formalism provides a powerful tool to recover the CMB signal without bias, and with minimal parameter addition \citep{Azzoni2021,RemazeillesmomentsILC,Desert2022, Vacher21,Sponseller2022,Azzoni2023,Wolz2024,Carones2024}. The moment expansion formalism can be extended to polarized dust emission using complex spin-2 fields, referred to as``spin-moment'' coefficients \citep{Vacher2023a,Vacher2023b}. In this work, we use this formalism to introduce a modeling framework that, within some simplifying assumptions, accounts for the combined variations of dust properties and magnetic field structure. Applying it to \Planck\ data, we demonstrate the relevance of this framework for analyzing dust polarization microwave observations for foreground modeling and component separation.

Our paper is structured as follows. 
In Sect.~\ref{sec:model}, we introduce the dust model and data modeling framework based on complex residual maps. We enumerate the predictions of our model in Sect.~\ref{sec:predictions}. 
In Sect.~\ref{sec:application}, building on the work of \cite{Ritacco2023}, we test the validity of our simplifying assumptions in relation to sky observations by analyzing the \Planck\ \srolltwo~maps \citep{Delouis2019}. 
In Sect.~\ref{sec:dust}, we interpret our results in terms of fluctuations in dust spectral properties and compare them with those obtained using the PR4 version of the \Planck\ data \citep{npipe}.
In section~\ref{sec:discussion}, we discuss our results, which we then summarize in Sect.~\ref{sec:conclusion}.

\section{Model definition}
\label{sec:model}

Our dust model and methodology, based on complex numbers, are designed to quantify the variance in the emission properties of grains aligned with magnetic field lines\footnote{To emit polarized thermal emission, dust grains must be aligned with the local magnetic field. Not all grains are aligned: it has been proven that grains smaller than $\sim 0.1\,\mu$m are not aligned \citep[e.g.,][]{KM95}, while the existence of a population of large, unaligned, possibly carbonaceous grains remains debated \citep{Chiar2006,DarkDustII,Hensley2023,THEMISII}.} which emit polarized radiation in the far-infrared and submm. In Sect.~\ref{sec:hypotheses}, we set out the assumptions of our dust signal model and in Sect.~\ref{sec:expansion} we show how these hypotheses enable accurate modeling of the polarized signal using complex numbers and the moment expansion. In Sect.~\ref{subsec:residuals}, we define the so-called polarization residual maps, from which we derive various covariances in Sect.~\ref{sec:covar}. These covariances are then used to build the predictions of our model in Sect.~\ref{sec:predictions}. 

\subsection{Dust model hypothesis}\label{sec:hypotheses}

Our dust model relies on simplifying hypotheses of two kinds: astrophysical (HA) and methodological (HM), listed below.
\makeatletter
\newcommand{\mylabel}[2]{#2\def\@currentlabel{#2}\label{#1}}
\makeatother

\begin{enumerate} 
\item \mylabel{HA1}{HA1}: The local dust polarized SED emitted at any given position in the Galaxy is a single MBB with local temperature $T$ and local spectral index $\beta$.
\item \mylabel{HA2}{HA2}: The variations of the dust polarization spectral parameters $T$ or $\beta$, from one point of the Galaxy to another, are not correlated with variations in the orientation of the magnetic field. 
\item \mylabel{HM3}{HM3}: Only two extreme cases are explored: the fluctuations of dust properties are either pure $\beta$ fluctuations over the whole sky or pure $T$ fluctuations. 
\item \mylabel{HA4}{HA4}: The grain alignment efficiency is uniform over the sky; only the structure of the magnetic field controls the dust polarization fraction. This hypothesis was shown by \cite{Planck18_XII} to be justified at the scale of a degree or more. 
\item \mylabel{HM5}{HM5}: Following \citet{Planck_XLVIII}, we modeled the structure of magnetic fields as a turbulent field close to equipartition between its ordered and random components.
\item \mylabel{HM6}{HM6}: Our data analysis methodology accounts for contamination of the dust polarization signal in \Planck\ data by CMB, noise, and systematics, but it neglects possible contribution from polarized CO emission or leakage of CO emission in the 217 and 100\GHz\ channels, as well as the presence of residual synchrotron emission after template removal.
\end{enumerate}
While these hypotheses are quite restrictive, our model is able to generate a frequency-dependence of polarization angles. The simplicity of the model and the clarity of its building hypotheses allow one to infer its predictions unambiguously and to identify its limits precisely. To assist the reader with the numerous notations, Table~\ref{tab-notations} in Appendix~\ref{A-dico} summarizes all the quantities used in our model.

\subsection{Spin-moment expansion of the complex polarized SED}\label{sec:expansion}

Following the work of \cite{Ritacco2023}, we developed an analytical framework to model the observed variations with the frequency of polarization intensity and angle. For this purpose, we did not use the $E$ and $B$ decomposition \citep{ZS97}. Instead, we quantified the total variance of the Stokes parameters. For a given frequency $\nu$, we define the dust complex polarized SED $\P_\nu$ from the Stokes parameters $Q_\nu$ and $U_\nu$ as follows (complex quantities are written in bold font):
\begin{equation}
\P_\nu \equiv Q_\nu + \i U_\nu = P_\nu\, \expf{\i2\psi_\nu}\,, 
\label{eq:spinor}
\end{equation}
where $\i$ is the imaginary unit ($\i^2=-1$), $P_\nu$ is the polarized intensity ($P_\nu = \sqrt{Q_\nu^2+U_\nu^2}=\sqrt{ \P_\nu\P_\nu^\star}$, with $^\star$ denoting complex conjugation), and $\psi_\nu = 0.5\arctan(U_\nu,Q_\nu)$ is the frequency-dependent polarization angle. 

Following \cite{Vacher2023a}, we note that the use of complex spin-moments is well suited to studying fluctuations in the dust complex polarized SED. The SED $\P_\nu$ observed over a region of the sky is given by the integral over all the local elementary SEDs $\d\P_\nu$ present within the observed light cone $\lc$: 
\begin{align}
\P_\nu = \int_\lc \d\P_\nu\,.
\label{eq:Pnu}
\end{align}
Variations in phases in this integral can partially or fully cancel the polarization signal, a phenomenon known as depolarization \citep{Planck_Int_XIX}. 
The so-called light cone, $\lc$, is a region of the sky over which all local signals $\d\P_\nu$ -- contained both in the plane of the sky and along the line of sight -- are averaged. Depending on the context of the analysis, $\lc$ can refer to a map pixel, the instrumental beam, or even the entire sky, possibly masked. In the present work, we considered $\lc$ to be the instrumental beam surrounding each pixel of the sky. 

We used an MBB model of dust emission (hypothesis \ref{HA1}, Sect.~\ref{sec:hypotheses}) characterized by a local spectral index $\beta$ and a local temperature $T$ to express the local dust emissivity $\varepsilon_\nu(T,\beta)$ 
at frequency $\nu$,
\begin{align}
\varepsilon_\nu(\beta,T) = 
B_\nu(T)\left(\frac{\nu}\nuref\right)^\beta\,,
\label{eq:emissivity}
\end{align}
where $B_\nu(T)$ is Planck's law at frequency $\nu$ and temperature $T$, and $\nuref$ is a reference frequency. Assuming that the local complex polarized intensity, $\d\P_\nuref$, at frequency $\nuref$ is well-defined in amplitude and phase everywhere in the light cone, the frequency-dependence of the local complex polarized SED, $\d\P_\nu$, is
\begin{align}
\d \P_\nu = \d\P_\nuref \frac{\varepsilon_\nu(\beta,T)}{\varepsilon_\nuref(\beta,T)} = \d\P_\nuref \frac{B_\nu(T)}{B_\nuref(T)}\left(\frac{\nu}{\nuref}\right)^\beta\,.
\label{eq:dPnu}
\end{align}
Because the emissivity $\varepsilon_\nu$ is a real quantity,  $\d\P_\nu$ has the same polarization angle at all frequencies. 

\begin{figure}
\includegraphics[width=0.9\columnwidth]{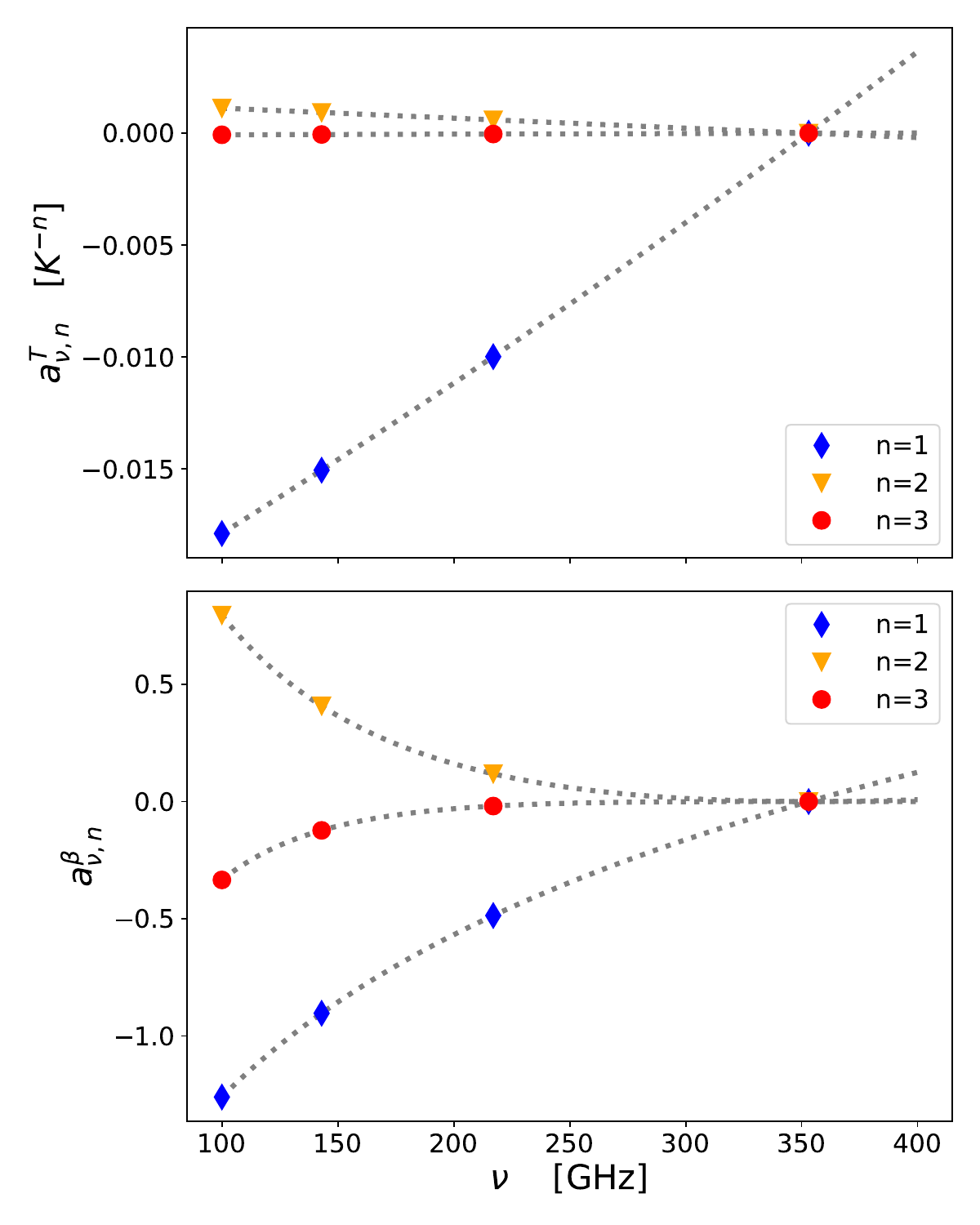}
\caption{Values of $\anu^{\,\beta}$ and $\anu^\T$ for orders $n=1$ (blue), $n=2$ (orange), and $n=3$ (red), shown for $\nuref=353$\GHz\ and pivot temperature $\mT=20\,$K. The points represent the four \Planck\ HFI channels.
}
\label{fig:a-nu}
\end{figure}

To estimate the impact of fluctuations in the spectral parameter $s\in\{\T,\beta\}$ on the observed complex polarized intensity $\P_\nu$  (under our hypothesis \ref{HM3}, which assumes that either $T$ or $\beta$ fluctuates over the whole sky but not both), we performed a Taylor expansion of the power-law term and the blackbody in $\d\P_\nu$ in Eq.~\eqref{eq:dPnu} around an arbitrary pivot spectral index $\mbeta$ and pivot temperature $\mT$. We find that the frequency dependence of the complex polarized intensity can be expressed as \citep{Vacher2023a}
\begin{align}
& \P_\nu = \P_\nuref\frac{\meps_\nu}{\meps_\nuref} \left(1 + \sum\nolimits_{n=1}^\infty \asn\Wsn \right)\,,
\label{eq:PnuW}
\end{align}
where 
\begin{align}
\meps_\nu \equiv \varepsilon_\nu(\mbeta,\mT)\label{eq:meps}
\end{align}
is the pivot SED, the function $\asn$ of $\nu$ denotes the $n^{\rm th}$ order derivative of the dust emissivity with respect to the spectral parameter $s$,
\begin{eqnarray}
\asn \equiv \frac{1}{n!}
\frac{\meps_\nuref}{\meps_\nu} 
\frac{\partial^n \left(\varepsilon_\nu/\varepsilon_\nuref\right)}{\partial s^n}(\mbeta,\mT)\,,
\label{eq:asn}
\end{eqnarray}
and $\Wsn$ is the $n^{\rm th}$ order spin-moment of $s$ \citep{Vacher2023a}, given\footnote{Here, the spin-moments are defined alternatively, as in\cite{Vacher2023a}, with the correspondence $\P_\nuref \equiv \mathcal{W}_0$ and $\bf{\mathcal{W}}_1^s \equiv \mathcal{W}^s_1/\mathcal{W}_0$. This redefinition of the first-order moment is more operational but valid only in the perturbative regime $\mathcal{W}_0 \gg \mathcal{W}^s_1$, a condition that which we assume here (for a discussion, see \cite{Vacher2023a,Vacher2023b}).} by 
\begin{align}
&\Wsn \equiv \frac{1}{\P_\nuref} \int_\lc \left(s-\ms\right)^n\, \d\P_\nuref\label{eq:Wsn}\,.
\end{align}
The spin-moments $\Wsn$ encode the departure of the spectral dependence of $\P_\nu$ from the pivot SED $\meps_\nu$. The pivot spectral parameters $\mbeta$ and $\mT$ are constant over $\mathcal{L}$. For simplicity, we further considered a single pair of pivots for all light cones and hence the whole sky. In this analysis, we used $\mT \equiv 20\,$K. The specific value for $\mbeta$ is not critical, as it does not appear in the data analysis. 

The frequency dependence of $\atn$ and $\abn$ is presented in Fig.~\ref{fig:a-nu} for $n=1,2$, and $3$. Their analytical expression for $n=1$ is as follows:
\begin{align}
 & \abun \equiv \ln\left({\nu/\nuref}\right)\,, \hspace{0.5cm}\label{eq:anu}\\ 
 & \atun \equiv \frac{h}{k \mT^2}\left[\frac{\nu\,\expf{h\nu/k\mT}}{\expf{h\nu/k\mT}-1} - \frac{\nuref\,\expf{h\nuref/k\mT}}{\expf{h\nuref/k\mT}-1}\right] \,. \label{eq:bnu}
\end{align}
The values of the $\asn$ functions -- hereafter referred to as the spectral gradients -- are listed explicitly up to $n=5$ for both $\beta$ and $\T$ in \cite{Chluba}. 

The polarization-weighted moment $\Wsn$ is distinct from the intensity-weighted moments $\Isn$ \citep{Chluba}:
\begin{align}
\Isn & \equiv \frac{1}{I}\int_\lc \left(s-\ms\right)^n \,\d I\,,\label{eq:Isn}
\end{align}
with $s\in\{\T,\beta\}$ as before. Like $\Wsn$, $\Isn$ characterizes the statistical properties of the spectral parameters $s$ of aligned grains. However, $\Isn$ is independent of the structure of the magnetic field, whereas $\Wsn$ is not. The difference between these two moments, 
\begin{align}
\Mosn \equiv \Wsn - \Isn\,, \,
\label{eq:Wbn-In-Mn}
\end{align}
is a complex quantity of interest. In Appendix \ref{sec:Mon}, we demonstrate that, under our hypothesis \ref{HA2}, which assumes no correlation between the fluctuations of dust spectral properties and variations in the magnetic field direction, $\Mosn$ has a statistically zero mean: $\left\langle\Mosn\right\rangle = 0$.

\subsection{Spectral rotation of polarization angles}\label{sec:rotatpsi}

\begin{figure}
\centering
\includegraphics[width=\half]{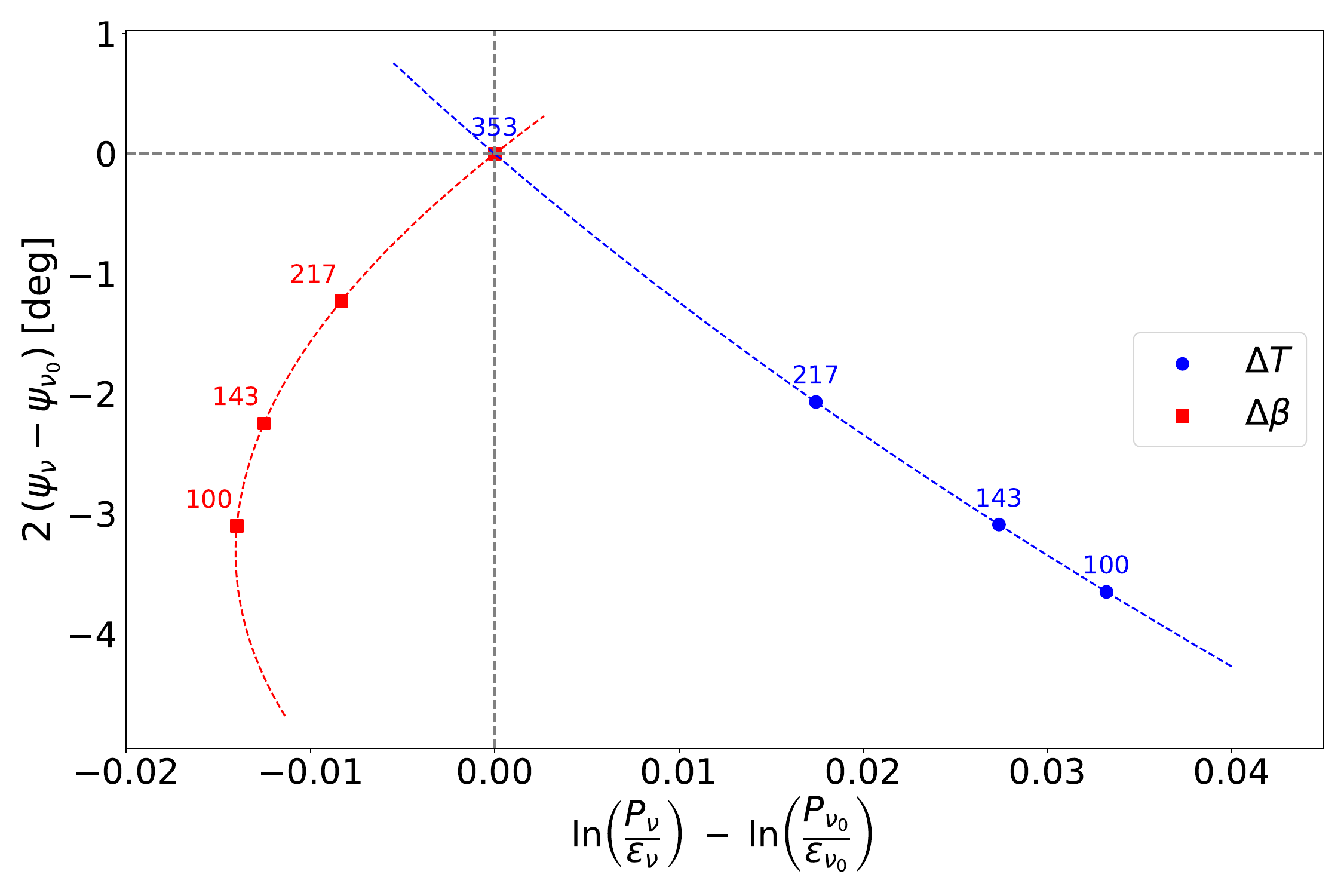}
\caption{Rotation of the SED in the complex plane for a sum of seven MBB SEDs with distinct spectral indices drawn from a Gaussian distribution with mean $1.5$ and standard deviation $0.1$ (red), or with temperatures drawn from a Gaussian distribution with mean $20$\,K and standard deviation $5$\,K (blue). The 3D orientation of the magnetic field is a random realization of the turbulent magnetic field inspired by \citet{Planck_XLVIII}. The dashed lines indicate frequencies from 40 to 400\GHz, and the four colored points represent the Planck HFI bands.}
\label{fig:schema}
\end{figure}

To gain further insight into how the spin-moments, $\Wn$, are related to the spectral dependence of $\P_\nu$, we first consider the specific case of small distortions of the polarized SED around the pivot SED ($|\sumn \asn \Wn| \ll 1$). Taking the logarithm of Eq.~\eqref{eq:PnuW} and using the approximation $\ln(1+x)\sim x$ for $x\ll 1$, we obtain 
\begin{equation}
\ln\left(\frac{\P_\nu}{\meps_\nu}\right) - \ln\left(\frac{\P_\nuref}{\meps_\nuref}\right) \simeq \sumn \asn \Wn \,,
\end{equation}
from which we take the real and imaginary parts and, using Eq.~\eqref{eq:spinor},  
\begin{align}
 \ln\left(\frac{P_\nu}{\meps_\nu}\right) - \ln\left(\frac{P_\nuref}{\meps_\nuref}\right) & \simeq \sumn \asn\, \RW_n \,,\\
 2\left(\psi_\nu - \psi_\nuref\right) & \simeq \sumn \asn \,\IW_n \,.\label{eq:rotatpsi}
 \end{align}
Equation~\eqref{eq:rotatpsi} describes the spectral rotation of the polarization angle with frequency, caused by fluctuations of the dust spectral parameters within the light cone. Its frequency dependence is governed by the $\asn$ parameters and its amplitude depends on the imaginary parts $\IW_n$ of the spin-moments. Figure~\ref{fig:schema} provides a numerical illustration of such a rotation in the log-complex plane for the sum of seven MBBs with Gaussian-distributed spectral parameters in a turbulent magnetic-field model inspired by \citet{Planck_XLVIII} (see also our description in Appendix~\ref{A-Nlayers}). We consider the two cases of our hypothesis \ref{HM3}: pure fluctuations in $T$ or pure fluctuations in $\beta$. 
Under the assumption of small variations in $T$ or $\beta$, the complex SED plotted in the log-complex plane $(\ln\left(P_\nu/\epsilon_\nu\right),2\psi_\nu)$ is close to a straight line for $T$ fluctuations and can be non-linear for $\beta$ fluctuations when the magnetic field direction is close to the line of sight (as shown in Fig.~\ref{fig:schema} for illustration). This shows that, in a given light cone, the rotation of the polarization angle with frequency is strongly correlated with the distortion of the polarized intensity relative to the pivot SED.

\subsection{Mean complex normalized SED}\label{subsec:meanSED}

From here on, we simplify our notation for channel $i$ of frequency $\nu_i$ and use $\P_i$ to denote the map of complex polarized intensity, $\meps_i$ for the pivot emissivity (constant over the sky), and $\asin$ for the spectral gradient.

Following \cite{Ritacco2023}, we cross-correlated the polarization maps to determine the mean complex SED $\mRP_i$ of dust polarization, averaged over the sky and normalized to our reference channel, given by
\begin{equation}
\mRP_i \equiv \frac{\left\langle \P_i\P_0^\star\right\rangle}{\left\langle \P_0\P_0^\star\right\rangle} 
= \frac{\meps_i}{\meps_0}\left(1 + \sumn \asin\mWsn\right)\,,
\label{eq:defmRPi}
\end{equation}
where $\mWsn$ is the mean value of $\Wsn$ over the sky, weighted by $P_0^2=\P_0\P_0^\star$ as
\begin{align}
\mWsn \equiv \left\langle P_0^2\Wsn\right\rangle\,/\,\Pms\,.\label{eq:mWsn}
\end{align}

\subsection{Residual maps for complex polarization \label{subsec:residuals}}

In this section, we use the pivot SED to construct complex residual maps at all frequencies and relate them to the $\Wn$ moments. We define the residual map $\Rcxi$ for the complex polarization $\P_i$:
\begin{align}
\Rcxi &\equiv P_0 \left(\frac{1}{\mRP_i}\frac{\P_i}{\P_0}-1\right) \,.
\label{eq:defRcxi}
\end{align}
\noindent
This complex map characterizes the residuals at frequency $\nu_i$ after removal of the components correlated with the polarization map at the reference frequency $\nuref$, and subsequently scaled to the reference frequency $\nuref$.
Here, $\Rcxi$ quantifies the complexity of the dust signal by measuring its deviations from the mean complex SED $\mRP_i$. 

Using Eqs.~\eqref{eq:PnuW} and \eqref{eq:defmRPi}, we obtain
\begin{align}
\Rcxi & = P_0\left(\frac{1+\sumn\asin\Wn}{1 + \sumn \asin\mWsn}-1\right) \,.
\end{align}
By choosing the pivot SED close to the mean SED, we ensure that $\left|\sumn\asin\,\mWsn\right| = \left|\mRP_i\frac{\meps_0}{\meps_i}-1\right| \ll 1$ (this hypothesis is verified in the data analysis in Sect.~\ref{sec:application}), 
so that, by expanding the previous equation to first order, we obtain
\begin{align}
\Rcxi & \simeq  P_0\sumn\asin\,\left(\Wsn-\mWsn\right)\,.
\label{eq:Rcxi-ai}
\end{align}
We decompose
$\Rcxi$ into its real and imaginary components: 
\begin{align}
\RPi & \equiv \Re\,\Rcxi = P_0\sumn \ain\,\Re\left(\Wsn-\mWsn\right) \,,\label{eq:RPi}\\
\Rti & \equiv \Im\,\Rcxi =
P_0\sumn \ain\,\Im\left(\Wsn-\mWsn\right) \label{eq:Rti}\,.
\end{align}

For $\Rti$ to be non-zero in a given light cone $\lc$, both the magnetic field and the spectral parameters should vary inside $\lc$, whereas a variation in spectral parameters alone suffices for $\RPi$ to be non-zero. We use the labels $P$ and $\psi$ for these map residuals because, as demonstrated in Sect.~\ref{sec:rotatpsi}, in the limit of small fluctuations, the real and imaginary parts of the complex polarized SED correspond directly to the polarized intensity $P$ and polarization angle $\psi$.

Similarly, we define the map residual $\Repsi$ of the dust polarized emissivity that corresponds to the moments $\In$ (Eq.~\eqref{eq:Isn}) as
\begin{align}
\Repsi \equiv P_0\sumn \asin\,\left(\Isn-\mIsn\right)\,.\label{eq:Repsi-ai}
\end{align}
Here, $\mIsn \equiv \left\langle P_0^2\Isn\right\rangle/\Pms$ denotes the sky-averaged value of $\Isn$. The map residual $\Repsi$ represents what would be observed if the polarization angles were constant across each light cone. It maps the variations of the spectral properties of aligned grains, independently of the magnetic field structure, while $\RP_i$ mixes both.

\subsection{Covariance matrices of residual maps}\label{sec:covar}

Consider a complex map $\bf X$ (resp. $\bf Y$) with sky-averaged values $\overline{\bf X}$ and $\overline{\bf Y}$, defined as
\begin{align}
\overline{\bf X} = \left\langle P_0^2\,{\bf X}\right\rangle\,/\,\Pms\,.
\end{align}
We use the notation $\Cov({\bf X},{\bf Y})$ for the covariance of $\bf X$ and $\bf Y$ weighted by $P_0^2$ and computed over the sky:
\begin{align} 
\Cov({\bf X},{\bf Y}) \equiv \left\langle P_0^2 \left({\bf X}-\overline{\bf X}\right) \left({\bf Y}-\overline{{\bf Y}}\right)^\star\right\rangle\,,\label{eq:defCov}
\end{align} 
where the quantity inside the brackets must be averaged over the partial sky after application of the mask. 

For all frequency pairs $(i,j)$, we define covariances for the polarization map residuals $\RP$ and $\Rt$:
\begin{align} 
\varPij & \equiv \left\langle \RPi{\RPj}\right\rangle = \sumallnm \asin\,\asjm\,\Cov\left(\Re\Wsn,\Re\Wsm\right)\,,\label{eq:covP}\\
\varpsiij &\equiv \left\langle\Rti{\Rtj}\right\rangle = \sumallnm \asin\,\asjm\,\Cov\left(\Im\Wsn,\Im\Wsm\right)\ \,. \label{eq:covpsi} 
\end{align}
While the residual maps $R$ represent values for each light cone of the map, the covariances $c$ are averaged quantities that sum up the statistics of all light cones that make up the masked sky. Because we are working at the map scale and not in Fourier space, this mask does not need to be continuous.

We also define the mixed-covariance of $\Rt$ with $\RP$:
\begin{align} 
\varpsiPij & \equiv \left\langle\Rti\RPj\right\rangle = \sumallnm \asin\,\asjm\,\Cov\left(\Im\Wsn,\Re\Wsm\right)\,, \label{eq:covpsiiPj}
\end{align} 
and its companion $\varPpsiij \equiv \varpsiP_{ji}$. 

Finally, we define the covariance matrix $\vareps$, which characterizes the fluctuations of the emissivity of aligned grains as
\begin{align} 
\varepsij & \equiv \left\langle\Repsi\Repsj\right\rangle = \sumallnm \asin\,\asjm\,\Cov\left(\Isn,\Ism\right) \,.\label{eq:vareps}
\end{align} 
This covariance is the only one not contaminated by variations in the magnetic field structure and so can be directly compared safely to dust models or observations in total intensity. 
 
All these quantities are $N\times N$ matrices of real numbers, with $N$ the number of channels. In contrast, the covariance $\varPcx$ of the complex map residual $\R$, 
\begin{align}
\varPcxij & \equiv \left\langle\Rcxi\Rcxj^\star\right\rangle = \sumallnm \asin\,\asjm\,\Cov\left(\Wsn,\Wsm\right), \label{eq:covPcx} 
\end{align}
is an $N\times N$ matrix of complex numbers, related to other real matrices through 
\begin{align} 
\varPcxij 
& = \left\langle\RPi{\RPj}+\Rti{\Rtj}\right\rangle + \i\left\langle\Rti{\RPj}-\RPi{\Rtj}\right\rangle\,,\label{eq:covPcx3}
\\
& = \left(\varPij + \varpsiij\right)+\i\left(\varpsiPij-\varpsiPji\right)\,.
\label{eq:covPcx2}
\end{align} 
We therefore have the trivial equalities, already expected from the definition of $\RP$ and $\Rt$ (Eq.~\eqref{eq:RPi} and \eqref{eq:Rti}):
\begin{align}
\Re\,\varPcx & = \varP + \varpsi \,,\label{eq:Pythagore}\\
\Im\,\varPcx & = \varpsiP-\varPpsi\,.\label{eq:Imc_cross}
\end{align}
By symmetry, $\Im\,\varPcxij=0$ if $i=j$. For $i\ne j$, we have
\begin{align} 
\Im\,\varPcxij & = \sum\nolimits_{m>n\ge 1} \left(\asin\,\asjm-\asim\asjn\right) \Im\,\Cov\left(\Wsn,\Wsm\right)\,.
\label{eq:ImvarPcxij}
\end{align}
\section{Predictions of our model}\label{sec:predictions}

Here and throughout this work, predictions of our model are highlighted by a boxed frame.

\subsection{Expected level of correlation between residual maps}\label{sec:firstorder}

Our dust model assumes that only a single spectral parameter, $\beta$ or $T$, fluctuates in the light cone, but not both simultaneously (hypothesis \ref{HM3}, Sect.~\ref{sec:hypotheses}). Fluctuations in $T$ or $\beta$ propagate into the residual maps. We next examine how well the residual maps correlate across frequencies.

Limiting the moment expansion of Eq.~\eqref{eq:Rcxi-ai} to first order gives
$\R_i\simeq\asio\,P_0\left(\Wone-\mWone\right)$ and $\R_j\simeq\asjo\,P_0\left(\Wone-\mWone\right)$. Residuals maps become simply proportional to each other. The correlation of $\R_i$ with $\R_j$ is perfect. This conclusion also holds for $\RP$ and for $\Rt$. Introducing higher-order $n$  weakens this correlation by adding the common signals $\Wn$ in channels $i$ and $j$, but are weighted with frequency-dependent coefficients $\ain$ and $\ajn$, which differ from $\aio$ and $\ajo$. From the values of $\ain$ (also displayed in Fig.~\ref{fig:a-nu}), we computed the Pearson correlation coefficients $\rho(\ain,\aim)$ across frequencies $\nu_i$ between the spectral parameters taken at two distinct orders $n$ and $m$. We find $\rho(\aiu^{\,\beta},\aid^{\,\beta})=-0.958$ and $\rho(\aiu^\T,\aid^\T)=-0.999$ between orders $n=1$ and $m=2$, and $\rho(\aiu^{\,\beta},\ait^\beta)=0.905$ and $\rho(\aiu^\T,\ait^\T)=0.998$ between orders $n=1$ and $m=3$. Despite changes in amplitude or sign from one order to another, the spectral gradients $\ain$ remain well correlated across the first, second, and third orders. This correlation is particularly strong for temperature fluctuations. Considering that the first order, which produces a perfect frequency-correlation, dominates the signal, we expect the sum 
$\sumn \asin\left(\Wsn-\mWsn\right)$
to remain very well correlated through frequencies $\nu_i$. Within the framework of our hypothesis \ref{HM3}, our model, even including orders higher than one in the moment expansion, predicts a perfect correlation of residual maps across frequencies if only the dust temperature varies, and a strong but not perfect correlation if only the dust spectral index varies. 
We therefore obtain the first prediction \ref{eq:P1} of our model:
\begin{align}
 \boxed{
 \rho_{ij}^\P(\R_i,\R_j)
 \simeq \rho_{ij}^P(\RP_i,\RP_j)
 \simeq \rho_{ij}^\psi(\Rt_i,\Rt_j) \simeq 1\,.
\label{eq:P1}\tag{\rm P1}
}
\end{align}

A word of caution is necessary. We are stating that the residuals maps at different frequencies are correlated with each other, not the maps of the dust signal themselves. As such, dust complexity creates frequency decorrelation in the sense of \cite{Planck2016-XXX}, meaning that the maps at different frequencies are no longer proportional. However, the complex distortions remain strongly correlated. This correlation can be understood geometrically as resulting from the rigid rotation of the complex polarized intensity in the complex plane (see Sect.~\ref{sec:rotatpsi} and Fig.~\ref{fig:schema}), which arises from beam and line-of-sight integration effects. 

\subsection{Properties of covariance matrices of residual maps}\label{sec:covar-2}

We now examine the properties of the covariances of residual maps across frequencies to provide clear predictions that can readily be compared with \Planck\ polarization data. 

Using our assumption of no correlation between the magnetic field structure and the fluctuations of dust properties (hypothesis \ref{HA2}), Eq.~\eqref{eq:defCov} yields $\Cov(\Isn,\Ism)\propto \Pms$. From Eq.~\eqref{eq:vareps} we therefore obtain the following scaling relation:
 \begin{align}
 \boxed{
 \vareps \propto \Pms \,.
\label{eq:P2}\tag{\rm P2}}
\end{align}
This prediction (\ref{eq:P2}) is natural: the variance of dust polarized emissivity scales with the square of the mean polarized intensity. 

To explore the properties of the variance of polarization angles $\varpsi$ and intensity $\varP$, we used the moment difference $\Mon = \Wn-\In$ (Eq.~\eqref{eq:Wbn-In-Mn}).
We defined its sky-averaged value as $\mMosn \equiv \left\langle P_0^2\Mosn\right\rangle/\Pms$. 
As $\Mon$ is statistically zero-mean (Sect.~\ref{sec:expansion} and Appendix~\ref{A-Nlayers}), so is $\Mosn-\mMosn$ so that Eq.~\eqref{eq:covP} and \eqref{eq:covpsi} become
\begin{align} 
\varPij & \simeq \sumallnm \asin\,\asjm\left(\Cov\left(\Isn,\Ism\right)+\Cov\left(\Re\Mosn,\Re\Mosm\right)\right)\,.\label{eq:covP-Mon}\\
\varpsiij &= \sumallnm \asin\,\asjm \,\Cov\left(\Im\Mosn,\Im\Mosm\right)\,.\label{eq:covpsi-Mon} 
\end{align}
Covariances of $\Re\Mon$ and $\Im\Mon$ involve the statistical properties of $\Mon$ that combine the fluctuations of the magnetic field and the optical properties of aligned grains within the light cone. In Appendix~\ref{A-Nlayers}, we present a toy model for a turbulent magnetic field close to equipartition, which successfully explains various statistical properties of dust polarization \citep{Planck18_XII}, and demonstrate that 
\begin{align}
 \left\langle \Re\Mon\Re\Mom\right\rangle\ &\simeq \mrat \left\langle \Im\Mon\,\Im\Mom\right\rangle \propto \frac{1}{\pref^2}\label{eq:ReDelta-eta}\,,
\end{align}
where $\pref = \Pref/\Iref$ is the polarization fraction at the reference frequency with $\Iref$ the total intensity, and $\mrat \simeq 0.7 \pm 0.2$ is a statistical parameter\footnote{The relatively large uncertainty in $\mrat$ accounts for the simplifications used in these estimations.} characterizing our turbulent magnetic field model. This property is the spectral analog of what is shown in \cite{Planck18_XII} for the spatial variations of polarization angles: dispersion in polarization angles anti-correlate with $\pref$.

In calculating the covariances $\Cov\left(\Im\Mosn,\Im\Mosm\right)$ through Eq.~\eqref{eq:defCov}, the scaling of $\Pref^2$ with $\pref^2$ exactly cancels the dependence of $\left\langle\Im\Mosn\times\Im\Mosm\right\rangle$ with $1/\pref^2$, yielding $\Cov\left(\Im\Mosn,\Im\Mosm\right) \propto \Ims$ and therefore
\begin{equation}
\boxed{
\varpsi 
\propto {\left\langle \Iref^2\right\rangle}\,. \label{eq:P3}\tag{\rm P3}
}
\end{equation}
Unlike $\vareps$, the variance $\varpsi$ of the polarization angle residuals is predicted, unexpectedly, to scale with the mean square of the total intensity in the mask, with no dependence on the mean polarization fraction. 

In the first term of the right-hand side of Eq.~\eqref{eq:covP-Mon}, we recognize, from Eq.~\eqref{eq:vareps}, the variance of dust emissivity $\varepsij$. The second term is, on average and according to Eqs.~\eqref{eq:covpsi-Mon} and \eqref{eq:ReDelta-eta}, proportional to $\varpsiij$, leading to the statistical relationship: 
\begin{align}
\boxed{
\varP \simeq \vareps + \mrat\,\varpsi\,.}\label{eq:P4}\tag{\rm P4}
\end{align}
The variance in polarized intensity $\varP$ not only contains the variance $\vareps$ of the dust polarization emissivity from one region of the sky to another, but also includes an additional variance due to distortions of the SED within the light cone, which can be estimated from the variance of the polarization angle residuals. Prediction \ref{eq:P4} implies that $\varP$, unlike $\vareps$, does not converge toward zero when the polarization fraction $\pref$ tends towards zero. It also suggests a recipe for estimating the unknown
variance $\vareps$ of the dust polarized emissivity:
\begin{equation}
 \hvareps \equiv \varP- \mrat\, \varpsi\label{eq:compute_vareps}\,.
\end{equation}
Since $\vareps\ge 0$, this implies an upper limit for the ratio $\varpsi/\varP$ (prediction \ref{eq:P5}):
\begin{align}
\boxed{
\frac{\varpsi}{\varP} \le \frac{1}{\mrat}\,.} \label{eq:P5}\tag{\rm P5}
\end{align}
We use this last property to test whether our hypothesis \ref{HM6}, which states that the signal is dominated by dust, holds, aside from contamination by CMB $E$-modes, noise, and systematics that we attempt to remove using simulations.

We conclude by analyzing the consequences for the covariances of the quasi-perfect correlation between the residual maps predicted in our model. Eqs.~\eqref{eq:covP} and \eqref{eq:covpsi} show that the frequency dependence of any covariance $\varij$ is encoded in the parameters $\ain$ and $\ajn$. These parameters are common to all covariances and remain well-correlated across frequencies, as shown in Sect.~\ref{sec:firstorder}.
This implies that the ratio between two distinct covariances sharing the same frequency pair is independent of the particular pair $(i,j$) of channels considered (prediction \ref{eq:P6}):
\begin{equation}
\boxed{
\frac{\varpsiij}{\varPij} = {\rm C^{te}}\hspace{0.2cm}\forall (i,j)\,.
\label{eq:P6}\tag{\rm P6}
}
\end{equation}
We conclude with our final prediction, a null test resulting from the strong correlation of the spectral gradients across frequencies. The matrix $\Im\,\varPcx$ (Eq.~\eqref{eq:ImvarPcxij}) involves the cross-product $\ain\,\ajm-\aim\ajn$, which is a small factor due to the strong correlation of the $\ain$ parameters across frequencies.
Hence, within the uncertainties, we expect (prediction \ref{eq:P7})
 \begin{align}
 \boxed{
\Im\,\varPcx \simeq 0\,.
}
\label{eq:P7}\tag{\rm P7}
\end{align}

In the following section, we compare our predictions with \Planck\ polarization data.

\section{Model validation using \Planck\ polarization data} \label{sec:application}

We applied our dust polarized emission model to analyze \Planck\ data. We extend the power-spectrum analysis presented by \citet{Ritacco2023} to our model framework by measuring the total variance of the residual maps at the map level. The \srolltwo\ version of the \Planck\ data is introduced in Sect.~\ref{subsec-Planck_data}, and its associated simulations, used to debias our estimations, are described in Sect.~\ref{subsec:debias}. In the subsequent sections, we present the mean complex polarized SED in Sect.~\ref{subsec:Mean_SED_Planck} and test our model predictions in Sect.~\ref{sec:checks}.

\subsection{Presentation of \Planck\ maps}
\label{subsec-Planck_data}

\citet{Ritacco2023} performed a power spectra analysis of the \Planck\ data to characterize spatial variations of the SED of dust polarized emission in terms of the $E$ and $B-$modes. We follow this analysis using \Planck\ full- and half-mission sky maps\footnote{\url{http://sroll20.ias.u-psud.fr/sroll20_data.html}} at 100, 143, 217, and $353$\GHz\ and end-to-end data simulations\footnote{ \url{http://sroll20.ias.u-psud.fr/sroll20_sim.html}} produced by the \srolltwo~software, which corrects the main systematic effects that impact polarization on large angular scales \citep{Delouis2019}. We applied the scaling factors $\tilde{\rho}_i$ at frequency $\nu_i$, listed in Table~1 of \citet{Ritacco2023}, which represent corrections to the polarization efficiencies. We also applied dust color corrections listed in this Table. 

We used {\tt HEALPix} pixelization \citep{healpix} at \Nside=32 (map pixel size of $1.8^{\circ}$). The resolution of the original maps was degraded applying a cosine filter in harmonic space \citep[see formula (1) in][]{Ritacco2023}. We followed the procedure described in Sect.~2.2 of \citet{Ritacco2023} to subtract the synchrotron polarized emission from the maps at 100 and $143$\GHz. We verified that uncertainties in the assumed spectral index of polarized synchrotron emission \citep[$\beta_S = -3.19\pm0.07$,][]{Ritacco2023} do not affect our results. We refer to the obtained maps as raw maps.

\begin{figure}
\includegraphics[width=\half,trim=0 1.5cm 0 1.5cm,clip]{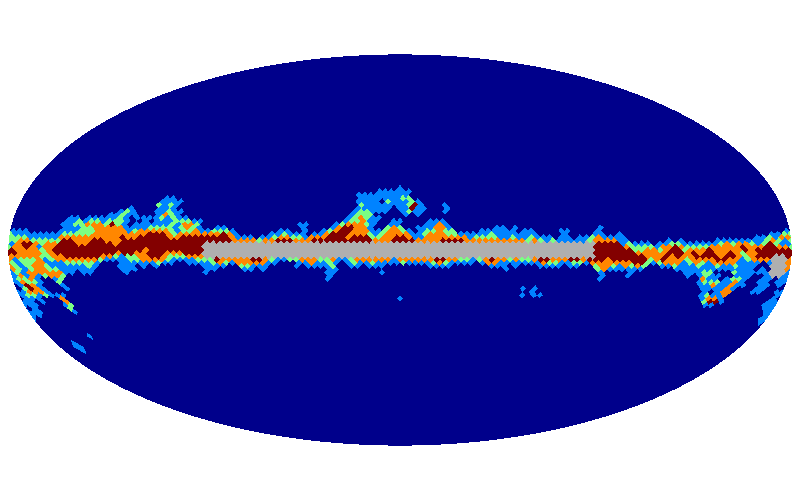}
\caption{{\tt HEALPix} masks with growing $\fsky$ ($\Nside=32$): 85\% (dark blue), 90\% (light blue), 92\% (green), 95\% (orange), and 97.3\% (brown). The 97.3\% mask additionally excludes the inner Galactic plane within 3 degrees of latitude and pixels within 4 degrees of the Crab pulsar (gray).}\label{fig:masks}
\end{figure}
We applied the complex polarization formalism of Sect.~\ref{sec:model} by computing the variances of residual maps. Our methodology follows that of \citet{Ritacco2023}.
{We used half-maps (two independent versions of the same map) to compute covariances, $\cvar_{\rm raw}$, for the \Planck\ data (see Appendix \ref{sec:half-maps}).}
Our results cannot be compared directly with those of \citet{Ritacco2023} because the transformation from $Q$ and $U$ to $E$ and $B$ is non-local. 
For most of our data analysis, we used a single mask representing $\fsky=97.3\%$ of the sky, similar to the 97\% mask of \cite{Ritacco2023}. Our mask at $\Nside=32$ (Fig.~\ref{fig:masks}) was obtained by removing pixels at Galactic latitude $|b| \le 3\deg$ in the inner Galaxy ($-90\deg \le l\le 90\deg$), as well as all pixels within $4\deg$ of the Crab pulsar. For certain plots, lower values of $\fsky$ were used (see Fig.~\ref{fig:masks}); these masks were obtained by removing pixels from the 97.3\% mask, ordered by decreasing optical depth, $\tau_{353}$, at 353\GHz.

In Section \ref{subsec:PR4}, we briefly comment on the comparison of our results obtained with \srolltwo\ maps and those obtained by applying the same methodology to the \Planck\ PR4 maps. The PR4 \Planck{} dataset is the latest and final \Planck{} data release, available from the \Planck{} Legacy Archive\footnote{\url{https://pla.esac.esa.int/}}. These \Planck{} maps were produced by applying the \npipe{} pipeline to the intensity and polarization data from both \Planck{}-LFI and \Planck{}-HFI \citep{npipe}, to create $I$, $Q$, and $U$ calibrated maps for each frequency band. The PR4 data release comes with a comprehensive set of ``end-to-end'' Monte Carlo simulations processed with \npipe{}, aimed at identifying and quantifying potential biases and the errors introduced by the pipeline.

\subsection{Data simulations, debiasing, and error-bars estimation}\label{subsec:debias}

To assess uncertainties associated with $Planck$ satellite noise, including systematics, and the CMB signal, we computed a set of simulated maps based on various models of dust polarization from \pysmthree\ \citep{Pysm,Zonca_2021,Pysm2025}. The \dmod\ model is built from the \Planck\ 353\GHz\ Legacy maps, extrapolated at lower frequencies with an MBB spectrum, using the spatially varying temperature and spectral index obtained from the \Planck\ intensity data with the \COMMANDER\ code \citep{Planck_2015_X}. The {\tt d10} model is a refined version of \dmod, in which the templates for amplitude and spectral parameters have been extracted from the GNILC analysis of the \Planck\ data \citep{Planck_XLVIII,Planck18_XII}, with smaller scales included using the so-called logarithm of the polarization fraction tensor formalism (for more details, see \cite{Pysm2025}). Finally, the {\tt d12} model consists of up to six superposed layers of modified black bodies, each with spectral parameters drawn from Gaussian distributions \citep{Martinez-Solaeche2018}.
The dust model maps were combined with the \srolltwo~simulations of data noise and systematics \citep{Delouis2019} and with CMB maps computed from the CMB power spectrum for the $\Lambda$CDM fiducial $Planck$ model \citep{planck2018VI} as described in Sect.~3.2 of \citet{Ritacco2023}. Overall, at each of the four \Planck\ frequencies, we obtained $\Nsims=200$ sets of simulated $Q$ and $U$ full and half-mission maps, with independent realizations of data uncertainties and CMB anisotropies. 
The simulations were treated identically to the \Planck\ data.

Our model is designed to study the frequency dependence of the variances, $\cvar$, of dust polarized emission. The \Planck\ data must be corrected for the bias introduced by noise, systematics, and the CMB in the raw data squared quantities, $\cvar_{\rm raw}$. For simulation number $k$, the bias introduced by noise, systematics, and the CMB on any squared quantity, $\cvar$, is estimated as
\begin{align}
\cvar_{\rm bias}(k) \equiv \cvar_{\rm sims}(k) - \cvar_{\rm dm}\,,\label{eq:bias}
\end{align}
where $\cvar_{\rm sims}(k)$ and $\cvar_{\rm dm}$ represent the covariances of simulation $k$ and of the chosen \pysmthree\ dust model, respectively.
Taking the raw biased data value $\cvar_{\rm raw}$, we obtain, for each simulation $k$, a debiased estimate, $\cdeb(k)$, for the covariance, $\cvar(k)$, in the \Planck\ \srolltwo\ data:
\begin{align}
\cdeb(k) \equiv \cvar_{\rm raw} - \cvar_{\rm bias}(k) = \cvar_{\rm raw} - \cvar_{\rm sims}(k) + \cvar_{\rm dm}\,.
\label{eq:debiais-cov}
\end{align}
The debiased value, $\cvar$, and its uncertainty, $\sigma_{\cvar}$, are taken as the mean and standard deviation of $\cdeb(k)$ over its 200 values, respectively.

From now on, we present only the results obtained using \dmod, our reference model. The results obtained using {\tt d10} and {\tt d12} are presented in the Appendix~\ref{A-otherdustmodels}; they do not differ significantly from those presented below, ensuring the robustness of our conclusions against the dust model used in our debiasing procedure.

\subsection{Mean complex polarized SED}\label{subsec:Mean_SED_Planck}

\begin{figure}
\includegraphics[width=\half]{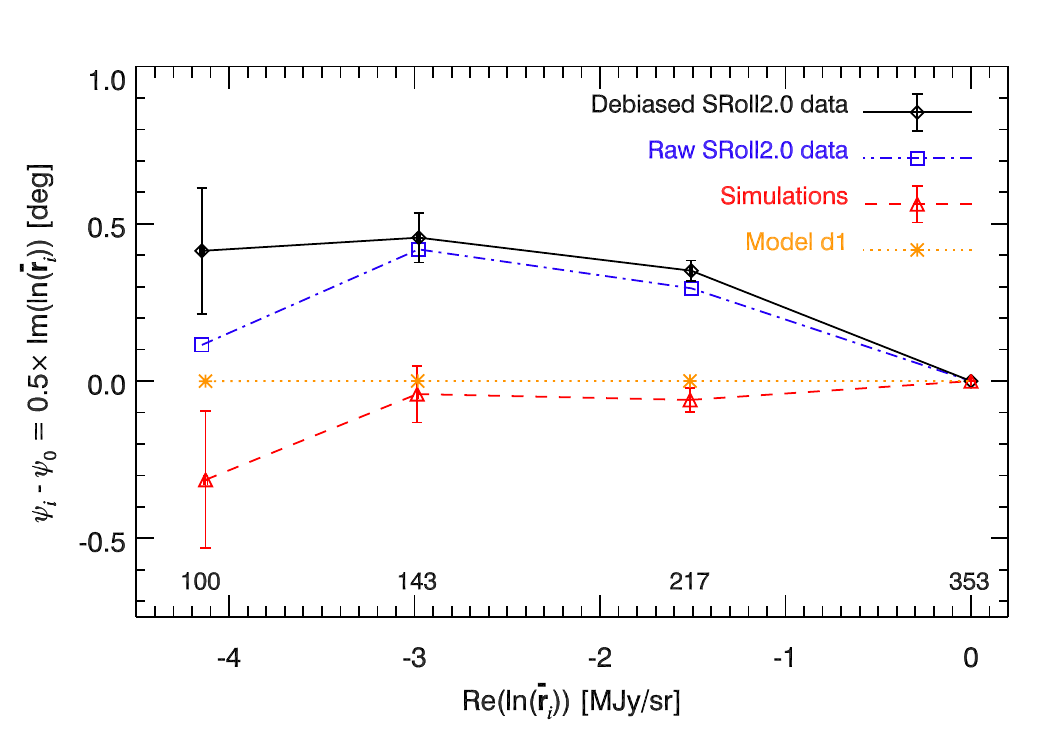}
\caption{Mean complex polarized SED $\mRP_i$ in the log complex plane for our 97\% mask: debiased \srolltwo~data (solid black line), raw \srolltwo~data (dashed-dotted blue line), simulations (mean shown by the dashed red line with error bars representing the standard deviation), and the model \dmod\ (dotted line).
Errors bars are based on 200 simulations (see Sect.~\ref{subsec:Mean_SED_Planck}). 
\Planck-HFI frequencies are indicated. 
}
\label{fig:meanSED}
\end{figure}

We extended earlier studies \citep{Planck2016-XXX,Planck18_XI,Ritacco2023} to the complex plane using the cross-correlation of \Planck\ maps, and determined the mean, debiased, complex SED $\mRP_i$ of dust polarization, normalized to our reference channel, 
\begin{equation}
\mRP_i \equiv 
\frac{\tilde{\rho}_i\tilde{\rho}_0\left\langle \P_i\P_0^\star\right\rangle_{\rm raw} 
-\frac{1}{\Nsims}\sum_{k=1}^\Nsims\left\langle \P_i\P_0^\star\right\rangle_{{\rm sim}\,k}
+\left\langle \P_i\P_0^\star\right\rangle_{\rm dm} }
{\tilde{\rho}_0^2\left\langle \P_0\P_0^\star\right\rangle_{\rm raw}-\frac{1}{\Nsims}\sum_{k=1}^\Nsims\left\langle \P_0\P_0^\star\right\rangle_{{\rm sim}\,k}+\left\langle \P_0\P_0^\star\right\rangle_{\rm dm}} 
\,,
\label{eq:mRPi_Planck}
\end{equation}
where $\tilde{\rho}_i$ is the correction to the polarization efficiency\footnote{There is no correction for the 353\GHz\ channel ($\tilde{\rho}_0=1$). The relative uncertainty on $\tilde{\rho}_i$ is $\sigma_{\tilde{\rho}}/\tilde{\rho} = 0.5\%$ for all channels \citep{Ritacco2023}. 
} 
at frequency $\nu_i$ \citep[][Table~1]{Ritacco2023}.
Similarly, we computed $\rP_i^{\rm dm}\equiv\left\langle \P_i\P_0^\star\right\rangle_{\rm dm}/\left\langle \P_0\P_0^\star\right\rangle_{\rm dm}$ for the dust model, $\rP_i^{\rm sim}(k)\equiv\left\langle \P_i\P_0^\star\right\rangle_k/\left\langle \P_0\P_0^\star\right\rangle_k$ for simulation $k$, and $\mRP_i^{\rm sims} \equiv \sum_{k=1}^\Nsims\left\langle \P_i\P_0^\star\right\rangle_k
/\sum_{k=1}^\Nsims\left\langle \P_0\P_0^\star\right\rangle_k$ for the mean of all simulations.

The uncertainty on the real and imaginary components of the mean simulated SED $\mRP_i^{\rm sims}$ is taken as the standard deviation of the corresponding real and imaginary components of the 200 simulation values $\rP^{\rm sim}_i(k)$.
The uncertainty in the observed SED $\mRP_i$ must also include the uncertainty in the recalibration factor, $\tilde{\rho}_i$, of the \Planck\ polarization data: $\sigma(\tilde{\rho_i})/\tilde{\rho}_i = 0.5\%$. We repeated the procedure described in Sect.~\ref{subsec:debias} 200 times, each time drawing a distinct recalibration factor for each channel, $i$, from a random Gaussian distribution of expectation $\tilde{\rho}_i$ and standard deviation $\sigma(\tilde{\rho}_i)=0.005\tilde{\rho}_i$. We debiased this realization with simulation $k$ alone to compute the $k^{\rm th}$ debiased estimate $\hrP_i(k)$ of $\mRP_i$. The uncertainty in the real and imaginary parts of $\mRP_i$ is then taken as the standard deviation of the real and imaginary components of this set of 200 values $\hrP_i(k)$, respectively. To test the assumption made in Sect.~\ref{subsec:residuals} to obtain Eq.~\eqref{eq:Rcxi-ai}, we computed the module of $\mRP_i\frac{\meps_0}{\meps_i} -1$ for a pivot SED, $\meps_i$, fitted to $\mRP_i$, and found it smaller than 0.03 at all frequencies.

Figure~\ref{fig:meanSED} shows the spectral dependence of the mean complex polarized SED in the log complex plane, for the raw \srolltwo\ data, the mean of simulations, the \dmod\ dust model used for debiasing and the resulting debiased SED $\mRP_i$. We observe a variation of the polarization angle with frequency in the data, which may appear significant.
However, our estimate for the uncertainty on $\mRP_i$ does not include the uncertainty in the orientation of the \Planck\ polarization bolometers. The most stringent estimates of these uncertainties come from the analysis of $EB$ cross-spectra carried out to constrain cosmic birefringence \citep{PIPXLIX_birefringence,Minami20,Diego2022BiRe,CosmicBirefringence}. \citet{Diego2022BiRe} list their results in Table~1, per bolometer for several Galactic masks. The scatter among Galactic masks and the statistical uncertainties both lie in the range 0.2 to $0.3^\circ$. This agrees with the systematic uncertainty of $0.28^\circ $ quoted by \citet{PIPXLIX_birefringence}. The magnitude of this uncertainty reduces the significance of the imaginary component of our mean SED.

While the mean complex SED, $\mRP_i$, depends on calibration errors in amplitude (correction factors $\tilde{\rho}_i$) and in phase (absolute orientation of bolometers in the focal plane), the definition of map residuals (Eq.~\eqref{eq:defRcxi}), involving the ratio of $\P_i$ to $\mRP_i$, renders the map residuals, and therefore their covariances, independent of any calibration error, whether in amplitude or in phase. This facilitates the comparison of different frequency maps or datasets with distinct systematics.

\subsection{Comparison of model predictions with \Planck\ polarization data}\label{sec:checks}

In this section, we compare our model predictions and hypotheses with the \Planck\ \srolltwo\ data.
Using simulations, we have verified that the 0.5\% in the recalibration factors $\tilde{\rho}_i$ does not have a noticeable effect on our results.

\begin{table} 
\caption{Test of prediction \ref{eq:P1}.}
\begin{tabular}{l l l l l}
\hline\hline
$(\nu_i,\nu_j)$ [GHz] & (217,143) & (217,100)& (143,100) \\
\hline
$\rho^P_{ij}$ & $1.04\pm0.06$ & $0.94\pm0.14$& $1.00\pm0.16$ \\
$\rho^\psi_{ij}$ & $0.8\pm0.2$ & $0.6\pm0.2$& $0.3\pm0.3$ \\ 
$\rho^\P_{ij}$ &$0.99\pm0.08$ & $0.81\pm0.13$& $0.77\pm0.17$ \\
\hline
\end{tabular}
\tablefoot{Values of the Pearson coefficient $\rho_{ij} = \varij/\sqrt{\varii\,\varjj}$ between all residual maps of the same kind $\varP$, $\varpsi$, and $\Re\,\varPcx$, for the 97.3\% mask. Error bars are from the simulations.}
\label{tab:rho}
\end{table}

First, we tested model prediction \ref{eq:P1} by computing the Pearson correlation coefficient $\rho_{ij}=\cvar_{ij}/\sqrt{\cvar_{ii}\,\cvar_{jj}}$ , between the residual maps $i$ and $j$. 
Our results for the $\R$, $\RP$, and $\Rt$ residual maps are presented in Table~\ref{tab:rho} for all frequency pairs, using our 97.3\% mask. The error bars on $\rho_{ij}$ are calculated from the standard deviation of the simulated values $\rho_{ij}(k)$, following the approach described in Sect.~\ref{subsec:debias}.
We find that the correlation coefficient between the $\RP$ residual maps is consistent with a perfect correlation. 
The correlation between the angle residual maps, $\Rt$, is strong for the (217,143) pair, but weak for the (143,100) and (100,217) pairs, although with large uncertainties. Therefore, prediction \ref{eq:P1} is validated for $\RP$ but not for $\Rt$. The correlation coefficient of $\Re\,\varPcx$ is high but only marginally compatible with 100\% for the (217,100) and (143,100) pairs. In the context of Eq.~\eqref{eq:Pythagore}, this loss of correlation between the residual maps $\Rcxi$ is a consequence of the loss of correlation between the residual maps $\Rti$. 

\begin{figure}
\includegraphics[width=\half]{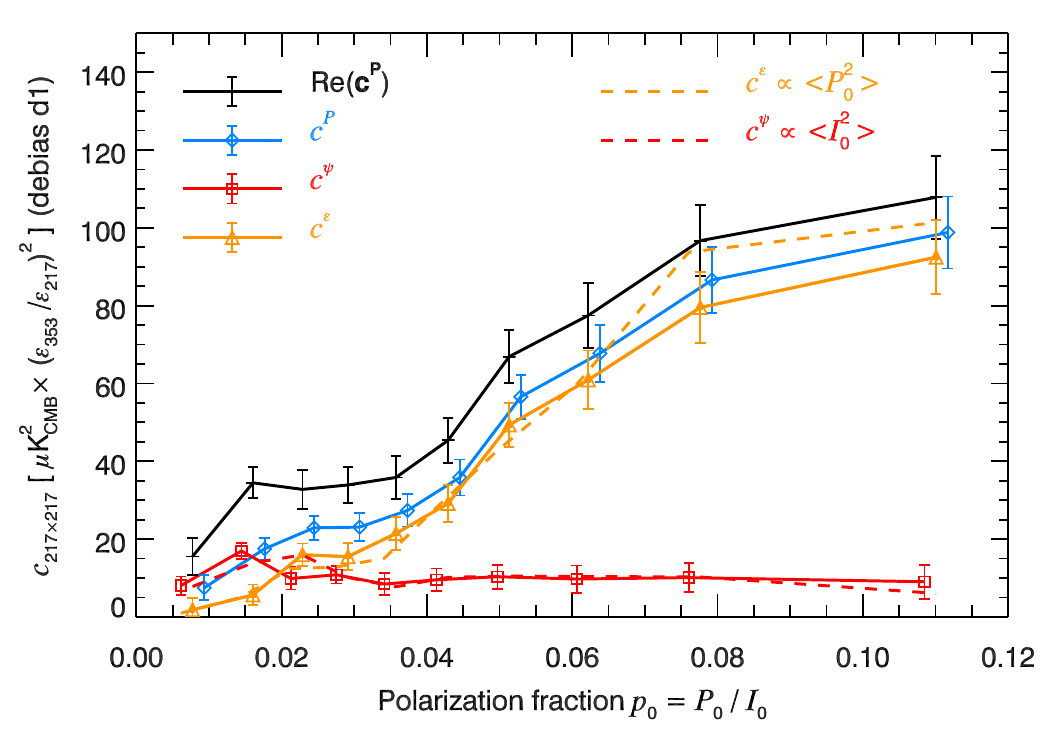}
\caption{Test of predictions \ref{eq:P2}, \ref{eq:P3}, and \ref{eq:P4} : dependence on the observed polarization fraction $\pref=P_0/\Iref$ at 353\GHz\ of the variances $\Re\,\varPcx$, $\varP$, $\varpsi$, and $\vareps$ computed at 217\GHz. 
Our models for $\vareps$ \eqref{eq:P2} and $\varpsi$ \eqref{eq:P3} are overplotted as dotted orange and red lines, respectively, using the mean value of the slope derived from these equations, multiplied by $\Pms$ and $\Ims$ in each bin of $\pref=P_0/\Iref$. 
}
\label{fig:cov_p}
\end{figure}

Second, we examined the validity of predictions \ref{eq:P2} and \ref{eq:P3}, which relate $\vareps$ and $\varpsi$ to $\Pms$ and $\Ims$, respectively. Fig.~\ref{fig:cov_p} shows how the variances, computed at 217\GHz\ for a high S/N ratio, depend on the polarization fraction, $\pref$, at 353\GHz. This correlation plot is obtained by binning pixels of the 97.3\% mask in bins of $\pref$ and computing, for each $\pref$ bin, the variances $c$ at 217\GHz and the mean squared values, $\Pms$ and $\Ims$ (a method only applicable at the map level). A model, $\vareps \propto \Pms$, shown as a dashed orange line, confirms our prediction \ref{eq:P2}. 
Another model, $\varpsi \propto \Ims$ (dashed red line), also confirms the surprising prediction \ref{eq:P3} of our model: $\varpsi$ is reasonably proportional to $\Ims$, rather than $\Pms$. 
Consequently, while $\vareps$ follows $\Pms$ as the polarization fraction tends toward zero, $\varP$ and $\Re\,\varPcx$ saturate at positive values (prediction \ref{eq:P4}) due to depolarization effects (the variance of which is quantified here by $\varpsi$) that persist even at low $\pref$. 

\begin{figure}
\includegraphics[width=\half]
{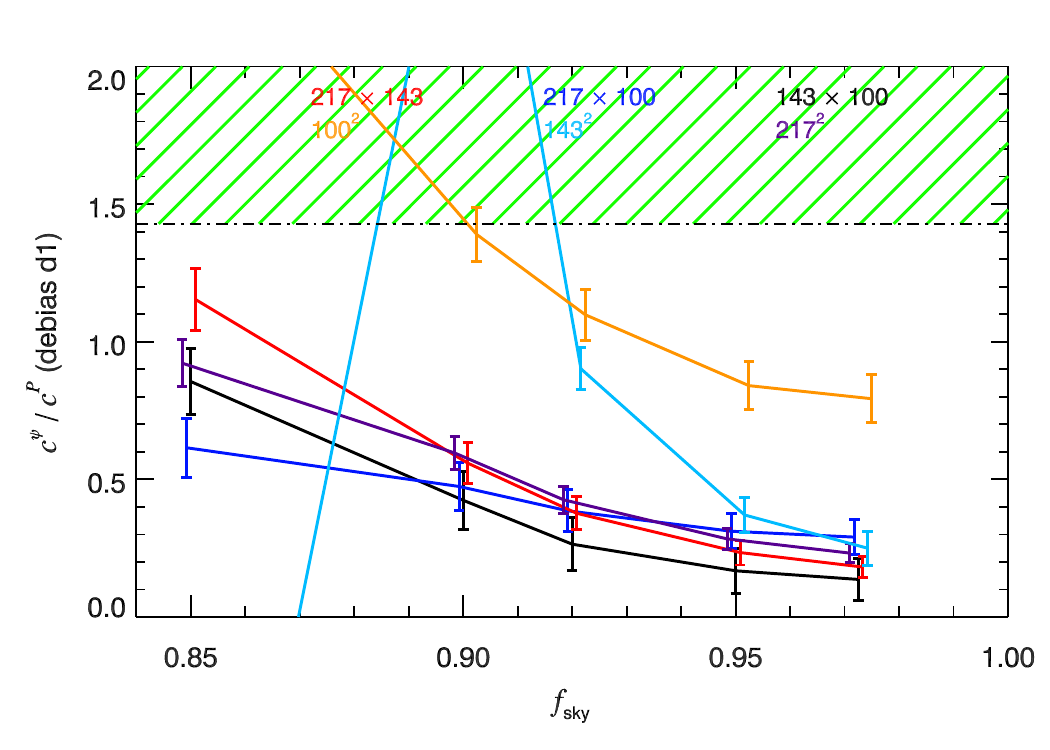}
\caption{
Test of predictions \ref{eq:P5} and \ref{eq:P6} as a function of $\fsky$. 
The hashed region represents forbidden values according to \ref{eq:P5}. 
}
\label{fig:checkratio}
\end{figure}

Predictions \ref{eq:P5} and \ref{eq:P6} are compared to \Planck\ \srolltwo~data in Fig.~\ref{fig:checkratio}, which shows how the ratio $\varpsiij/\varPij$ varies with $\fsky$, for all pairs ($i,j$) of residual maps. The error bars for this ratio are computed from simulations, following the approach described in Sect.~\ref{subsec:debias}. 
Prediction \ref{eq:P5} ($\varpsi/\varP\le 1/\mrat$, unhashed region) is verified for all frequency pairs except for the variances at 100 and 143\GHz\ at lower $\fsky$. This indicates that either the hypotheses of our model or the debiasing method break down at lower $\fsky$. For the 95\% and 97.3\% masks, the $\varpsi/\varP$ ratio is, within error bars, almost independent of the frequency pair considered (prediction \ref{eq:P6}), with the notable exception of the variance at 100\GHz. 

Overall, Table~\ref{tab:rho} and Fig.~\ref{fig:checkratio} highlight a limitation of our approach for the 100\GHz\ channel. In Sect.~\ref{sec:discussion}, we discuss to what extent this distinctive behavior can be attributed to dust physics beyond our hypotheses, to residual contamination by a component other than dust and CMB (e.g., synchrotron or CO emission), or to limitations of our debiasing method.

 \begin{table} 
\caption{Test of prediction \ref{eq:P7}.}
 \begin{tabular}{l l l l l}
 \hline\hline
$(\nu_i,\nu_j)$ [GHz] & (217,143) & (217,100) & (143,100) \\
 \hline
 $\Im\,\varpsiPij \left(\unitij\right)$ & $-19\pm5$ & $-11\pm12$& $16\pm28$ \\ 
 \hline
 \end{tabular}
 \tablefoot{Values of $\Im\,\varPcxij$ for the 97.3\% mask and the three pairs of \Planck\ frequencies.}
 \label{tab:P7}
 \end{table}

Finally, in Table~\ref{tab:P7}, we test prediction \ref{eq:P7}, which states that $\Im\,\varPcx_{ij}$ should be, within the uncertainties, close to zero for all frequency pairs. This prediction is not verified, even for the frequency pair (217,143) for which the residual maps show the strongest correlation (see Table~\ref{tab:rho}). This result is puzzling, as our model does not predict any correlation between $\Rt$ and $\RP$, except for chance correlation. This issue is examined in more detail in the following section.

\section{Inferring the variance of the spectral properties of aligned grains}
\label{sec:dust}

Polarization observations trace the spectral properties of a specific population of dust grains that are aligned with the magnetic field lines. They can be compared with the overall spectral properties inferred from dust far-infrared (FIR) and submm total emission, which trace all grains, both aligned and unaligned. 
Assuming a common temperature\footnote{Lacking high-frequency channels in polarization, the temperature of aligned grains could not be measured from \Planck\ polarization data.} for aligned and unaligned grains, \cite{Planck18_XI} found close values for the mean spectral index in polarization and intensity: $\beta_P - \beta_I = 0.05 \pm 0.03$ in the high-latitude diffuse ISM. In this section, we extend this work by quantifying the variance of the fluctuations in the spectral properties of aligned grains for the 97.3\% mask, where the signal is dominated by the Galactic plane. 
We trace these variations back to their origin in the fluctuations of $\beta$ or $T$, and correlate them with the spectral fluctuations observed in total intensity.

\subsection{Disentangling fluctuations of dust temperature and spectral index}\label{sec:Tbeta}

\begin{figure}
\includegraphics[width=\half]{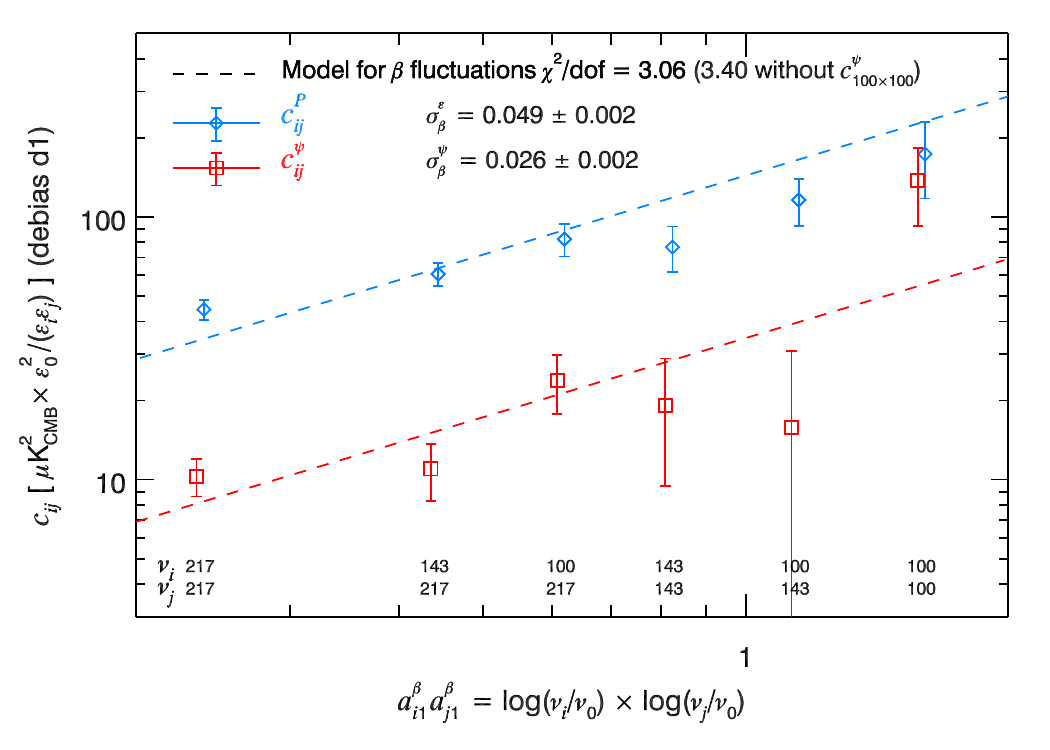}
\includegraphics[width=\half]{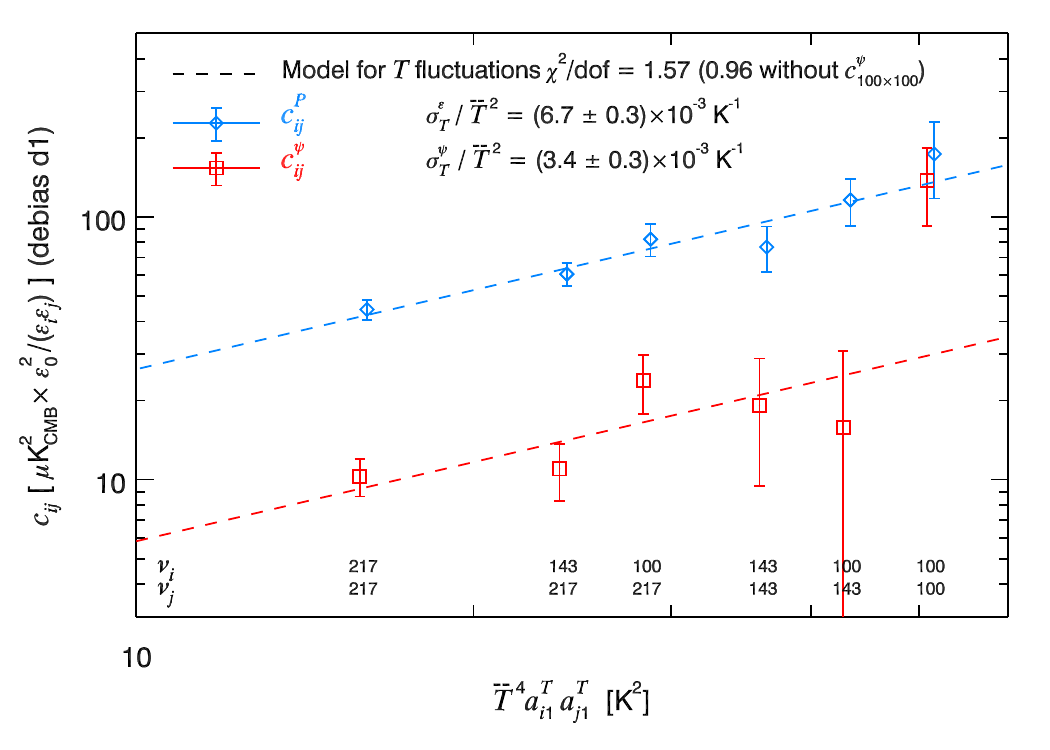}
\caption{Comparison of \srolltwo~data with the predictions of our model 
for $\varP$ (blue) and $\varpsi$ (red), for pure $\beta$ fluctuations \tlab, and pure $T$ fluctuations \blab. 
}
\label{fig:varbT}
\end{figure}

As demonstrated in Sect.~\ref{sec:firstorder}, the spectral gradients $\asin$ for $n=2$ and $n=3$ are 
strongly correlated with $\aio^s$. 
By correlating the covariance matrices $\varPij$ and $\varpsiij$ with $\aio^{\,\beta}\ajo^{\,\beta}$ and $\mT^4\aio^\T\ajo^\T$ separately\footnote{While $\abun$ is independent of $\mbeta$, $\atun$ depends on $\mT$. The expression $\mT^2\atun$ depends only weakly on $\mT$, with less than 10\% variation when $\mT$ spans a range between 15 and 25\,K.}, we aimed to identify whether these  fluctuations originated mainly from variations in $\beta$ or $T$, while simultaneously estimating their amplitude. 
This analysis is illustrated in Fig.~\ref{fig:varbT} for pure $\beta$ fluctuations (labeled \tlab) and pure $T$ fluctuations (labeled \blab). The spectral trends observed for $\varP$ and $\varpsi$ are not compatible with a scenario in which they originate from fluctuations in $\beta$, as this scenario predicts a steeper slope. 
A better agreement is obtained if only the dust temperature $T$ fluctuates. In the Galactic plane, starlight heating may indeed dominate the fluctuations of dust emission properties. For the variance in angles ($\varpsi$), fluctuations in $T$ may also be favored, although the significant contamination observed in the 100\GHz\ residual angle map prevents a firm conclusion.
To quantify these trends, we performed a $\chi^2$ minimization using the ${\tt MPFITCOVAR}$ IDL routine \citep{MPFIT2009}. Our data combine the two data sets, $\varPij$ and $\varpsiij$, into an array of 12 values. The corresponding $12\times12$ noise covariance matrix was computed from the covariances between the 200 simulated $\varPij(k)$ and $\varpsiij(k)$ covariance matrices (see Sect.~\ref{subsec:debias}). Using predictions \ref{eq:P2}, \ref{eq:P3}, and \ref{eq:P4}, we define a linear model for fluctuations in the spectral parameter $s\in\{\T,\beta\}$, where $\varpsiij = \aio^s\ajo^s\,\left(\varspsi\right)^2\Pms$ and $\varPij = \aio^s\ajo^s\,\left(\varseps\right)^2\Pms + \mrat\,\varpsiij$, with $\varspsi \equiv \sqrt{\Cov\left(\Im\Wso,\Im\Wso\right)}$ and $\varseps \equiv \sqrt{\Cov\left(\Iso,\Iso\right)}$.
When fitting with pure fluctuations in $T$, we obtain a $\chi^2$ of $15.7$ (or $1.57$ per degree of freedom). For pure $\beta$ fluctuations, $\chi^2$ increases to 31. Removing $\varpsi_{100\times100}$ from the dataset clearly improves the fit for fluctuations in $T$ ($\chi^2/$dof = 0.96), but not for fluctuations in $\beta$. Changing the value of $\mrat$ does not affect the $\chi^2$ of the fit' it modifies only the value of $\varseps$, while keeping $\left(\varseps\right)^2+\mrat\left(\varspsi\right)^2$ constant. 
Our best fit provides a reasonable value for the typical dispersion in dust temperature in the Galactic plane, corresponding to $\varTeps=(2.7\pm0.1)\,$K for a reference temperature $\mT=20\,$K.

\begin{figure}
\centering
\includegraphics[width=\half]{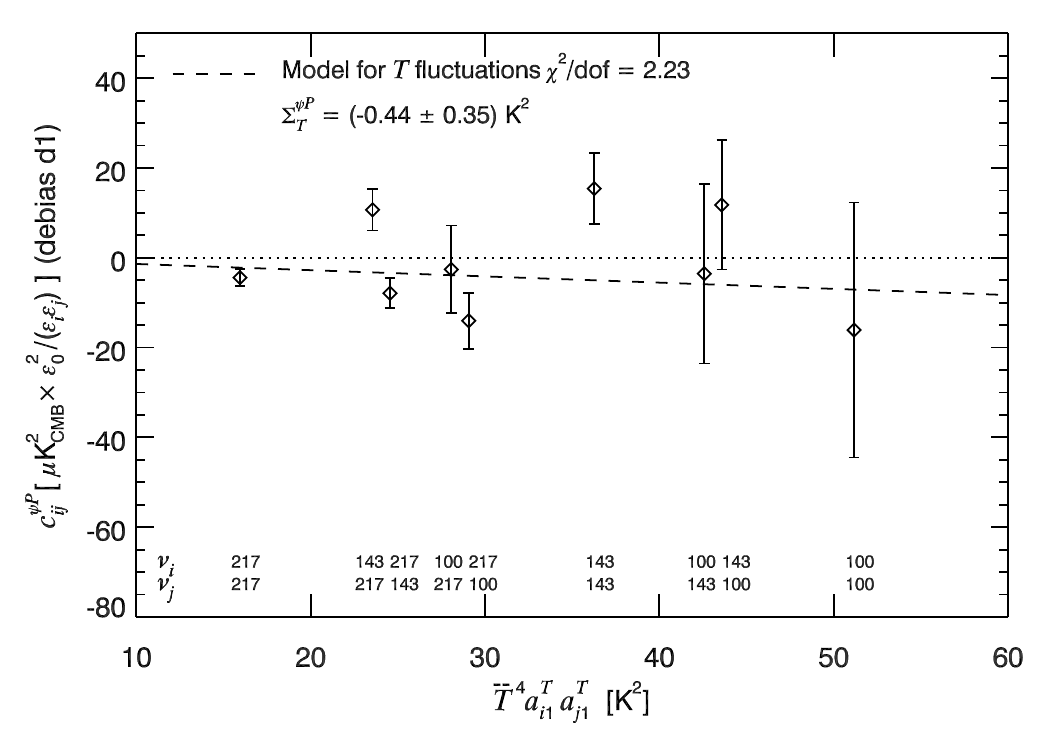}
\caption{Covariances $\varpsiPij$ compared to a first-order model (dashed line) based on fluctuations of the dust temperature alone.}
\label{fig:cov_psiP}
\end{figure}
Finally, Fig.~\ref{fig:cov_psiP} shows the frequency dependence of the mixed covariances $\varpsiPij$ (Eq.~\eqref{eq:covpsiiPj}). This graph confirms that the 143\GHz\ channel behaves differently from the 217 and 100\GHz\ channels, possibly indicating contamination by CO emission in these two channels (see Sect.~\ref{sec:discussion}). A first-order model (dashed line) based on fluctuations of the dust temperature, $\varpsiPij = \aio^\T\ajo^\T \,\varTpsiP\left\langle P_0^2\right\rangle$, with $\varTpsiP \equiv \Cov\left(\Im\Wto,\Re\Wto\right)$, cannot reproduce the observations. A model based on fluctuations in $\beta$ also fails. 
The frequency dependence of the mixed covariances $\varpsiPij$, together with the values of $\Im\,\varPcx_{ij}$ (Table~\ref{tab:P7}), warrants further detailed analysis. We defer this to future work.

\subsection{Correlating fluctuations in intensity and polarization}

\begin{figure}
\centering
\includegraphics[width=\half]{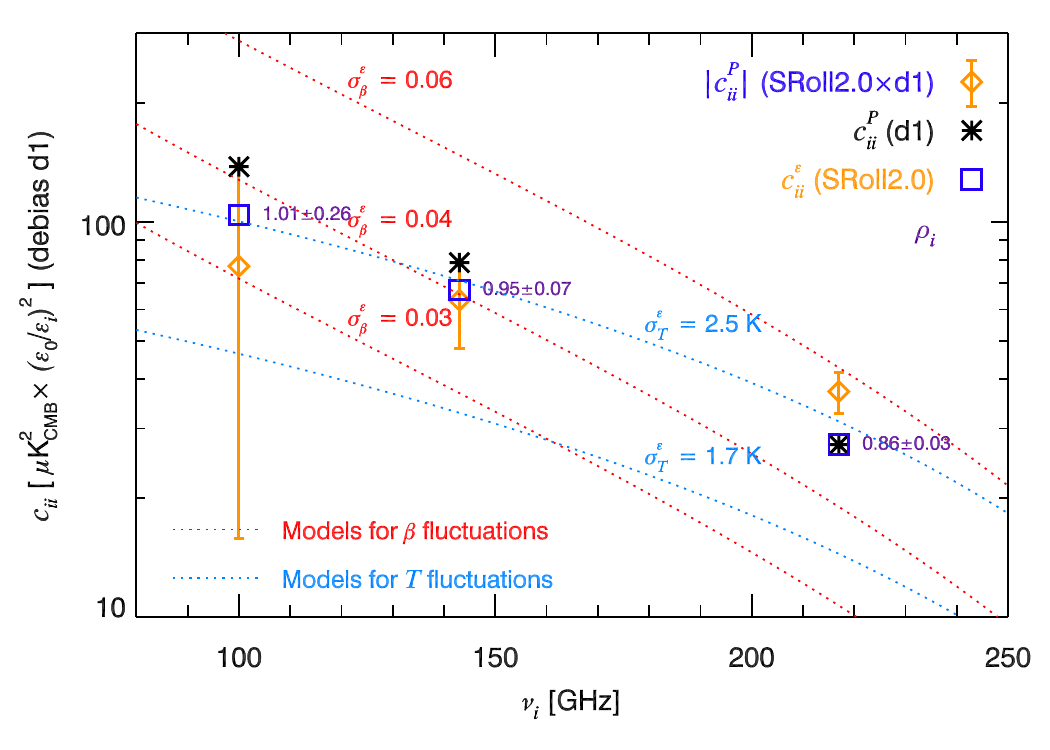}
\caption{Comparison of the spectral dependence of covariances $\vareps$ from \srolltwo~data, $\varP$ for the dust model \dmod, and $\varP$ for the cross-product of \srolltwo\ and \dmod. The Pearson correlation coefficient $\rho(\dmod,\sr)$ is indicated (purple). Predictions for models with $T$ or $\beta$ fluctuations are overplotted.}
\label{fig:PvsI}
\end{figure}

We next correlated the fluctuations observed in the dust polarized emissivity
with the fluctuations observed in total intensity. We used the \dmod\ dust model, based on \COMMANDER\ total intensity maps, as a proxy for the dust emissivity fluctuations in total intensity. This model does not include any significant decorrelation in angles (see Sect.~\ref{subsec-Planck_data}), which implies that $\Rt(\dmod) \simeq 0$ and $\varP(\dmod) \simeq \vareps(\dmod)$. Correlating $\RP(\sr)$ from the \srolltwo\ data with $\RP(\dmod)$ from the \dmod\ dust model therefore quantifies the variance of dust emissivity common to both residual maps, independently of the variance in angles present in the \Planck\ \srolltwo\ maps $\RP(\sr)$. Figure \ref{fig:PvsI} compares the dependence with the frequency $\nu_i$ of the variance $\vareps_{ii}(\sr)$ of the \srolltwo~data, of the variance $\vareps_{ii}(\dmod)=\varP_{ii}(\dmod)$ of the \dmod\ dust model, and their cross-covariance,
\begin{equation}
\varP_{ii}(\dmod\times\sr) \equiv \left\langle \RPi(\dmod)\times\RPi(\sr)\right\rangle \,.
\end{equation}
Fig.~\ref{fig:PvsI} shows that the variance of emissivity of aligned grains (as traced by $\vareps(\sr)$) and that of all grains (aligned or unaligned, traced by $\varP(\dmod)$), are of compatible amplitude, except at 217\GHz. The $\srolltwo$ polarization data follow a $T$ trend, while the $\dmod$ intensity data lie between the pure $T$ and pure $\beta$ trends. The data-model correlation coefficients, 
\begin{equation}
\rho_i^\varepsilon(\dmod\times\sr)=\varP_{ii}(\dmod\times\sr)\,/\,\sqrt{\vareps_{ii}(\sr)\times\varP_{ii}(\dmod)},
\end{equation}
are relatively high (above 85\%) at all frequencies. However, the value at 100\GHz\ must be interpreted with caution, as the value derived for $\vareps_{100\times100}$ through Eq.~\eqref{eq:compute_vareps} is underestimated due to contamination by $\varpsi_{100\times100}$, as observed in Fig.~\ref{fig:varbT}. Overall, this indicates that the fluctuations observed in the total SED of a given light cone correspond closely to the fluctuations present in the polarized SED of the same light cone. 

\subsection{Results obtained with the \Planck\ PR4 data release}\label{subsec:PR4}

In Appendix~\ref{A-PR4}, we present figures and tables obtained using the same methodology but applied to the last official release of \Planck\ data, PR4. Our goal was not to show a preference for one data release over another, but to demonstrate that our methodology can  quantify differences between datasets and track their origin. Despite the clear differences in the mean complex SED, our assessment of the seven predictions remains the same. The main difference between these two data releases appears in the spectral dependence of covariances. The PR4 and \srolltwo\ covariances, $\varP$ and $\varpsi$, are compatible within uncertainties, except for $\varP_{217\times217}$ and $\varpsi_{100\times100}$, possibly indicating a distinct handling of contamination by CO emission. The PR4 and \srolltwo\ $\varpsiP$ mixed covariances are compatible for all  channel pairs.
Contamination of the 100\GHz\ channel appears to be significant in $\varP$ in the PR4 data, unlike in $\srolltwo$. 
Finally, while the $\srolltwo$ data shows a preference for pure fluctuations in $T$, PR4 data, with $\chi^2/{\rm dof}\simeq 14$, are not compatible with a dust model assuming either pure fluctuations in $T$ or in $\beta$, even when excluding$\varpsi_{100\times100}$ and $\varP_{100\times100}$. Our methodology reveals clear differences between the $\srolltwo$ and PR4 data, particularly in the mean properties and variances of the 217 and 100\GHz\ channels contaminated by CO leakage.

\subsection{Dependence on the dust model used for debiasing}
The influence of our debiasing method (Sect.~\ref{subsec:debias}), applied with the $\dmod$ dust model, is tested in Appendix~\ref{A-otherdustmodels} using the {\tt d10} and {\tt d12} \pysmthree\ dust models. This effect is weak and does not alter our conclusions. However, the fact that no dust model includes a frequency-rotation of polarization angles at the pixel level (with the exception of {\tt d12}, which does not appear to behave realistically, as indicated by the negative Pearson coefficients in Fig.~\ref{fig:othermodels-PvsI}) is clearly problematic. One of the goals of this article is to propose new diagnostics for developing such models. Coupling these diagnostics with model-building tools based on moment expansion, as proposed by \cite{Vacher2025}, could provide a powerful combination to provide new, realistic, and complex foreground models anchored in data.

\section{Discussion}\label{sec:discussion}

Our model, based on simple hypotheses, 
makes strong predictions and testable predictions. 
Its success enables the measure of the variance in the spectral properties of aligned grains.
Its limitations provide new insights into the origins of the imperfect correlation observed in polarization angle residual maps. 

\subsection{insights into the properties of aligned grains \label{sec:prop-al-grains}}

Correlating fluctuations in the polarized SED with those observed in total intensity, we find a strong, though not perfect, correlation ($\rho > 85\%$; Fig.~\ref{fig:PvsI}). This behavior is expected. What we call dust grains are actually a set of grain populations with distinct composition, optical properties, and alignment properties \citep{Shiu2024,DarkDustII}.
Typically, small grains and possibly unaligned large carbonaceous grains do not emit in polarization \citep{KM95,Chiar2006}. Maps of total intensity collect the thermal emission from all grains, whether aligned or not with the magnetic field. Because the impact of magnetic fields on fluctuations in the polarized SED is statistically mitigated within our methodology, the strong correlation between intensity and polarization residual maps leads us to conclude that the emission from aligned grains is highly dominant in \Planck\ channels, compared with that from unaligned grains.

As a second result on dust properties, we find that the frequency-dependence of the variance of these fluctuations is consistent only with variations in dust temperature $T$, showing no indication of a variation in the dust spectral index $\beta$. This result may be specific to the low Galactic latitudes that dominate our signal, where the light cone samples the systematic increase in the radiation-field intensity toward the molecular, star-forming ring of the Galaxy. However, we do not draw firm conclusions from this preliminary result. Firstly, it is at odds with those drawn from analyses of FIR and submm total-intensity SEDs \citep{Ysard2013,Planck13_XI,Planck2014dust} and from a recent evolution of dust models \citep{THEMISII}, and secondly, because it is not confirmed using the PR4 version of the \Planck\ data (see Sect.~\ref{subsec:PR4}). This apparent contradiction requires further investigation.

\subsection{Investigating the origin of the imperfect correlation of polarization angle residuals \label{sec:disc-decor-ang}}
Our model predicts a perfect correlation between the polarization SEDs and angles residuals (but not of the polarized SEDs themselves).
Analyzing the \Planck\ \srolltwo\ data with this model, we find that fluctuations of the submm polarized SED are indeed strongly correlated ($\rho^P > 90\%$; Table~\ref{tab:rho}) across frequencies for the polarized intensity, $P_\nu$, but only weakly correlated for the polarization angles, $\psi_\nu$. The level of correlation between residual maps is lower with PR4, but remains similar (see Appendix~\ref{A-PR4}). This raises the question of which physical processes or components could be responsible for the weaker-than-expected correlation observed between the polarization angle residual maps.

Dust itself could be the origin of part of this loss of correlation. 
First, we restricted our MBB modeling to the simple situation (hypothesis \ref{HM3}), where only $T$ or $\beta$ vary in the sky, not both. Observations show that, in total intensity, both $T$ and $\beta$ vary within any given region of the sky. The possible correlation between $T$ and $\beta$ has also been the subject of a long series of studies \citep{Dupac2003,Anderson2010,Planck_XXV,Planck13_XI}. All of these studies are based on intensity rather than polarization maps, which motivated our simple hypothesis \ref{HM3}. Nevertheless, the impact of  simultaneous variations, possibly correlated or anti-correlated, of $T$ and $\beta$  on our analysis remains to be explored. 
Second, the predicted 100\% correlation between frequencies for fluctuations in $T$ derives from the law of blackbody thermal emission and is therefore as strict as a physical law. 
In contrast, fluctuations in $\beta$ represent a simplification of what may truly occur. This simplification assumes that variations in the FIR and submm dust emissivity, $Q_{\rm abs}(\nu)$, are perfectly correlated across frequencies. Laboratory data do not support this assumption \citep{Demyk2022}. The prediction of a 100\% correlation for fluctuations in $\beta$ is therefore not derived from solid physics, but from our simplifying hypothesis \ref{HA1}, the MBB model. 
Finally, through the same hypothesis \ref{HA1}, our study neglects refinements of dust models, such as variations in dust cooling efficiency \citep{THEMISII}, the magnetic dipole emission of dust grains \citep{DraineHensley2013}, which may produce polarized emission around 100\GHz\ with a polarization direction perpendicular to the magnetic field, or the polarized emission by a population of very large (up to $1\,\mu$m) grains that may have distinct dust properties \citep{Siebenmorgen2023}. 
In summary, the predicted 100\% correlation of map residuals between frequencies strongly relies on our hypothesis \ref{HA1}, and is therefore questionable.

Within the framework of our hypothesis \ref{HM6}, we also ignore any component other than dust, the CMB, and \Planck\ noise and systematics. 
The removal of synchrotron polarized emission from the 143 and 100\GHz\ maps (see Sect.~\ref{subsec-Planck_data}) may have left some residuals. There is also a significant component that is not included in the simulations, as it is difficult to model: the possible leakage of CO line emission into polarized Stokes maps $Q$ and $U$, in the 100 and 217\GHz\ channels. In the Galactic plane, CO emission arises from various regions of the Galaxy, with distinct radial velocities and therefore distinct CO line emission frequencies. Because each \Planck\ HFI bolometer that measures linear polarization has its own spectral transfer efficiency, a leakage from intensity to polarization is observed. This artificial polarization, absent from the 143\GHz\ channel, contributes to the observed decorrelation of polarization angles, as well as to the distortion in the spectral dependence of their covariances shown in Fig.~\ref{fig:varbT}. It can also affect the spectral dependence of the mean SED (Fig.~\ref{fig:meanSED}). Finally, CO molecules may also have an unexpected effect on the measurement of dust polarization by \Planck. Anisotropic resonant scattering of background dust polarized emission by foreground aligned CO molecules \citep{Houde2013a,Houde2013b} can induce a rotation of the  dust polarization signal, potentially affecting the correlation of angles in the 217 and 100\GHz\ channels, while leaving the 143\GHz\ channel uncontaminated.

\section{Summary and conclusions \label{sec:conclusion}}

We have presented an analytical framework for interpreting variations in the submillimeter polarized SED in terms of fluctuations in spectral properties, $T$ and $\beta$, of aligned grains across the sky and within the light cones of observation. Following \cite{Vacher2023a}, our approach is based on complex quantities, allowing for a consistent treatment of the fluctuations in the spectral parameters of the polarized SED and of the spectral rotation of angles, isolated in distinct residual maps. 
When either $T$ or $\beta$ fluctuates, our model predicts an almost perfect correlation across frequencies between all dust polarization residual maps. The decorrelation observed in the Stokes parameters $Q_\nu$ and $U_\nu$ translates into a perfect frequency-correlation of $\P_\nu$ in the complex plane.

Following \cite{Ritacco2023}, we validated our model using \Planck\ \srolltwo\ data \citep{Delouis2019}, with a $\fsky=97.3\%$ mask excluding the inner Galactic plane. 
To characterize fluctuations in dust polarization properties, covariances of residual maps were computed at the map level, and debiased from contamination by the CMB, noise, and systematics using 200 end-to-end simulations. 
We find that polarized-intensity residual maps are strongly correlated across frequencies, while polarization angle residual maps are weakly correlated for the (217,143) and (217,100)\GHz\ pairs and uncorrelated for the (143,100) \GHz\ pair. This departure from the predicted perfect correlation between residual maps is discussed in the context of dust physics and data analysis.
Polarized SED fluctuations are dominated by variations in the dust temperature ($\sigma_T \sim 2.7\,K$, assuming a mean temperature $\mT=20\,K$ for aligned grains), excluding any significant contribution from fluctuations in the dust polarization spectral index, $\beta$. In our analysis, dominated by the Galactic plane, fluctuations in the dust emissivity observed in the polarization SED at all frequencies are strongly correlated with those traced by \COMMANDER\ PR2 maps characterizing the dust SED in total intensity \citep{Planck_2015_X}, implying that emission from aligned grains greatly exceeds emission by non-aligned grains in the submillimeter. These results, based on $\srolltwo$ maps, are compatible with those obtained using the latest official \Planck\ release, PR4, following the same methodology, with the notable exception that the spectral dependence of the covariances computed with PR4 data is not compatible with a dust model assuming pure fluctuations in either $T$ or $\beta$.

Observations of dust emission in polarization provide access to a new dimension of dust evolution, inaccessible to unpolarized observations: the statistics of dust properties variations within the light cone (2D within the beam and 3D along the line of sight).
This information appears in what we call the ``spectral rotation of polarization angles.''
Our model captures a significant fraction of the decorrelation in angles found in \Planck\ data, as well as the variance observed in dust unpolarized-emission SEDs. The spatial variation of the SED induced by rotation of the polarization angle along the line of sight is a critical aspect that could strongly affect the detection of CMB $B$-modes and any other attempts to detect cosmic birefringence.
Our methodology, which is independent of calibration errors and can be used with any mask, enables correlation analyses with tracers of the ISM to trace the origin of decorrelation.
This work is useful and timely for the development of component-separation methods for current and future generations of CMB experiments. It will aid the development of new, more realistic sky simulations that include SED decorrelation with amplitudes consistent with \Planck\ data. 

\begin{acknowledgements}
V.G. thanks M. Houde for interesting discussions.
L.V. acknowledges partial support by the Italian Space Agency LiteBIRD Project (ASI Grants No. 2020-9-HH.0 and 2016-24-H.1-2018), as well as the RadioForegroundsPlus Project HORIZON-CL4-2023-SPACE-01, GA 101135036 and through the Project SPACE-IT-UP by the Italian Space Agency and Ministry of University and Research, Contract Number 2024-5-E.0.
A.R. acknowledges financial support from the Italian Ministry of University and Research - Project Proposal CIR01\_00010.
A.B. acknowledges financial support from the INAF initiative "IAF Astronomy Fellowships in Italy" (grant name MEGASKAT). 
\end{acknowledgements}

\bibliographystyle{aa}
\bibliography{bi}

@ARTICLE{Anderson2010,
       author = {{Anderson}, L.~D. and {Zavagno}, A. and {Rod{\'o}n}, J.~A. and {Russeil}, D. and {Abergel}, A. and {Ade}, P. and {Andr{\'e}}, P. and {Arab}, H. and {Baluteau}, J. -P. and {Bernard}, J. -P. and {Blagrave}, K. and {Bontemps}, S. and {Boulanger}, F. and {Cohen}, M. and {Compi{\`e}gne}, M. and {Cox}, P. and {Dartois}, E. and {Davis}, G. and {Emery}, R. and {Fulton}, T. and {Gry}, C. and {Habart}, E. and {Huang}, M. and {Joblin}, C. and {Jones}, S.~C. and {Kirk}, J.~M. and {Lagache}, G. and {Lim}, T. and {Madden}, S. and {Makiwa}, G. and {Martin}, P. and {Miville-Desch{\^e}nes}, M. -A. and {Molinari}, S. and {Moseley}, H. and {Motte}, F. and {Naylor}, D.~A. and {Okumura}, K. and {Pinheiro Gon{\c{c}}alves}, D. and {Polehampton}, E. and {Saraceno}, P. and {Sauvage}, M. and {Sidher}, S. and {Spencer}, L. and {Swinyard}, B. and {Ward-Thompson}, D. and {White}, G.~J.},
        title = "{The physical properties of the dust in the RCW 120 H II region as seen by Herschel}",
      journal = {\aap},
         year = 2010,
        month = jul,
       volume = {518},
          eid = {L99},
        pages = {L99},
          doi = {10.1051/0004-6361/201014657},
archivePrefix = {arXiv},
       eprint = {1005.1565},
 primaryClass = {astro-ph.GA},
       adsurl = {https://ui.adsabs.harvard.edu/abs/2010A&A...518L..99A}
}

@ARTICLE{Dupac2003,
       author = {{Dupac}, X. and {Bernard}, J. -P. and {Boudet}, N. and {Giard}, M. and {Lamarre}, J. -M. and {M{\'e}ny}, C. and {Pajot}, F. and {Ristorcelli}, I. and {Serra}, G. and {Stepnik}, B. and {Torre}, J. -P.},
        title = "{Inverse temperature dependence of the dust submillimeter spectral index}",
      journal = {\aap},
         year = 2003,
        month = jun,
       volume = {404},
        pages = {L11-L15},
          doi = {10.1051/0004-6361:20030575},
archivePrefix = {arXiv},
       eprint = {astro-ph/0304253},
 primaryClass = {astro-ph},
       adsurl = {https://ui.adsabs.harvard.edu/abs/2003A&A...404L..11D}
}

@ARTICLE{Planck_XXV,
author = {{Planck Collaboration 2011 XXIV}},
title = "{\textit{Planck} early results. XXV. Thermal dust in nearby molecular
 clouds}",
journal = {\aap},
archivePrefix = "arXiv",
eprint = {1101.2037},
year = 2011,
volume = 536,
pages = {A25},
doi = {10.1051/0004-6361/201116483}
}

@ARTICLE{Desert2022,
       author = {{D{\'e}sert}, Fran{\c{c}}ois-Xavier},
        title = "{The interstellar dust emission spectrum. Going beyond the single-temperature grey body}",
      journal = {\aap},
         year = 2022,
        month = mar,
       volume = {659},
          eid = {A70},
        pages = {A70},
          doi = {10.1051/0004-6361/202142617},
archivePrefix = {arXiv},
       eprint = {2111.05046},
 primaryClass = {astro-ph.GA},
       adsurl = {https://ui.adsabs.harvard.edu/abs/2022A&A...659A..70D}
}

@ARTICLE{Chiar2006,
       author = {{Chiar}, J.~E. and {Adamson}, A.~J. and {Whittet}, D.~C.~B. and {Chrysostomou}, A. and {Hough}, J.~H. and {Kerr}, T.~H. and {Mason}, R.~E. and {Roche}, P.~F. and {Wright}, G.},
        title = "{Spectropolarimetry of the 3.4 {\ensuremath{\mu}}m Feature in the Diffuse ISM toward the Galactic Center Quintuplet Cluster}",
      journal = {\apj},
         year = 2006,
        month = nov,
       volume = {651},
       number = {1},
        pages = {268-271},
          doi = {10.1086/507462},
archivePrefix = {arXiv},
       eprint = {astro-ph/0607245},
 primaryClass = {astro-ph},
       adsurl = {https://ui.adsabs.harvard.edu/abs/2006ApJ...651..268C}
}

@ARTICLE{Siebenmorgen2023,
       author = {{Siebenmorgen}, R.},
        title = "{Dark dust. II. Properties in the general field of the diffuse ISM}",
      journal = {\aap},
         year = 2023,
        month = feb,
       volume = {670},
          eid = {A115},
        pages = {A115},
          doi = {10.1051/0004-6361/202243860},
archivePrefix = {arXiv},
       eprint = {2211.10146},
 primaryClass = {astro-ph.GA},
       adsurl = {https://ui.adsabs.harvard.edu/abs/2023A&A...670A.115S}
}

@ARTICLE{DraineHensley2013,
       author = {{Draine}, B.~T. and {Hensley}, Brandon},
        title = "{Magnetic Nanoparticles in the Interstellar Medium: Emission Spectrum and Polarization}",
      journal = {\apj},
         year = 2013,
        month = mar,
       volume = {765},
       number = {2},
          eid = {159},
        pages = {159},
          doi = {10.1088/0004-637X/765/2/159},
archivePrefix = {arXiv},
       eprint = {1205.7021},
 primaryClass = {astro-ph.GA},
       adsurl = {https://ui.adsabs.harvard.edu/abs/2013ApJ...765..159D}
}

@ARTICLE{Houde2013b,
       author = {{Hezareh}, Talayeh and {Wiesemeyer}, Helmut and {Houde}, Martin and {Gusdorf}, Antoine and {Siringo}, Giorgio},
        title = "{Non-Zeeman circular polarization of CO rotational lines in SNR IC 443}",
      journal = {\aap},
         year = 2013,
        month = oct,
       volume = {558},
          eid = {A45},
        pages = {A45},
          doi = {10.1051/0004-6361/201321900},
archivePrefix = {arXiv},
       eprint = {1308.3683},
 primaryClass = {astro-ph.GA},
       adsurl = {https://ui.adsabs.harvard.edu/abs/2013A&A...558A..45H}
}

@ARTICLE{Houde2013a,
       author = {{Houde}, Martin and {Hezareh}, Talayeh and {Jones}, Scott and {Rajabi}, Fereshte},
        title = "{Non-Zeeman Circular Polarization of Molecular Rotational Spectral Lines}",
      journal = {\apj},
         year = 2013,
        month = feb,
       volume = {764},
       number = {1},
          eid = {24},
        pages = {24},
          doi = {10.1088/0004-637X/764/1/24},
archivePrefix = {arXiv},
       eprint = {1212.2237},
 primaryClass = {astro-ph.GA},
       adsurl = {https://ui.adsabs.harvard.edu/abs/2013ApJ...764...24H}
}

@INPROCEEDINGS{MPFIT2009,
       author = {{Markwardt}, C.~B.},
        title = "{Non-linear Least-squares Fitting in IDL with MPFIT}",
    booktitle = {Astronomical Data Analysis Software and Systems XVIII},
         year = 2009,
       editor = {{Bohlender}, D.~A. and {Durand}, D. and {Dowler}, P.},
       series = {Astronomical Society of the Pacific Conference Series},
       volume = {411},
        month = sep,
        pages = {251},
          doi = {10.48550/arXiv.0902.2850},
archivePrefix = {arXiv},
       eprint = {0902.2850},
 primaryClass = {astro-ph.IM},
       adsurl = {https://ui.adsabs.harvard.edu/abs/2009ASPC..411..251M}
}

@ARTICLE{Martinez-Solaeche2018,
       author = {{Mart{\'i}nez-Solaeche}, Gin{\'e}s and {Karakci}, Ata and {Delabrouille}, Jacques},
        title = "{A 3D model of polarized dust emission in the Milky Way}",
      journal = {\mnras},
         year = 2018,
        month = may,
       volume = {476},
       number = {1},
        pages = {1310-1330},
          doi = {10.1093/mnras/sty204},
archivePrefix = {arXiv},
       eprint = {1706.04162},
 primaryClass = {astro-ph.CO},
       adsurl = {https://ui.adsabs.harvard.edu/abs/2018MNRAS.476.1310M}
}

@ARTICLE{Pysm2025,
  author = {Group, The Pan-Experiment Galactic Science},
  title = {Full-sky Models of Galactic Microwave Emission and Polarization at Sub-arcminute Scales for the Python Sky Model},
  journal = {submitted to ApJ, arXiv preprint arXiv:2502.20452},
  year = {2025},
  eprint = {2502.20452},
  archiveprefix = {arXiv},
  url = {https://arxiv.org/abs/2502.20452},
  primaryclass = {astro-ph.CO},
}

@ARTICLE{THEMISII,
       author = {{Ysard}, N. and {Jones}, A.~P. and {Guillet}, V. and {Demyk}, K. and {Decleir}, M. and {Verstraete}, L. and {Choubani}, I. and {Miville-Desch{\^e}nes}, M. -A. and {Fanciullo}, L.},
        title = "{THEMIS 2.0: A self-consistent model for dust extinction, emission, and polarisation}",
      journal = {\aap},
         year = 2024,
        month = apr,
       volume = {684},
          eid = {A34},
        pages = {A34},
          doi = {10.1051/0004-6361/202348391},
archivePrefix = {arXiv},
       eprint = {2401.07739},
 primaryClass = {astro-ph.GA},
       adsurl = {https://ui.adsabs.harvard.edu/abs/2024A&A...684A..34Y}
}

@ARTICLE{Sullivan2025,
       author = {{Sullivan}, Raelyn M. and {Abghari}, Arefe and {Diego-Palazuelos}, Patricia and {Hergt}, Lukas T. and {Scott}, Douglas},
        title = "{Planck PR4 (NPIPE) map-space cosmic birefringence}",
      journal = {\jcap},
         year = 2025,
        month = jun,
       volume = {2025},
       number = {6},
          eid = {025},
        pages = {025},
          doi = {10.1088/1475-7516/2025/06/025},
archivePrefix = {arXiv},
       eprint = {2502.07654},
 primaryClass = {astro-ph.CO},
       adsurl = {https://ui.adsabs.harvard.edu/abs/2025JCAP...06..025S}
}

@ARTICLE{Remazeilles2016,
       author = {{Remazeilles}, M. and {Dickinson}, C. and {Eriksen}, H.~K.~K. and {Wehus}, I.~K.},
        title = "{Sensitivity and foreground modelling for large-scale cosmic microwave background B-mode polarization satellite missions}",
      journal = {\mnras},
         year = 2016,
        month = may,
       volume = {458},
       number = {2},
        pages = {2032-2050},
          doi = {10.1093/mnras/stw441},
archivePrefix = {arXiv},
       eprint = {1509.04714},
 primaryClass = {astro-ph.CO},
       adsurl = {https://ui.adsabs.harvard.edu/abs/2016MNRAS.458.2032R}
}

@ARTICLE{Vacher2025,
       author = {{Vacher}, L. and {Carones}, A. and {Aumont}, J. and {Chluba}, J. and {Krachmalnicoff}, N. and {Ranucci}, C. and {Remazeilles}, M. and {Rizzieri}, A.},
        title = "{How bad could it be? Modelling the 3D complexity of the polarised dust signal using moment expansion}",
      journal = {\aap},
         year = 2025,
        month = may,
       volume = {697},
          eid = {A212},
        pages = {A212},
          doi = {10.1051/0004-6361/202453066},
archivePrefix = {arXiv},
       eprint = {2411.11649},
 primaryClass = {astro-ph.CO},
       adsurl = {https://ui.adsabs.harvard.edu/abs/2025A&A...697A.212V}
}

@ARTICLE{Hensley2023,
       author = {{Hensley}, Brandon S. and {Draine}, B.~T.},
        title = "{The Astrodust+PAH Model: A Unified Description of the Extinction, Emission, and Polarization from Dust in the Diffuse Interstellar Medium}",
      journal = {\apj},
         year = 2023,
        month = may,
       volume = {948},
       number = {1},
          eid = {55},
        pages = {55},
          doi = {10.3847/1538-4357/acc4c2},
archivePrefix = {arXiv},
       eprint = {2208.12365},
 primaryClass = {astro-ph.GA},
       adsurl = {https://ui.adsabs.harvard.edu/abs/2023ApJ...948...55H}
}

@ARTICLE{Planck_Int_XIX,
       author = {{Planck Collaboration Int. XIX}},
        title = "{Planck intermediate results. XIX. An overview of the polarized thermal emission from Galactic dust}",
      journal = {\aap},
         year = 2015,
        month = apr,
       volume = {576},
          eid = {A104},
        pages = {A104},
          doi = {10.1051/0004-6361/201424082},
archivePrefix = {arXiv},
       eprint = {1405.0871},
 primaryClass = {astro-ph.GA},
       adsurl = {https://ui.adsabs.harvard.edu/abs/2015A&A...576A.104P}
}

@ARTICLE{ZS97,
       author = {{Zaldarriaga}, Matias and {Seljak}, Uro{\v{s}}},
        title = "{All-sky analysis of polarization in the microwave background}",
      journal = {\prd},
         year = 1997,
        month = feb,
       volume = {55},
       number = {4},
        pages = {1830-1840},
          doi = {10.1103/PhysRevD.55.1830},
archivePrefix = {arXiv},
       eprint = {astro-ph/9609170},
 primaryClass = {astro-ph},
       adsurl = {https://ui.adsabs.harvard.edu/abs/1997PhRvD..55.1830Z}
}

@ARTICLE{DarkDustII,
       author = {{Siebenmorgen}, R.},
        title = "{Dark dust. II. Properties in the general field of the diffuse ISM}",
      journal = {\aap},
         year = 2023,
        month = feb,
       volume = {670},
          eid = {A115},
        pages = {A115},
          doi = {10.1051/0004-6361/202243860},
archivePrefix = {arXiv},
       eprint = {2211.10146},
 primaryClass = {astro-ph.GA},
       adsurl = {https://ui.adsabs.harvard.edu/abs/2023A&A...670A.115S}
}

@ARTICLE{Shiu2024,
       author = {{Shiu}, Corwin and {Benton}, Steven J. and {Filippini}, Jeffrey P. and {Fraisse}, Aur{\'e}lien A. and {Jones}, William C. and {Nagy}, Johanna M. and {Padilla}, Ivan L. and {Soler}, Juan D.},
        title = "{Evidence for Spatially Distinct Galactic Dust Populations}",
      journal = {\apj},
         year = 2024,
        month = jul,
       volume = {970},
       number = {1},
          eid = {43},
        pages = {43},
          doi = {10.3847/1538-4357/ad46f6},
archivePrefix = {arXiv},
       eprint = {2310.04410},
 primaryClass = {astro-ph.GA},
       adsurl = {https://ui.adsabs.harvard.edu/abs/2024ApJ...970...43S}
}

@ARTICLE{KM95,
       author = {{Kim}, Sang-Hee and {Martin}, P.~G.},
        title = "{The Size Distribution of Interstellar Dust Particles as Determined from Polarization: Spheroids}",
      journal = {\apj},
         year = 1995,
        month = may,
       volume = {444},
        pages = {293},
          doi = {10.1086/175604},
       adsurl = {https://ui.adsabs.harvard.edu/abs/1995ApJ...444..293K}
}

@ARTICLE{Planck_XLVIII,
       author = {{Planck Collaboration Int. XLVIII}},
        title = "{Planck intermediate results. XLVIII. Disentangling Galactic dust emission and cosmic infrared background anisotropies}",
      journal = {\aap},
         year = 2016,
        month = dec,
       volume = {596},
          eid = {A109},
        pages = {A109},
          doi = {10.1051/0004-6361/201629022},
archivePrefix = {arXiv},
       eprint = {1605.09387},
 primaryClass = {astro-ph.CO},
       adsurl = {https://ui.adsabs.harvard.edu/abs/2016A&A...596A.109P}
}

@ARTICLE{npipe,
       author = {{Planck Collaboration Int. LVII}},
        title = "{Planck intermediate results. LVII. Joint Planck LFI and HFI data processing}",
      journal = {\aap},
         year = 2020,
        month = nov,
       volume = {643},
          eid = {A42},
        pages = {A42},
          doi = {10.1051/0004-6361/202038073},
archivePrefix = {arXiv},
       eprint = {2007.04997},
 primaryClass = {astro-ph.CO},
       adsurl = {https://ui.adsabs.harvard.edu/abs/2020A&A...643A..42P}
}

@ARTICLE{Clark19,
       author = {{Clark}, S.~E. and {Hensley}, Brandon S.},
        title = "{Mapping the Magnetic Interstellar Medium in Three Dimensions over the Full Sky with Neutral Hydrogen}",
      journal = {\apj},
         year = 2019,
        month = dec,
       volume = {887},
       number = {2},
          eid = {136},
        pages = {136},
          doi = {10.3847/1538-4357/ab5803},
archivePrefix = {arXiv},
       eprint = {1909.11673},
 primaryClass = {astro-ph.GA},
       adsurl = {https://ui.adsabs.harvard.edu/abs/2019ApJ...887..136C}
}

@ARTICLE{Planck13_XI,
        author = "{{Planck Collaboration 2013 XI}}",
        title = "{Planck 2013 results. XI. All-sky model of thermal dust emission}",
      journal = {\aap},
         year = 2014,
        month = nov,
       volume = {571},
          eid = {A11},
        pages = {A11},
          doi = {10.1051/0004-6361/201323195},
archivePrefix = {arXiv},
       eprint = {1312.1300},
 primaryClass = {astro-ph.GA},
       adsurl = {https://ui.adsabs.harvard.edu/abs/2014A&A...571A..11P}
}

@ARTICLE{Jenkins09,
       author = {{Jenkins}, Edward B.},
        title = "{A Unified Representation of Gas-Phase Element Depletions in the Interstellar Medium}",
      journal = {\apj},
         year = 2009,
        month = aug,
       volume = {700},
       number = {2},
        pages = {1299-1348},
          doi = {10.1088/0004-637X/700/2/1299},
archivePrefix = {arXiv},
       eprint = {0905.3173},
 primaryClass = {astro-ph.GA},
       adsurl = {https://ui.adsabs.harvard.edu/abs/2009ApJ...700.1299J}
}

@ARTICLE{Draine03,
       author = {{Draine}, B.~T.},
        title = "{Interstellar Dust Grains}",
      journal = {\araa},
         year = 2003,
        month = jan,
       volume = {41},
        pages = {241-289},
          doi = {10.1146/annurev.astro.41.011802.094840},
archivePrefix = {arXiv},
       eprint = {astro-ph/0304489},
 primaryClass = {astro-ph},
       adsurl = {https://ui.adsabs.harvard.edu/abs/2003ARA&A..41..241D}
}

@ARTICLE{Minami20,
       author = {{Minami}, Yuto and {Komatsu}, Eiichiro},
        title = "{New Extraction of the Cosmic Birefringence from the Planck 2018 Polarization Data}",
      journal = {\prl},
         year = 2020,
        month = nov,
       volume = {125},
       number = {22},
          eid = {221301},
        pages = {221301},
          doi = {10.1103/PhysRevLett.125.221301},
archivePrefix = {arXiv},
       eprint = {2011.11254},
 primaryClass = {astro-ph.CO},
       adsurl = {https://ui.adsabs.harvard.edu/abs/2020PhRvL.125v1301M}
}

@ARTICLE{PIPXLIX_birefringence,
       author = {{Planck Collaboration Int. XLIX}},
        title = "{Planck intermediate results. XLIX. Parity-violation constraints from polarization data}",
      journal = {\aap},
         year = 2016,
        month = dec,
       volume = {596},
          eid = {A110},
        pages = {A110},
          doi = {10.1051/0004-6361/201629018},
archivePrefix = {arXiv},
       eprint = {1605.08633},
 primaryClass = {astro-ph.CO},
       adsurl = {https://ui.adsabs.harvard.edu/abs/2016A&A...596A.110P}
}

@ARTICLE{CosmicBirefringence,
       author = {{Diego-Palazuelos}, P. and {Mart{\'\i}nez-Gonz{\'a}lez}, E. and {Vielva}, P. and {Barreiro}, R.~B. and {Tristram}, M. and {de la Hoz}, E. and {Eskilt}, J.~R. and {Minami}, Y. and {Sullivan}, R.~M. and {Banday}, A.~J. and {G{\'o}rski}, K.~M. and {Keskitalo}, R. and {Komatsu}, E. and {Scott}, D.},
        title = "{Robustness of cosmic birefringence measurement against Galactic foreground emission and instrumental systematics}",
      journal = {\jcap},
         year = 2023,
        month = jan,
       volume = {2023},
       number = {1},
          eid = {044},
        pages = {044},
          doi = {10.1088/1475-7516/2023/01/044},
archivePrefix = {arXiv},
       eprint = {2210.07655},
 primaryClass = {astro-ph.CO},
       adsurl = {https://ui.adsabs.harvard.edu/abs/2023JCAP...01..044D}
}

@ARTICLE{Jost2023,
       author = {{Jost}, Baptiste and {Errard}, Josquin and {Stompor}, Radek},
        title = "{Characterizing cosmic birefringence in the presence of Galactic foregrounds and instrumental systematic effects}",
      journal = {\prd},
         year = 2023,
        month = oct,
       volume = {108},
       number = {8},
          eid = {082005},
        pages = {082005},
          doi = {10.1103/PhysRevD.108.082005},
archivePrefix = {arXiv},
       eprint = {2212.08007},
 primaryClass = {astro-ph.CO},
       adsurl = {https://ui.adsabs.harvard.edu/abs/2023PhRvD.108h2005J}
}

@article{planck2018VI,
	author = {{Planck Collaboration 2018 VI}},
	title = {Planck 2018 results - VI. Cosmological parameters},
	DOI= "10.1051/0004-6361/201833910",
	url= "https://doi.org/10.1051/0004-6361/201833910",
	journal = {A\&A},
	year = 2020,
	volume = 641,
	pages = "A6",
}

@ARTICLE{Planck_2015_X,
       author = {{Planck Collaboration 2015 X}},
        title = "{Planck 2015 results. X. Diffuse component separation: Foreground maps}",
      journal = {\aap},
         year = 2016,
        month = sep,
       volume = {594},
          eid = {A10},
        pages = {A10},
          doi = {10.1051/0004-6361/201525967},
archivePrefix = {arXiv},
       eprint = {1502.01588},
 primaryClass = {astro-ph.CO},
       adsurl = {https://ui.adsabs.harvard.edu/abs/2016A&A...594A..10P}
}

@ARTICLE{Pysm,
       author = {{Thorne}, B. and {Dunkley}, J. and {Alonso}, D. and {N{\ae}ss}, S.},
        title = "{The Python Sky Model: software for simulating the Galactic microwave sky}",
      journal = {\mnras},
         year = 2017,
        month = aug,
       volume = {469},
       number = {3},
        pages = {2821-2833},
          doi = {10.1093/mnras/stx949},
archivePrefix = {arXiv},
       eprint = {1608.02841},
 primaryClass = {astro-ph.CO},
       adsurl = {https://ui.adsabs.harvard.edu/abs/2017MNRAS.469.2821T}
}

@article{Zonca_2021,
  doi = {10.21105/joss.03783},
  url = {https://doi.org/10.21105/joss.03783},
  year = {2021},
  publisher = {The Open Journal},
  volume = {6},
  number = {67},
  pages = {3783},
  author = {Andrea Zonca and Ben Thorne and Nicoletta Krachmalnicoff and Julian Borrill},
  title = {{The Python Sky Model 3 software}},
  journal = {Journal of Open Source Software}
}

@ARTICLE{Sponseller2022,
       author = {{Sponseller}, Danielle and {Kogut}, Alan},
        title = "{Mitigating Bias in CMB B-modes from Foreground Cleaning Using a Moment Expansion}",
      journal = {\apj},
         year = 2022,
        month = sep,
       volume = {936},
       number = {1},
          eid = {8},
        pages = {8},
          doi = {10.3847/1538-4357/ac846f},
archivePrefix = {arXiv},
       eprint = {2207.13109},
 primaryClass = {astro-ph.CO},
       adsurl = {https://ui.adsabs.harvard.edu/abs/2022ApJ...936....8S}
}

@ARTICLE{McBride2023,
       author = {{McBride}, Lisa and {Bull}, Philip and {Hensley}, Brandon S.},
        title = "{Characterizing line-of-sight variability of polarized dust emission with future CMB experiments}",
      journal = {\mnras},
         year = 2023,
        month = mar,
       volume = {519},
       number = {3},
        pages = {4370-4383},
          doi = {10.1093/mnras/stac3754},
archivePrefix = {arXiv},
       eprint = {2207.14213},
 primaryClass = {astro-ph.CO},
       adsurl = {https://ui.adsabs.harvard.edu/abs/2023MNRAS.519.4370M}
}

@article{vansyngel-et-al-2017,
	Adsurl = {http://adsabs.harvard.edu/abs/2017A%26A...603A..62V},
	Archiveprefix = {arXiv},
	Author = {{Vansyngel}, F. and {Boulanger}, F. and {Ghosh}, T. and {Wandelt}, B. and {Aumont}, J. and {Bracco}, A. and {Levrier}, F. and {Martin}, P.~G. and {Montier}, L.},
	Doi = {10.1051/0004-6361/201629992},
	Eid = {A62},
	Eprint = {1611.02577},
	Journal = {\aap},
	Month = jul,
	Pages = {A62},
	Title = {{Statistical simulations of the dust foreground to cosmic microwave background polarization}},
	Volume = 603,
	Year = 2017}

@article{ghosh-et-al-2017,
	Adsurl = {http://adsabs.harvard.edu/abs/2017A%26A...601A..71G},
	Archiveprefix = {arXiv},
	Author = {{Ghosh}, T. and {Boulanger}, F. and {Martin}, P.~G. and {Bracco}, A. and {Vansyngel}, F. and {Aumont}, J. and {Bock}, J.~J. and {Dor{\'e}}, O. and {Haud}, U. and {Kalberla}, P.~M.~W. and {Serra}, P.},
	Doi = {10.1051/0004-6361/201629829},
	Eid = {A71},
	Eprint = {1611.02418},
	Journal = {\aap},
	Month = may,
	Pages = {A71},
	Title = {{Modelling and simulation of large-scale polarized dust emission over the southern Galactic cap using the GASS Hi data}},
	Volume = 601,
	Year = 2017}

@ARTICLE{planck2016-XLIV,
author = {{Planck Collaboration Int. XLIV}},
title = "{Planck intermediate results. XLIV. The structure of the
 Galactic magnetic field from dust polarization maps of the southern Galactic
 cap}",
journal = {\aap},
archivePrefix = "arXiv",
eprint = {1604.01029},
year = 2016,
volume = 596,
pages = {A105},
doi = {10.1051/0004-6361/201628636}
}

@ARTICLE{Delouis2019,
       author = {{Delouis}, J. -M. and {Pagano}, L. and {Mottet}, S. and {Puget}, J. -L. and {Vibert}, L.},
        title = "{SRoll2: an improved mapmaking approach to reduce large-scale systematic effects in the Planck High Frequency Instrument legacy maps}",
      journal = {\aap},
         year = 2019,
        month = sep,
       volume = {629},
          eid = {A38},
        pages = {A38},
          doi = {10.1051/0004-6361/201834882},
archivePrefix = {arXiv},
       eprint = {1901.11386},
 primaryClass = {astro-ph.CO},
       adsurl = {https://ui.adsabs.harvard.edu/abs/2019A&A...629A..38D}
}

@ARTICLE{Demyk2022,
       author = {{Demyk}, K. and {Meny}, C. and {Lu}, X. -H. and {Papatheodorou}, G. and {Toplis}, M.~J. and {Leroux}, H. and {Depecker}, C. and {Brubach}, J. -B. and {Roy}, P. and {Nayral}, C. and {Ojo}, W. -S. and {Delpech}, F. and {Paradis}, D. and {Gromov}, V.},
        title = "{Low temperature MIR to submillimeter mass absorption coefficient of interstellar dust analogues. I. Mg-rich glassy silicates (Corrigendum)}",
      journal = {\aap},
         year = 2022,
        month = oct,
       volume = {666},
          eid = {C3},
        pages = {C3},
          doi = {10.1051/0004-6361/201629711e},
       adsurl = {https://ui.adsabs.harvard.edu/abs/2022A&A...666C...3D}
}

@ARTICLE{Diego2022BiRe,
       author = {{Diego-Palazuelos}, P. and {Eskilt}, J.~R. and {Minami}, Y. and {Tristram}, M. and {Sullivan}, R.~M. and {Banday}, A.~J. and {Barreiro}, R.~B. and {Eriksen}, H.~K. and {G{\'o}rski}, K.~M. and {Keskitalo}, R. and {Komatsu}, E. and {Mart{\'\i}nez-Gonz{\'a}lez}, E. and {Scott}, D. and {Vielva}, P. and {Wehus}, I.~K.},
        title = "{Cosmic Birefringence from the Planck Data Release 4}",
      journal = {\prl},
         year = 2022,
        month = mar,
       volume = {128},
       number = {9},
          eid = {091302},
        pages = {091302},
          doi = {10.1103/PhysRevLett.128.091302},
archivePrefix = {arXiv},
       eprint = {2201.07682},
 primaryClass = {astro-ph.CO},
       adsurl = {https://ui.adsabs.harvard.edu/abs/2022PhRvL.128i1302D}
}

@ARTICLE{vardustdisk2,
       author = {{Schlafly}, E.~F. and {Meisner}, A.~M. and {Stutz}, A.~M. and {Kainulainen}, J. and {Peek}, J.~E.~G. and {Tchernyshyov}, K. and {Rix}, H. -W. and {Finkbeiner}, D.~P. and {Covey}, K.~R. and {Green}, G.~M. and {Bell}, E.~F. and {Burgett}, W.~S. and {Chambers}, K.~C. and {Draper}, P.~W. and {Flewelling}, H. and {Hodapp}, K.~W. and {Kaiser}, N. and {Magnier}, E.~A. and {Martin}, N.~F. and {Metcalfe}, N. and {Wainscoat}, R.~J. and {Waters}, C.},
        title = "{The Optical-infrared Extinction Curve and Its Variation in the Milky Way}",
      journal = {\apj},
         year = 2016,
        month = apr,
       volume = {821},
       number = {2},
          eid = {78},
        pages = {78},
          doi = {10.3847/0004-637X/821/2/78},
archivePrefix = {arXiv},
       eprint = {1602.03928},
 primaryClass = {astro-ph.GA},
       adsurl = {https://ui.adsabs.harvard.edu/abs/2016ApJ...821...78S}
}

@ARTICLE{Fanciullo2015,
       author = {{Fanciullo}, L. and {Guillet}, V. and {Aniano}, G. and {Jones}, A.~P. and {Ysard}, N. and {Miville-Desch{\^e}nes}, M. -A. and {Boulanger}, F. and {K{\"o}hler}, M.},
        title = "{Dust models post-Planck: constraining the far-infrared opacity of dust in the diffuse interstellar medium}",
      journal = {\aap},
         year = 2015,
        month = aug,
       volume = {580},
          eid = {A136},
        pages = {A136},
          doi = {10.1051/0004-6361/201525677},
archivePrefix = {arXiv},
       eprint = {1506.07011},
 primaryClass = {astro-ph.GA},
       adsurl = {https://ui.adsabs.harvard.edu/abs/2015A&A...580A.136F}
}

@ARTICLE{pelgrims2021,
       author = {{Pelgrims}, V. and {Clark}, S.~E. and {Hensley}, B.~S. and {Panopoulou}, G.~V. and {Pavlidou}, V. and {Tassis}, K. and {Eriksen}, H.~K. and {Wehus}, I.~K.},
        title = "{Evidence for line-of-sight frequency decorrelation of polarized dust emission in Planck data}",
      journal = {\aap},
         year = 2021,
        month = mar,
       volume = {647},
          eid = {A16},
        pages = {A16},
          doi = {10.1051/0004-6361/202040218},
archivePrefix = {arXiv},
       eprint = {2101.09291},
 primaryClass = {astro-ph.CO},
       adsurl = {https://ui.adsabs.harvard.edu/abs/2021A&A...647A..16P}
}

@ARTICLE{Chluba,
       author = {{Chluba}, Jens and {Hill}, James Colin and {Abitbol}, Maximilian H.},
        title = "{Rethinking CMB foregrounds: systematic extension of foreground parametrizations}",
      journal = {\mnras},
         year = 2017,
        month = nov,
       volume = {472},
       number = {1},
        pages = {1195-1213},
          doi = {10.1093/mnras/stx1982},
archivePrefix = {arXiv},
       eprint = {1701.00274},
 primaryClass = {astro-ph.CO},
       adsurl = {https://ui.adsabs.harvard.edu/abs/2017MNRAS.472.1195C}
}

@ARTICLE{Azzoni2021,
       author = {{Azzoni}, S. and {Abitbol}, M.~H. and {Alonso}, D. and {Gough}, A. and {Katayama}, N. and {Matsumura}, T.},
        title = "{A minimal power-spectrum-based moment expansion for CMB B-mode searches}",
      journal = {\jcap},
         year = 2021,
        month = may,
       volume = {2021},
       number = {5},
          eid = {047},
        pages = {047},
          doi = {10.1088/1475-7516/2021/05/047},
archivePrefix = {arXiv},
       eprint = {2011.11575},
 primaryClass = {astro-ph.CO},
       adsurl = {https://ui.adsabs.harvard.edu/abs/2021JCAP...05..047A}
}

@ARTICLE{Vacher21,
       author = {{Vacher}, L. and {Aumont}, J. and {Montier}, L. and {Azzoni}, S. and {Boulanger}, F. and {Remazeilles}, M.},
        title = "{Moment expansion of polarized dust SED: A new path towards capturing the CMB B-modes with LiteBIRD}",
      journal = {\aap},
         year = 2022,
        month = apr,
       volume = {660},
          eid = {A111},
        pages = {A111},
          doi = {10.1051/0004-6361/202142664},
archivePrefix = {arXiv},
       eprint = {2111.07742},
 primaryClass = {astro-ph.CO},
       adsurl = {https://ui.adsabs.harvard.edu/abs/2022A&A...660A.111V}
}

@ARTICLE{RemazeillesmomentsILC,
       author = {{Remazeilles}, Mathieu and {Rotti}, Aditya and {Chluba}, Jens},
        title = "{Peeling off foregrounds with the constrained moment ILC method to unveil primordial CMB B modes}",
      journal = {\mnras},
         year = 2021,
        month = may,
       volume = {503},
       number = {2},
        pages = {2478-2498},
          doi = {10.1093/mnras/stab648},
archivePrefix = {arXiv},
       eprint = {2006.08628},
 primaryClass = {astro-ph.CO},
       adsurl = {https://ui.adsabs.harvard.edu/abs/2021MNRAS.503.2478R}
}

@ARTICLE{Planck18_XI,
       author = {{Planck Collaboration 2018 XI}},
        title = "{Planck 2018 results. XI. Polarized dust foregrounds}",
      journal = {\aap},
         year = 2020,
        month = sep,
       volume = {641},
          eid = {A11},
        pages = {A11},
          doi = {10.1051/0004-6361/201832618},
archivePrefix = {arXiv},
       eprint = {1801.04945},
 primaryClass = {astro-ph.GA},
       adsurl = {https://ui.adsabs.harvard.edu/abs/2020A&A...641A..11P}
}

@ARTICLE{Planck18_XII,
       author = {{Planck Collaboration 2018 XII}},
        title = "{Planck 2018 results. XII. Galactic astrophysics using polarized dust emission}",
      journal = {\aap},
         year = 2020,
        month = sep,
       volume = {641},
          eid = {A12},
        pages = {A12},
          doi = {10.1051/0004-6361/201833885},
archivePrefix = {arXiv},
       eprint = {1807.06212},
 primaryClass = {astro-ph.GA},
       adsurl = {https://ui.adsabs.harvard.edu/abs/2020A&A...641A..12P}
}

@ARTICLE{Planck2014dust,
       author = {{Planck Collaboration Int. XVII}},
        title = "{Planck intermediate results. XVII. Emission of dust in the diffuse interstellar medium from the far-infrared to microwave frequencies}",
      journal = {\aap},
         year = 2014,
        month = jun,
       volume = {566},
          eid = {A55},
        pages = {A55},
          doi = {10.1051/0004-6361/201323270},
archivePrefix = {arXiv},
       eprint = {1312.5446},
 primaryClass = {astro-ph.GA},
       adsurl = {https://ui.adsabs.harvard.edu/abs/2014A&A...566A..55P}
}

@ARTICLE{Planck2016-XXX,
       author = {{Planck Collaboration Int. XXX}},
        title = "{Planck intermediate results. XXX. The angular power spectrum of polarized dust emission at intermediate and high Galactic latitudes}",
      journal = {\aap},
         year = 2016,
        month = feb,
       volume = {586},
          eid = {A133},
        pages = {A133},
          doi = {10.1051/0004-6361/201425034},
archivePrefix = {arXiv},
       eprint = {1409.5738},
 primaryClass = {astro-ph.CO},
       adsurl = {https://ui.adsabs.harvard.edu/abs/2016A&A...586A.133P}
        }

@ARTICLE{Ritacco2023,
       author = {{Ritacco}, Alessia and {Boulanger}, Fran{\c{c}}ois and {Guillet}, Vincent and {Delouis}, Jean-Marc and {Puget}, Jean-Loup and {Aumont}, Jonathan and {Vacher}, L{\'e}o},
        title = "{Dust polarization spectral dependence from Planck HFI data. Turning point for cosmic microwave background polarization-foreground modeling}",
      journal = {\aap},
         year = 2023,
        month = feb,
       volume = {670},
          eid = {A163},
        pages = {A163},
          doi = {10.1051/0004-6361/202244269},
archivePrefix = {arXiv},
       eprint = {2206.07671},
 primaryClass = {astro-ph.CO},
       adsurl = {https://ui.adsabs.harvard.edu/abs/2023A&A...670A.163R}
}

@ARTICLE{PlanckL,
       author = {{Planck Collaboration Int. L}}, 
        title = "{Planck intermediate results. L. Evidence of spatial variation of the polarized thermal dust spectral energy distribution and implications for CMB B-mode analysis}",
      journal = {\aap},
         year = 2017,
        month = mar,
       volume = {599},
          eid = {A51},
        pages = {A51},
          doi = {10.1051/0004-6361/201629164},
archivePrefix = {arXiv},
       eprint = {1606.07335},
 primaryClass = {astro-ph.CO},
       adsurl = {https://ui.adsabs.harvard.edu/abs/2017A&A...599A..51P}
}

@INPROCEEDINGS{Ptep,
       author = {{LiteBIRD Collaboration}},
        title = "{Probing Cosmic Inflation with the LiteBIRD Cosmic Microwave Background Polarization Survey}",
    booktitle = {PTEP},
         year = {2022},
       series = {PTEP},
       volume = {11443},
        month = dec,
          eid = {114432F},
        pages = {114432F},
          doi = {10.1117/12.2563050},
archivePrefix = {arXiv},
       eprint = {2101.12449},
 primaryClass = {astro-ph.IM},
       adsurl = {https://ui.adsabs.harvard.edu/abs/2020SPIE11443E..2FH}
}

@ARTICLE{healpix,
       author = {{G{\'o}rski}, K.~M. and {Hivon}, E. and {Banday}, A.~J. and {Wand
        elt}, B.~D. and {Hansen}, F.~K. and {Reinecke}, M. and {Bartelmann}, M.},
        title = "{HEALPix: A Framework for High-Resolution Discretization and Fast Analysis of Data Distributed on the Sphere}",
      journal = {\apj},
         year = 2005,
        month = apr,
       volume = {622},
       number = {2},
        pages = {759-771},
          doi = {10.1086/427976},
archivePrefix = {arXiv},
       eprint = {astro-ph/0409513},
 primaryClass = {astro-ph},
       adsurl = {https://ui.adsabs.harvard.edu/abs/2005ApJ...622..759G}
}

@ARTICLE{Tassis2015,
       author = {{Tassis}, K. and {Pavlidou}, V.},
        title = "{Searching for inflationary B modes: can dust emission properties be extrapolated from 350 GHz to 150 GHz?}",
      journal = {\mnras},
         year = 2015,
        month = jul,
       volume = {451},
        pages = {L90-L94},
          doi = {10.1093/mnrasl/slv077},
archivePrefix = {arXiv},
       eprint = {1410.8136},
 primaryClass = {astro-ph.CO},
       adsurl = {https://ui.adsabs.harvard.edu/abs/2015MNRAS.451L..90T}
}

@ARTICLE{Ysard2013,
       author = {{Ysard}, N. and {Abergel}, A. and {Ristorcelli}, I. and {Juvela}, M. and {Pagani}, L. and {K{\"o}nyves}, V. and {Spencer}, L. and {White}, G. and {Zavagno}, A.},
        title = "{Variation in dust properties in a dense filament of the Taurus molecular complex (L1506)}",
      journal = {\aap},
         year = 2013,
        month = nov,
       volume = {559},
          eid = {A133},
        pages = {A133},
          doi = {10.1051/0004-6361/201322066},
archivePrefix = {arXiv},
       eprint = {1309.6489},
 primaryClass = {astro-ph.GA},
       adsurl = {https://ui.adsabs.harvard.edu/abs/2013A&A...559A.133Y}
}

@ARTICLE{Vacher2023a,
       author = {{Vacher}, L. and {Chluba}, J. and {Aumont}, J. and {Rotti}, A. and {Montier}, L.},
        title = "{High precision modeling of polarized signals: Moment expansion method generalized to spin-2 fields}",
      journal = {\aap},
         year = 2023,
        month = jan,
       volume = {669},
          eid = {A5},
        pages = {A5},
          doi = {10.1051/0004-6361/202243913},
archivePrefix = {arXiv},
       eprint = {2205.01049},
 primaryClass = {astro-ph.CO},
       adsurl = {https://ui.adsabs.harvard.edu/abs/2023A&A...669A...5V}
}

@ARTICLE{Vacher2023b,
       author = {{Vacher}, L. and {Aumont}, J. and {Boulanger}, F. and {Montier}, L. and {Guillet}, V. and {Ritacco}, A. and {Chluba}, J.},
        title = "{Frequency dependence of the thermal dust E/B ratio and EB correlation: Insights from the spin-moment expansion}",
      journal = {\aap},
         year = 2023,
        month = apr,
       volume = {672},
          eid = {A146},
        pages = {A146},
          doi = {10.1051/0004-6361/202245292},
archivePrefix = {arXiv},
       eprint = {2210.14768},
 primaryClass = {astro-ph.CO},
       adsurl = {https://ui.adsabs.harvard.edu/abs/2023A&A...672A.146V}
}

@ARTICLE{Azzoni2023,
       author = {{Azzoni}, S. and {Alonso}, D. and {Abitbol}, M.~H. and {Errard}, J. and {Krachmalnicoff}, N.},
        title = "{A hybrid map-C$_{{\ensuremath{\ell}}}$ component separation method for primordial CMB B-mode searches}",
      journal = {\jcap},
         year = 2023,
        month = mar,
       volume = {2023},
       number = {3},
          eid = {035},
        pages = {035},
          doi = {10.1088/1475-7516/2023/03/035},
archivePrefix = {arXiv},
       eprint = {2210.14838},
 primaryClass = {astro-ph.CO},
       adsurl = {https://ui.adsabs.harvard.edu/abs/2023JCAP...03..035A}
}

@ARTICLE{Wolz2024,
       author = {{Wolz}, Kevin and {Azzoni}, Susanna and {Herv{\'\i}as-Caimapo}, Carlos and {Errard}, Josquin and {Krachmalnicoff}, Nicoletta and {Alonso}, David and {Baccigalupi}, Carlo and {Baleato Lizancos}, Ant{\'o}n and {Brown}, Michael L. and {Calabrese}, Erminia and {Chluba}, Jens and {Dunkley}, Jo and {Fabbian}, Giulio and {Galitzki}, Nicholas and {Jost}, Baptiste and {Morshed}, Magdy and {Nati}, Federico},
        title = "{The Simons Observatory: Pipeline comparison and validation for large-scale B-modes}",
      journal = {\aap},
         year = 2024,
        month = jun,
       volume = {686},
          eid = {A16},
        pages = {A16},
          doi = {10.1051/0004-6361/202346105},
archivePrefix = {arXiv},
       eprint = {2302.04276},
 primaryClass = {astro-ph.CO},
       adsurl = {https://ui.adsabs.harvard.edu/abs/2024A&A...686A..16W}
}

@ARTICLE{Carones2024,
       author = {{Carones}, A. and {Remazeilles}, M.},
        title = "{Optimization of foreground moment deprojection for semi-blind CMB polarization reconstruction}",
      journal = {\jcap},
         year = 2024,
        month = jun,
       volume = {2024},
       number = {6},
          eid = {018},
        pages = {018},
          doi = {10.1088/1475-7516/2024/06/018},
archivePrefix = {arXiv},
       eprint = {2402.17579},
 primaryClass = {astro-ph.CO},
       adsurl = {https://ui.adsabs.harvard.edu/abs/2024JCAP...06..018C}
}

\appendix

\section{Dictionary of defined quantities}\label{A-dico}

Table~\ref{tab-notations} presents all our defined quantities, with references to equations, sections, figures and tables.
\begin{table*}[h!]
\caption{Dictionary of all defined quantities, with complex quantities in bold.}
\begin{tabular}{l l l l l}
\hline\hline
Notation & Meaning & Reference \\
\hline
$\P_i = Q_i+\i U_i$ & complex polarized intensity in channel $i$ & Eq.~\eqref{eq:spinor}, Sect.~\ref{sec:expansion}\\
$\psi_i $ & polarization angle in channel $i$ & Sect.~\ref{sec:expansion} \\
$\nu_0$ & reference frequency & Sect.~\ref{sec:expansion} \\
$\Pref$ & polarized intensity map at the reference frequency & Sect.~\ref{subsec:residuals} \\
$\Iref$ & total intensity map at the reference frequency & Sect.~\ref{sec:covar-2} \\
$\pref$ & polarization fraction $\pref=\Pref/\Iref$ at the reference frequency & Sect.~\ref{sec:covar-2} \\
\hline
$T$, $\beta$, $\varepsilon_\nu$ & temperature, spectral index and emissivity of aligned grains 
& Eq.~\eqref{eq:emissivity}\\
$\mT$, $\mbeta$ & pivot temperature and spectral index defining the pivot SED & Eq.~\eqref{eq:meps}, Sect.~\ref{sec:expansion} \\
$\meps_i=\meps_{\nu_i}(\mbeta,\mT)$ & value of the pivot SED in channel $i$ of frequency $\nu_i$ & Eq.~\eqref{eq:meps}\\
\hline
$\ain^s, \ain^\T, \ain^\beta$ & order $n$ of spectral gradients, for $s\in\{\T,\beta\}$ and for $T$ and $\beta$ & Eq.~\eqref{eq:asn} \\
$\Wn, \Wbn,\Wtn$ & polarization-weighted complex spin-moments of order $n$ & Eq.~\eqref{eq:Wsn}\\
$\In, \Ibn, \Itn$& intensity-weighted moments of order $n$ & Eq.~\eqref{eq:Isn} \\
$\Mosn = \Wn - \In $ & difference between moments characterizing depolarization effects & Eq.~\eqref{eq:Wbn-In-Mn} \\
$\mWsn$, $\mIsn$, $\mMosn$ & mean values over the sky weighted by $P_0^2$ & Eq.~\eqref{eq:mWsn}, Sects.~\ref{subsec:meanSED}, \ref{subsec:residuals} \\
\hline
$\tilde{\rho}_i$ & correction to the polarization efficiency in channel $i$ & Sects.~\ref{subsec-Planck_data} \& \ref{subsec:Mean_SED_Planck}\\
$\mRP_i$ & mean complex SED normalized to the reference frequency & Eq.~\eqref{eq:defmRPi} and \eqref{eq:mRPi_Planck}, Sects.~\ref{subsec:meanSED} \& \ref{subsec:Mean_SED_Planck} \\ 
$\Rcxi$ & complex residual map for channel $i$ & Eq.~\eqref{eq:defRcxi} \\
$\RPi$ & residual map of polarized intensity in channel $i$ & Eq.~\eqref{eq:RPi} \\
$\Rti$ & residual map of polarization angle in channel $i$ & Eq.~\eqref{eq:Rti} \\
$\Repsi$ & residual map of dust emissivity in polarization & Eq.~\eqref{eq:Repsi-ai} \\
$\Cov(X,Y)$ & covariance of $X$ and $Y$ maps, weighted by $P_0^2$ & Eq.~\eqref{eq:defCov}, Sect.~\ref{sec:covar}\\
$\varPij$ & covariance of $\RPi$ and $\RPj$ &Eq.~\eqref{eq:covP} \\
$\varpsiij$ & covariance of $\Rti$ and $\Rtj$ & Eq.~\eqref{eq:covpsi}\\
$\varpsiPij$, $\varPpsiij$ & mixed covariance of $\Rti$ and $\RPj$ & Eq.~\eqref{eq:covpsiiPj}\\
$\varepsij$, $\hvareps_{ij}$ & covariance of $\Repsi$ and $\Repsj$, and its estimate & Eqs.\eqref{eq:vareps} and \eqref{eq:compute_vareps}\\
$\varPcxij$ & complex covariance of $\Rcxi$ and $\Rcxj$ &Eq.~\eqref{eq:covPcx} \\
$\rho^\P_{ij}$, $\rho^P_{ij}$, $\rho^\psi_{ij}$ & Pearson correlation coefficients between residual maps ($i\ne j$) & Table~\ref{tab:rho}, Sect.~\ref{sec:checks}\\
\hline
$\varspsi, \varTpsi, \varbpsi$ & standard deviations fitting the spectral dependence of $\varpsiij$ & Sect.~\ref{sec:Tbeta} \\ 
$\varseps, \varTeps, \varbeps$ & standard deviations fitting the spectral dependence of $\varepsij$ & Sect.~\ref{sec:Tbeta} \\ 
$\varTpsiP$ & mixed covariance fitting the spectral dependence of $\varpsiPij$ & Sect~\ref{sec:prop-al-grains} \\
\hline
$\mrat$, $\eta(p)$ & parameters characteristic of the magnetic field model & Appendix~\ref{A-Nlayers} \\
\hline
\end{tabular}
\label{tab-notations}
\end{table*}

\section{Relation between the intensity-weighted and polarization-weighted moments of spectral parameters}
\label{sec:Mon}

The polarization-weighted $\Wsn$ and intensity-weighted $\Isn$ spin-moments quantify the same statistics within the light cone, the moments of the fluctuations of spectral parameter $s$ of aligned grains, but with a distinct weight. In this section, we show that $\Mosn \equiv \Wsn - \Isn$ is of zero mean.

We first introduce the local (unobserved) complex polarization fraction 
\begin{align}
\hat{\p} \equiv \frac{\d\P}{\d I}\,,
\label{eq:hatp}
\end{align}
 covered by $\,\hat{}\,$ to be distinguished from the observed complex polarization fraction
 \begin{align}
\p \equiv \frac{\P}{I}\,.
\label{eq:p}
\end{align}
We have 
 \begin{align}
 \P = \int_\lc \d \P = \int_\lc \hat{\p}\,\d I\,. 
 \end{align}
Using Eq.~\eqref{eq:hatp} and \eqref{eq:p}, Eq.~\eqref{eq:Wsn} defining $\Wsn$ becomes, expressed this time in terms of an integral over the total intensity $I$: 
\begin{align}
\Wsn = \frac{1}{I}\int_\lc \left(s-\ms\right)^n \frac{\hat{\p}}{\p}\,\d I \,,
\label{eq:here}
\end{align}
so that
\begin{align}
\Mosn \equiv \Wsn-\Isn = \frac{1}{I}\int_\lc \left(s-\ms\right)^n \deltap\,\d I \,,
\label{eq:there}
\end{align}
where
\begin{align}
 \deltap \equiv \left(\frac{\hat{\p}}{\p}-1\right) \label{eq:dp0}
\end{align}
is the complex quantity characterizing the departure of the local complex polarization fraction $\hat{\p}$ from the observed complex polarization fraction $\p$.

When integrated over the light cone, the quantity $\deltap$ has by definition a mean value of zero: 
\begin{align}
& \int_\lc \deltap\,\d I = 
 \left(\frac{1}{\p}\int_\lc \hat{\p}\,\d I \right) - I = \frac{1}{\p} \left(\int_\lc \d\P\right) - I = {\mathbf 0} \,.
\label{eq:dpzeromean}
\end{align}
As a consequence, from Eq.~\eqref{eq:there} we conclude that $\Mosn$ is also of zero mean under our hypothesis \ref{HA2} of no correlation between the structure of the magnetic field and the variations of dust temperature and spectral index.

\section{Toy model for line of sight integration}
\label{A-Nlayers}

In this appendix, we make again use of the phenomenological model of the submillimeter polarized thermal dust emission, developed in \citet{planck2016-XLIV}, \citet{ghosh-et-al-2017}, and \citet{vansyngel-et-al-2017}. This simple model proved successful to explain different statistical properties of polarization quantities \citep{planck2016-XLIV,Planck18_XII}. 

We decompose the Galactic sky into $N$ layers of equal intensity $\Iref/N$. Each layer $k$ ($k \in [1\dots N]$) has a uniform magnetic field $\vecB_k=\vecB_0+\vecDB_k$, sum of an ordered component $\vecB_0$ (a uniform Galactic magnetic field) and of a random component $\vecDB_k$, following our hypothesis \ref{HM5}. The turbulent component $\Delta\vecB_k$ in each layer is a realization of an isotropic Gaussian random field in three dimensions, of variance $\sigma^2_{B}$. The random and ordered components are taken close to equipartition ($f_M \equiv \sigma_{B}/B_0\simeq 1$) in order to match observational properties \citep{Planck18_XII}. 

Our model provides two maps of angles for each layer $k$: one $\gamma^{(k)}$ for the inclination angle of the magnetic field with respect to the plane of the sky, and one $\psi^{(k)}$ for the projection of the magnetic field onto the plane of the sky, rotated by $90\deg$.
We use this model to quantify the variance of the real and imaginary parts of $\deltap^{(k)}$ (Eq.~\eqref{eq:dp0}), where $\p$ is the observed, light cone-integrated, complex polarization fraction, while $\hat{\p}^{(k)}$ is the local uniform polarization fraction in layer $k$. Assuming that the grain alignment efficiency is uniform in the sky (our hypothesis \ref{HA4}) with a maximal polarization fraction $\pmax$, we have the following expressions:
\begin{align}
& \hat{\p}^{(k)} = \pmax \cos^2\gamma^{(k)} \expf{\i2\psi^{(k)}}\,, \\
& \p = \frac{1}{N}\sum_{k=1}^N \hat{\p}^{(k)}\,.
\end{align}
Fig.~\ref{fig:vardp-simu} shows the dependency, with the normalized polarization fraction $p/\pmax$, of the mean value of the variance of $\Re\deltap^{(k)}$, of the variance of $\Im\deltap^{(k)}$ and of their ratio, for three values of the number $N$ of layers and five values of $\sigma_B/B_0$ comprised between 0.7 and 1.1 (all combinations compatible within uncertainties with \Planck\ data). 
In the top and middle panel, the variance $\sigma^2_{\Re\deltap}$ and $\sigma^2_{\Im\deltap}$ are found to be inversely correlated with $p^2$ at low values of $p$, but with a departure from this trend at high $p$. To characterize the distinct dependence of $\sigma^2_{\Re\deltap}$ and $\sigma^2_{\Im\deltap}$ with $p$, we study how their ratio $\eta$ varies with $p$
\begin{equation}
\eta(p) \equiv \frac{\sigma^2_{\Re\deltap}(p)}{\sigma^2_{\Im\deltap}(p)}\,.
\end{equation}
The ratio $\eta(p)$ shows a clear decreasing trend with $p$ (Fig.~\ref{fig:vardp-simu}, bottom panel). The higher the polarization fraction, the more the spectral variations in angles dominate over the spectral variation in the polarized intensity SED. 
This effect is strong at low polarization fractions (as $\eta(0)\simeq 1)$, and weak or undetected at high polarization fractions ($\eta(\pmax)= 0$). Modeling the dependence of $\eta$ with $p$ is beyond the scope of this paper. 
Our first goal is to take into account the dependence of $\sigma^2_{\Im\deltap}$ and $\sigma^2_{\Re\deltap}$ in $1/p^2$, not their second order dependence through their ratio $\eta(p)$.
Therefore, as a reference value, we ignore our mask and simply take the mean value of $\eta(p)$ over the full-sky: $\mrat\simeq 0.7\pm0.2$. The significant uncertainty is here to acknowledge for the crude approximation we have by ignoring the $p$-dependence of $\eta$. Our mask is dominated by regions with low $p$ ($p_{353} < 0.4\pmax$) at $4\deg$ of resolution, where $\eta(p)$ does not vary too much, making our approximation reasonable.
In this frame, we have the approximate relation
\begin{align}
\left\langle P_0^2 \,\Re\Mon\,\Re\Mom\right\rangle\ & \simeq \mrat \left\langle P_0^2\,\Im\Mon\,\Im\Mom\right\rangle\,.
\end{align}

\begin{figure*}[h!]
\includegraphics[width=\third]{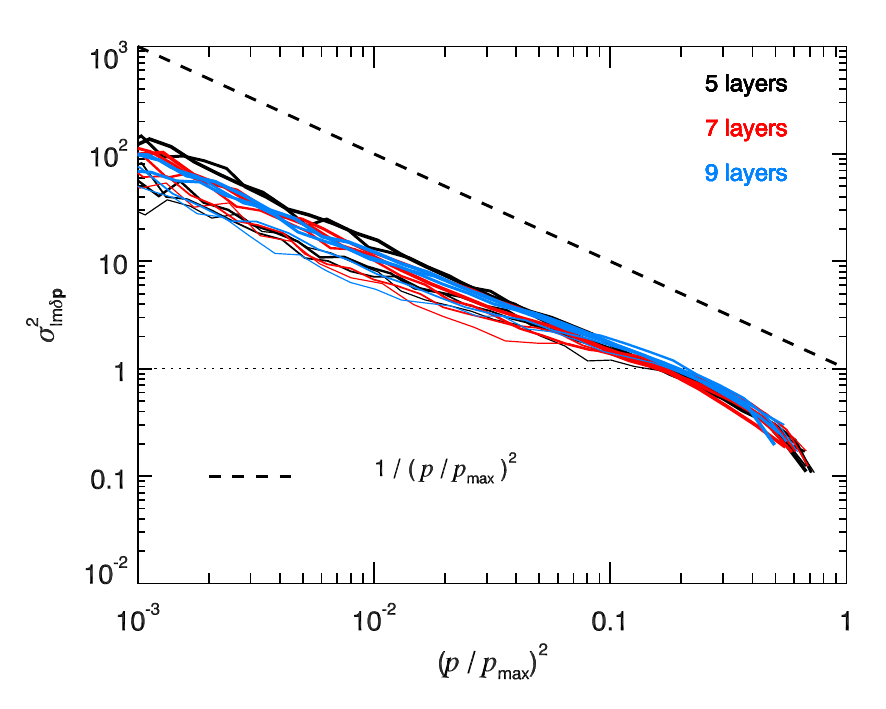}
\includegraphics[width=\third]{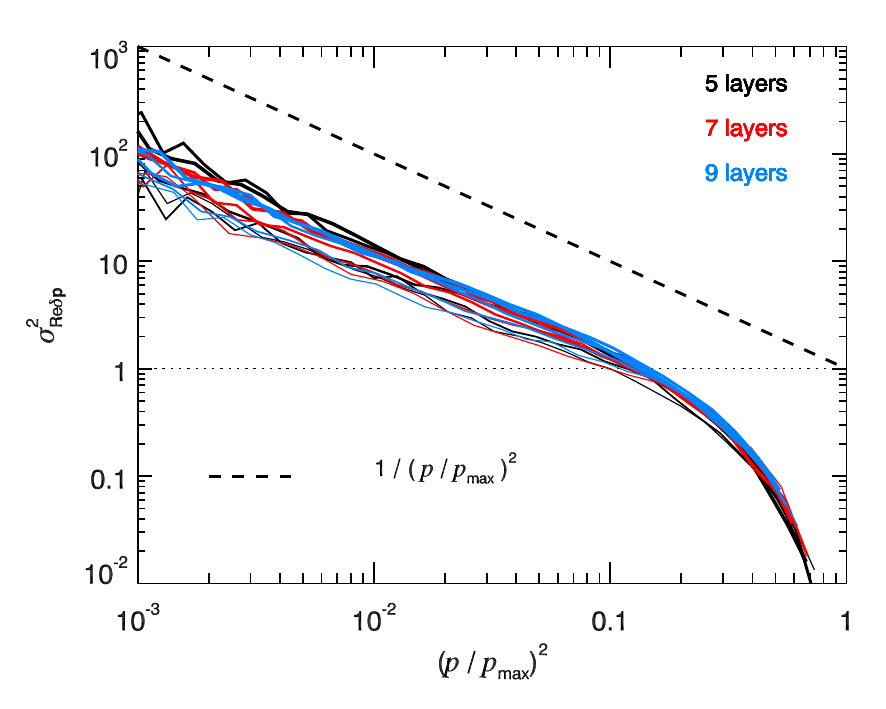}
\includegraphics[width=\third]{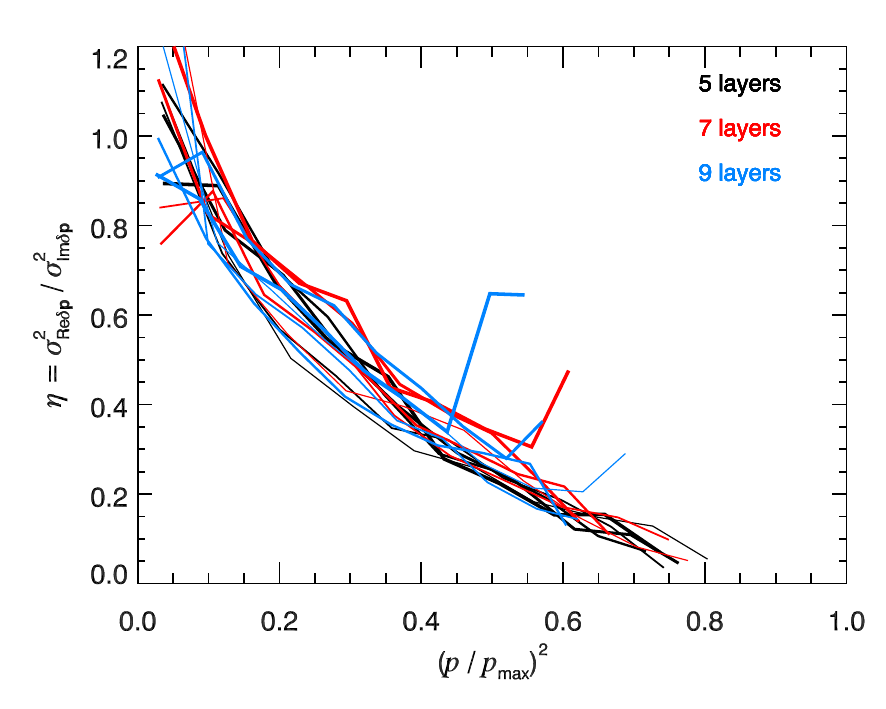}
\caption{Mean value for the variance of the imaginary part of $\deltap$ \llab, real part of $\deltap$ \clab, and their ratio \rlab, as a function of the polarization fraction normalized to the maximal polarization fraction $\pmax$, for our toy model made of $N$ independent layers with a magnetic field having both an ordered and a random component close to equipartition. The number of layers is varied between 5, 7 and 9 to emphasize uncertainties, with the corresponding ratio $f_M = \sigma_B/B_0$ of turbulent to ordered magnetic field varying from 0.7 (thinner lines) to 1.1 (thicker lines) in steps of 0.1 (all values compatible with \Planck\ data). Simulations are produced at $1\deg$ of resolution ($\Nside=128$) and then smoothed to $4\deg$ ($\Nside=32)$ as in our study.}
\label{fig:vardp-simu}
\end{figure*}

We are aware that our toy model does not include a description of the Galactic plane that dominates our statistics. When the Galactic latitude $|b|$ decreases, the increased number of independent components in the light cone will decrease the variance of $\Re\deltap^{(k)}$ and $\Im\deltap^{(k)}$. Their ratio $\eta$, however, should not change too much as both variances will be affected the same way. The dependence of $\eta$ with $p$ and $|b|$ would deserve more investigations. Entering such detailed modeling is however premature for our model which relies on simple hypotheses (see Sect.~\ref{sec:hypotheses}).

\section{Computing unbiased variances with \Planck\ half-maps}\label{sec:half-maps}

Our methodology is based on the calculation of covariances of \Planck\ residual maps. Residual maps $\Rcxi$, $\RPi$ and $\Rti$, as computed from Eqs.~\eqref{eq:defRcxi}, \eqref{eq:RPi} and \eqref{eq:Rti} involve a combination of the complex polarized intensity maps $\P_i$ for channel $i$ and $\P_0$ for the reference channel. In this section, we describe how to compute in practice variances of residual maps.

The variance $\varPcx_{ii}=\left\langle \Rcxi\Rcxi^\star\right\rangle$ being the mean value of a squared quantity, it will be biased by noise unless one uses the product of two uncorrelated versions of the same map $\P_i^{\tt HM1}$ and $\P_i^{\tt HM2}$ called half-maps. 
From Eq.~\eqref{eq:defRcxi} we define the two complex residuals half-maps $\Rcxi^\hmo$ and $\Rcxi^\hmt$ for channel $i$ by replacing the complex maps $\P_i$ and $\P_0$ by their corresponding half-maps, and the real map $P_0$ by its unbiased\footnote{The value of $P_0$ for $\sim 1\%$ of pixels having $\Re\left(\P_0^\hmo\times\left(\P_0^\hmt\right)^\star\right)<0$ has been forced to zero, with no noticeable impact on our analysis.} estimate $\sqrt{\Re(\P_0^\hmo\times\left(\P_0^\hmt\right)^\star}$.
The expression for the variance $\varPcx_{ii}$ (Eq.~\eqref{eq:covPcx}) then simply reads:
 \begin{align}
 \varPcx_{ii} = \left\langle\Rcxi^\hmo\Rcxi^{\star\hmt}\right\rangle \,.
\label{eq:covPii-hm}
 \end{align}
For $i\ne j$, the covariance $\varPcxij=\left\langle \Rcxi\Rcxj^\star\right\rangle$ involves the product of maps with uncorrelated noise\footnote{The subtraction of synchrotron template map at 100 and 143\GHz\ (see Sect.~\ref{subsec-Planck_data}) correlate those maps. For the calculation of covariances for the ($143\times100$) couple, we therefore use expression \eqref{eq:covPii-hm}. }. It is therefore possible to keep the full maps $\P_i$ and $\P_j$ in this calculation, but not for $\P_0$. We build two mixed residual maps $\Rcxi^\fullo$ and $\Rcxi^\fullt$ from Eq.~\eqref{eq:defRcxi} by keeping the full map $\P_i$ but still replacing the reference maps $\P_0$ by its corresponding half-map (the map $P_0$, playing the role of a weight, remains the same). With these notations, 
\begin{align}
\varPcxij = \left\langle\Rcxi^\fullo\Rcxj^{\star\fullt}\right\rangle \hspace{0.3cm}\forall\, i \ne j \,.
\label{eq:covPij-mx}
\end{align}
A small bias will remain in each component of the covariance matrix $\varPcx$ due to chance correlation, which we attempt to suppress by using simulations (see Sect.~\ref{subsec:Mean_SED_Planck} and our discussion in Sect.~\ref{sec:discussion}). The same procedure is applied to compute the real covariance matrices $\varPij$, $\varpsiij$ and $\varpsiPij$.

\section{Results with the \Planck\ PR4 release}\label{A-PR4}

\begin{figure*}
\includegraphics[width=\half]{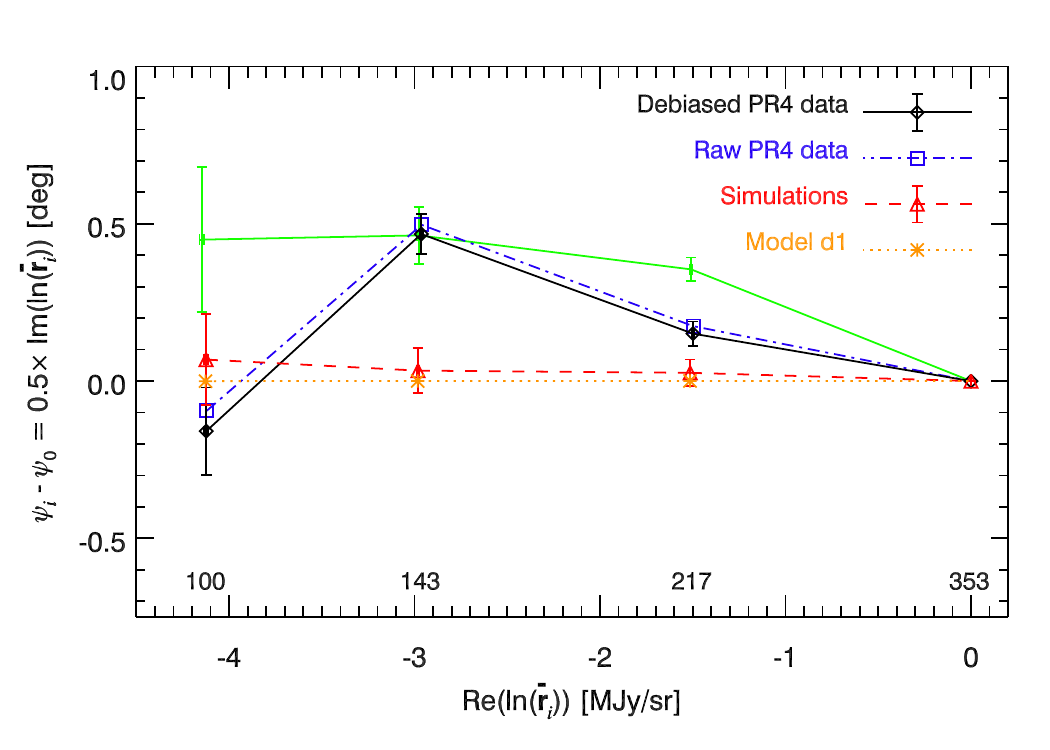}
\includegraphics[width=\half]{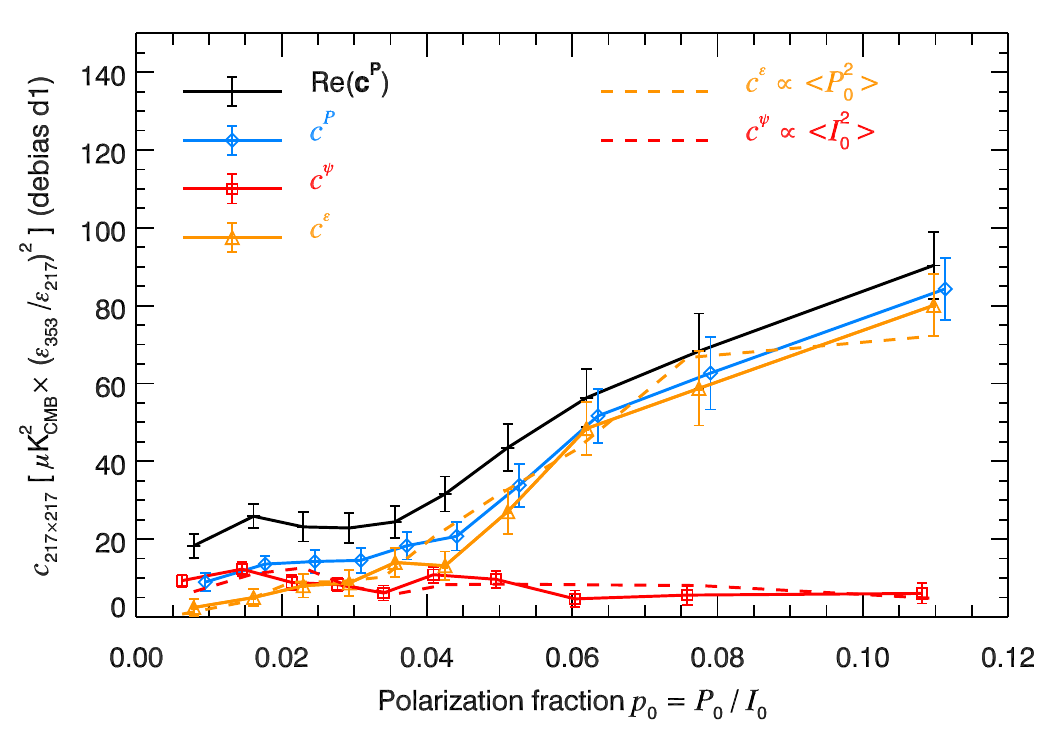}
\includegraphics[width=\half]{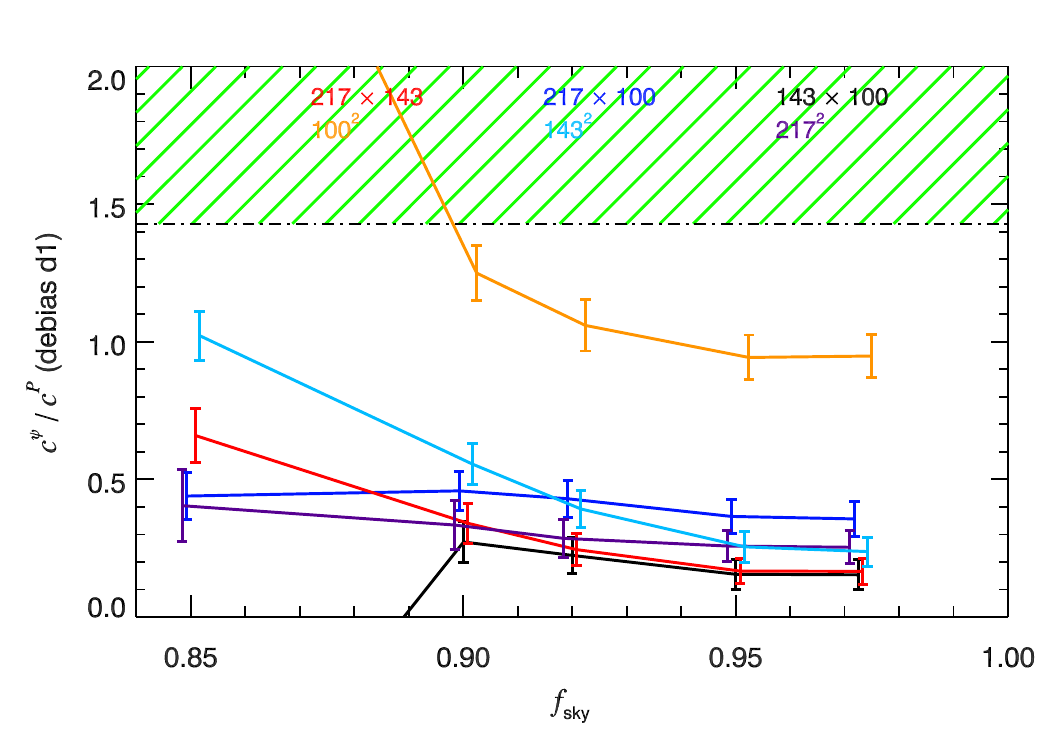}
\includegraphics[width=\half]{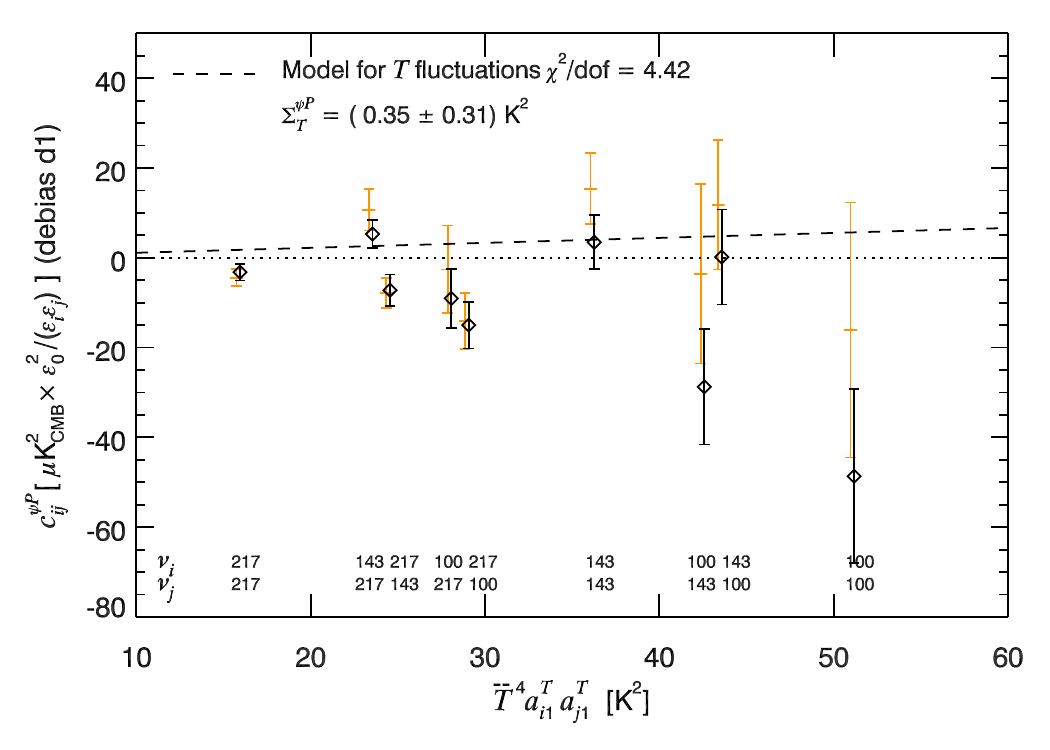}
\caption{Same as Figs.~\ref{fig:meanSED}, \ref{fig:cov_p}, \ref{fig:checkratio} and \ref{fig:cov_psiP}, but for the \Planck\ PR4 official release. To facilitate the comparison between PR4 and $\srolltwo$ in the top left figure, the mean \srolltwo\ debiased SED is overplotted in green.}\label{fig:PR4_a}
\end{figure*}

\begin{figure*}
\includegraphics[width=\half]{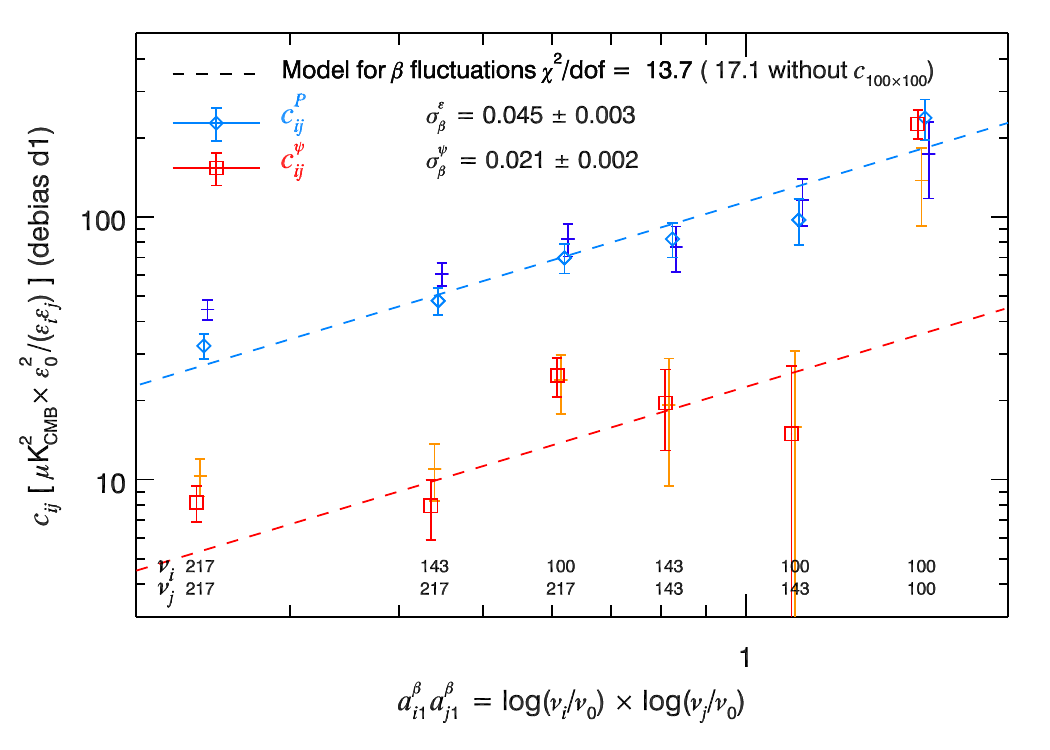}
\includegraphics[width=\half]{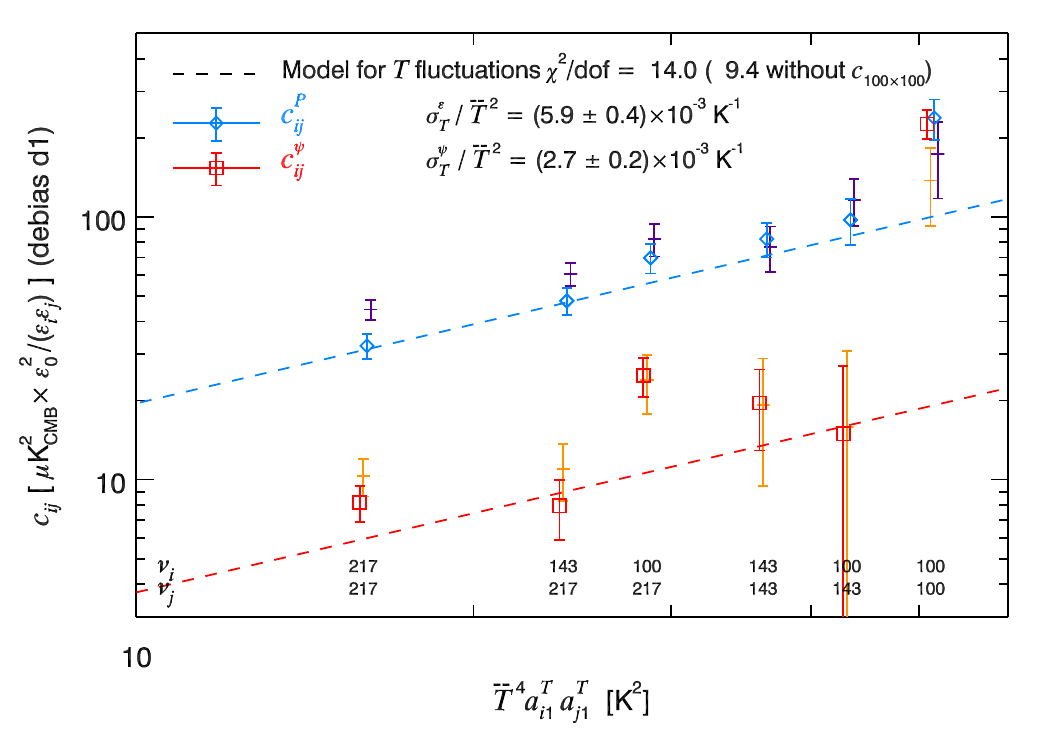}
\caption{Same as Fig.~\ref{fig:varbT}, but for the \Planck\ PR4 official release. To facilitate comparison between PR4 and $\srolltwo$, data with errors bars for \srolltwo\ are overplotted with no symbols.}\label{fig:PR4_b}
\end{figure*}
\begin{figure}
\includegraphics[width=\half]{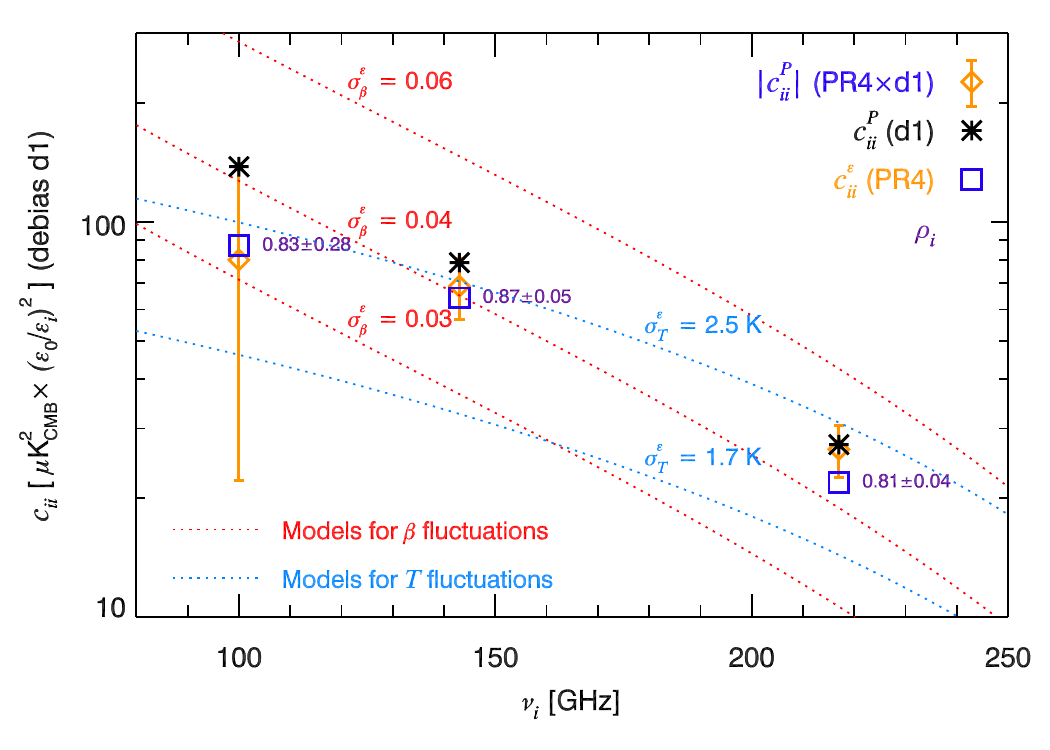}
\caption{Same as Fig.~\ref{fig:PvsI}, but for the \Planck\ PR4 official release.}\label{fig:PR4_c}
\end{figure}

\begin{table} 
\caption{Same as Table~\ref{tab:rho} but for the \Planck\ PR4 official release.}
\begin{tabular}{l l l l l}
\hline\hline
$(\nu_i,\nu_j)$ [GHz] & (217,143) & (217,100)& (143,100) \\
\hline
$\rho^P_{ij}$ & $0.93\pm0.03$ & $0.80\pm0.05$& $0.70\pm0.06$ \\
$\rho^\psi_{ij}$ & $0.63\pm0.09$ & $0.58\pm0.07$& $0.22\pm0.14$ \\ 
$\rho^\P_{ij}$ &$0.87\pm0.03$ & $0.69\pm0.04$& $0.52\pm0.07$ \\
\hline
\end{tabular}
\label{tab:rho_PR4}
\end{table}

 \begin{table} 
 \caption{Same as Table~\ref{tab:P7} but for the \Planck\ PR4 official release.}
 \begin{tabular}{l l l l l}
 \hline\hline
$(\nu_i,\nu_j)$ [GHz] & (217,143) & (217,100) & (143,100) \\
 \hline
 $\Im\,\varpsiPij \left(\unitij\right)$ & $-13\pm3$ & $-6\pm7$& $29\pm13$ \\ 
 \hline
 \end{tabular}
 \label{tab:P7_PR4}
 \end{table}

In this section, we present the results obtained by replacing \Planck\ \srolltwo\ data by the \Planck\ PR4 maps and \npipe{} simulations, and compare them with those obtained with \Planck\ \srolltwo\ data.
Figures \ref{fig:PR4_a}, \ref{fig:PR4_b}, \ref{fig:PR4_c} and Tables ~\ref{tab:rho_PR4} and \ref{tab:P7_PR4} present the results we obtain replacing the $\srolltwo$ version of \Planck\ data by the last official \Planck\ release PR4. We remind the reader that all our calculated covariances are independent of any possible calibration error in the polarization data, either in amplitude or in phase (see Sect.~\ref{subsec:Mean_SED_Planck}), thereby facilitating the comparison between the properties of these two versions of \Planck\ data.

The mean complex SED for PR4 (Fig.~\ref{fig:PR4_a}, top left panel, black solid line) differs in angle from that of \srolltwo\ (green solid line) at 100 and 217\GHz, but not at 143\GHz. This reveals the distinct treatment of CO-contaminated channels between these two versions of \Planck\ data. Predictions \ref{eq:P2}, \ref{eq:P3}, and \ref{eq:P4} (Fig.~\ref{fig:PR4_a}, top right panel) are verified using PR4, like for \srolltwo. Inspection of polarization ratios $\varpsi/\varP$ (predictions \ref{eq:P5} and \ref{eq:P6}, Fig.~\ref{fig:PR4_a}, bottom left panel) lead to the same conclusions like for \srolltwo. Like for \srolltwo\, prediction \ref{eq:P7} (Fig.~\ref{fig:PR4_a}, bottom right panel and Table~\ref{tab:P7_PR4}) is not verified for PR4, with a higher discrepancy than with $\srolltwo$ owing to the smaller error bars from PR4.

The main difference between PR4 and \srolltwo\ arises in the level of correlation between residual maps (Table~\ref{tab:rho_PR4}) and in the spectral dependence of covariances (Fig.~\ref{fig:PR4_b}). Table~\ref{tab:rho_PR4} shows that prediction \ref{eq:P1} is not verified with PR4, even for $\RP$ residual maps: the Pearson coefficient between residual maps is systematically lower with PR4 than with \srolltwo\ (owing to the larger error bars for $\srolltwo$ these values remain compatible, however). 
Comparing PR4 and \srolltwo\ covariances in Fig.~\ref{fig:PR4_b}, we see that $\varpsi$ covariances are, within error bars, compatible. This is remarkable as the mean SEDs are different. This probably means that the main difference regarding polarization angles between these two version of \Planck\ data lies in the mean complex SED, not in the fluctuations around this mean. Covariances $\varP$ are compatible, except for $\varP_{217\times217}$ for which the discrepancy by far exceeds uncertainties.
The $\chi^2$ fits to the PR4 data are bad for our two scenarios (whether $T$ or $\beta$ fluctuate, hypothesis \ref{HM3}). The covariances $\varpsi_{100\times100}$ and $\varP_{100\times100}$ clearly depart from the general trend. Removing them from the fitted values does not however improve the fit dramatically. 

Finally, the correlation between SEDs in polarization (PR4) and intensity (\dmod) in Fig.~\ref{fig:PR4_c} reveals a good level of correlation, but not as strong as with \srolltwo\ (Fig.~\ref{fig:PvsI}). This is probably a consequence of the demonstration by Fig.~\ref{fig:PR4_b} that the fluctuations around the mean SED in PR4 do not follow a spectral behavior corresponding to fluctuations of $\beta$ or $T$, unlike for \dmod\ and \srolltwo.

\section{Using other \pysmthree\ dust models for debiasing}\label{A-otherdustmodels}

\begin{figure*}
\includegraphics[width=\half]{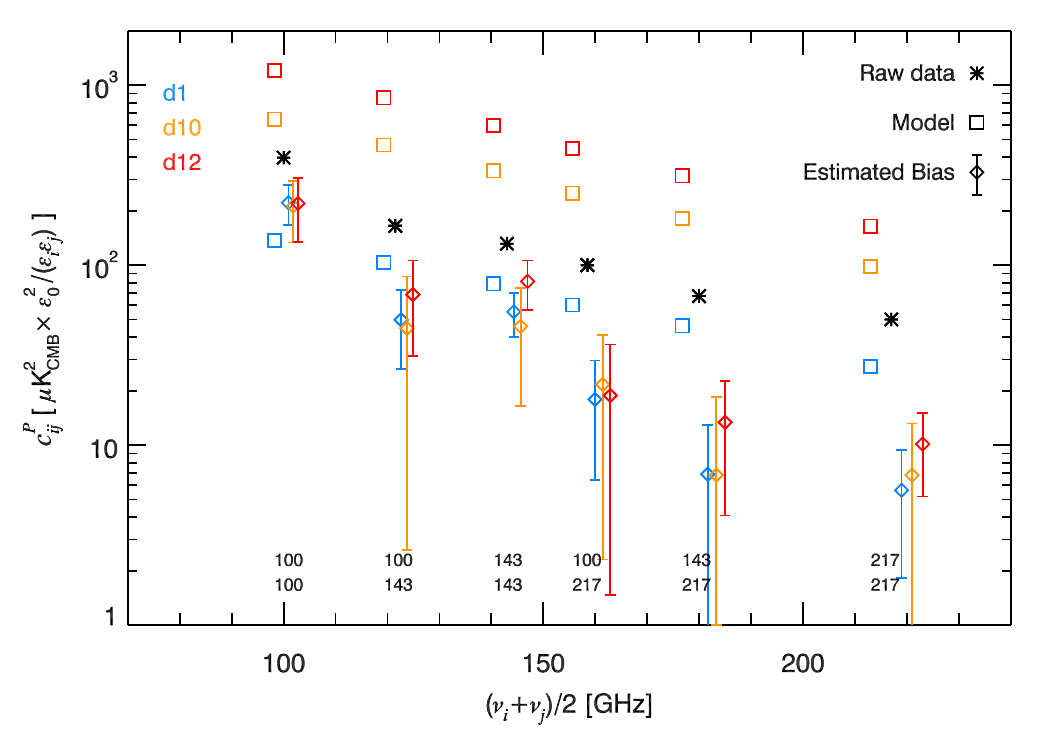}
\includegraphics[width=\half]{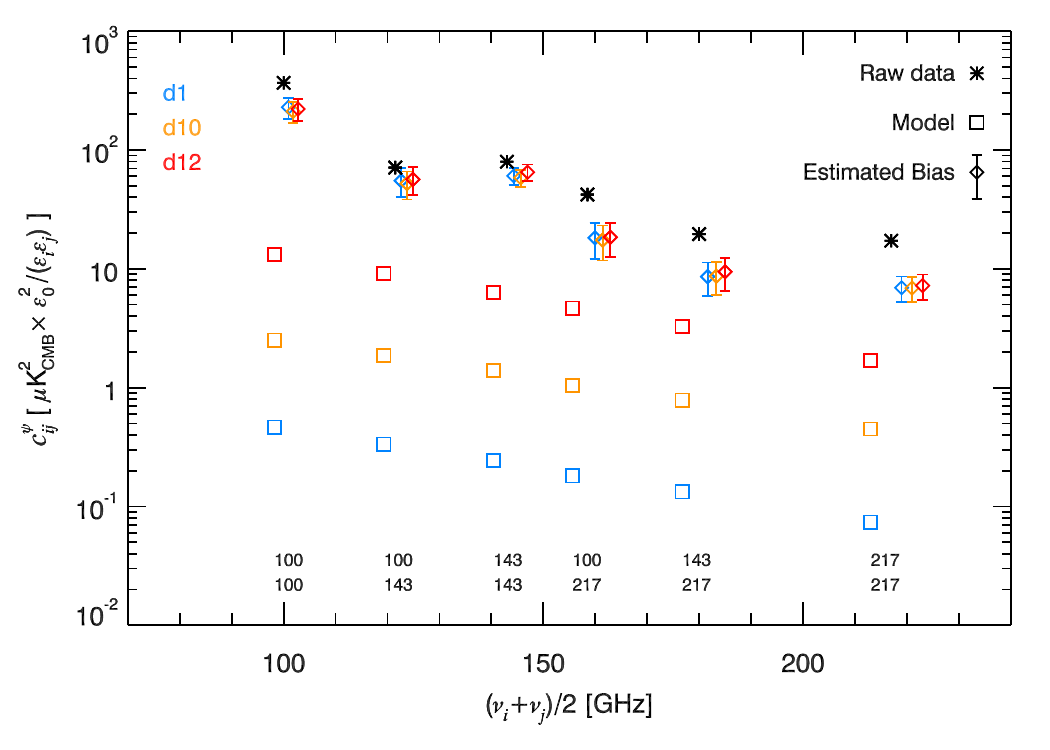}
\caption{Effect of the dust model on the amplitude of bias and error-bars estimates for $\varP$ \llab\ and $\varpsi$ \rlab.}\label{fig:othermodels-bias}
\end{figure*}

Figure~\ref{fig:othermodels-bias} presents how our estimate for the bias produced on covariances by CMB and noise and systematics depends on the dust model chosen to compute this estimate. From Eq.~\eqref{eq:bias} we compute, for each model and for each member of the covariance matrix, the mean bias $c_{\rm bias} \equiv \frac{1}{\Nsims}\sum_{k=1}^\Nsims c_{\rm bias}(k)$ together with its uncertainty (the standard deviation of $c_{\rm bias}(k)$).
We compare the amplitude and error bars of $c_{\rm bias}$ for the three \pysmthree\ dust models \dmod, {\tt d10} and {\tt d12}, for $\varP$ \llab\ and $\varpsi$ \rlab. Although the covariances of models strongly differ, all biases estimated are compatible for all couple of channels, within the uncertainties computed from the 200 \srolltwo\ simulations.

\begin{figure*}
\includegraphics[width=\half]{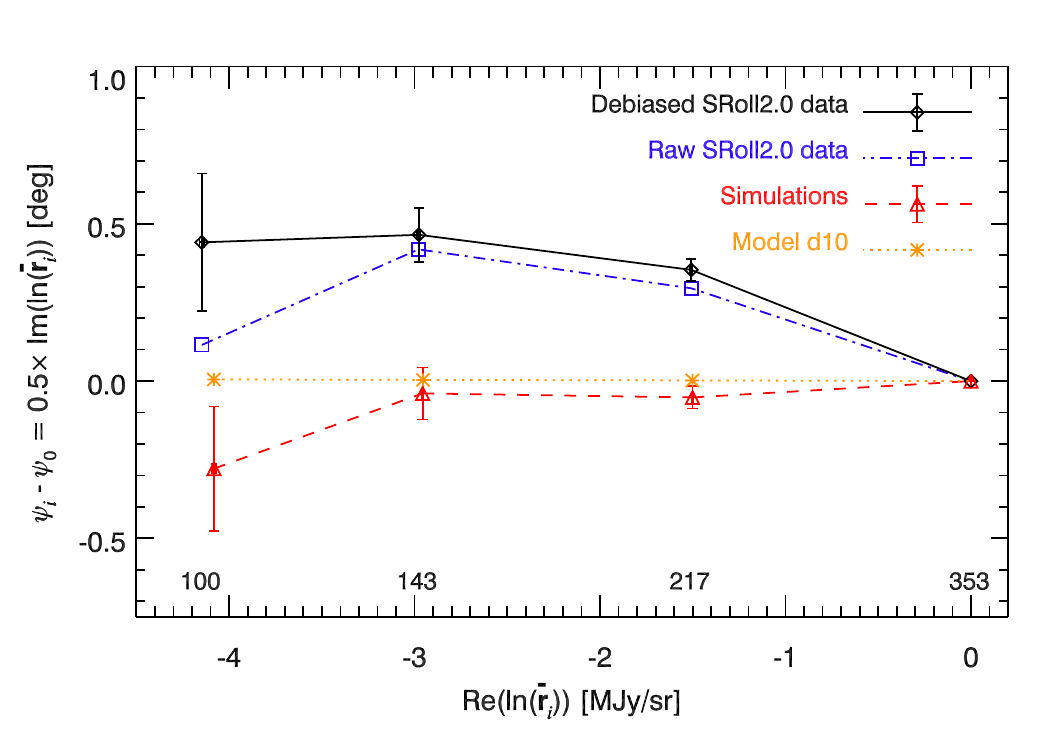}
\includegraphics[width=\half]{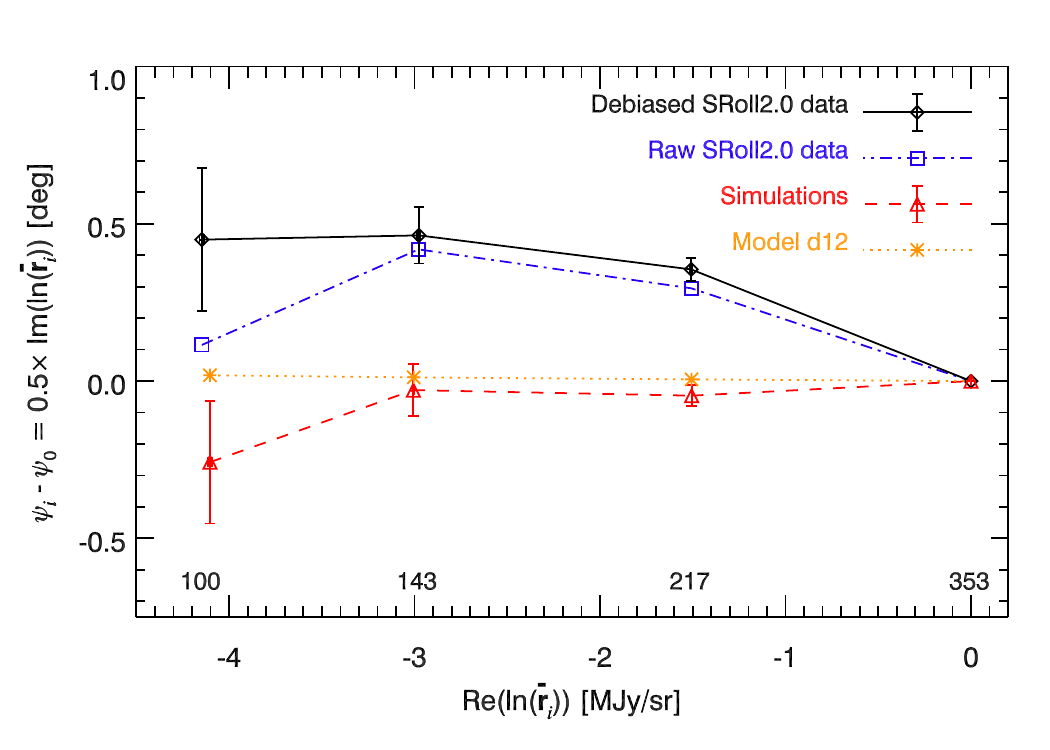}
\caption{Same as Fig.~\ref{fig:meanSED}, but for {\tt d10} \llab\ and {\tt d12} \rlab\ \pysmthree\ dust models.
}
\label{fig:othermodels-meanSED}
\end{figure*}

Figure~\ref{fig:othermodels-meanSED} presents the mean SED derived from \Planck\ \srolltwo\ data using the dust models {\tt d10} \llab\ and {\tt d12} \rlab\ to debias covariances from CMB, noise and systematics. There is no significant difference with the one obtained with model \dmod\ (Fig.~\ref{fig:meanSED}).

\begin{figure*}
\includegraphics[width=\half]
{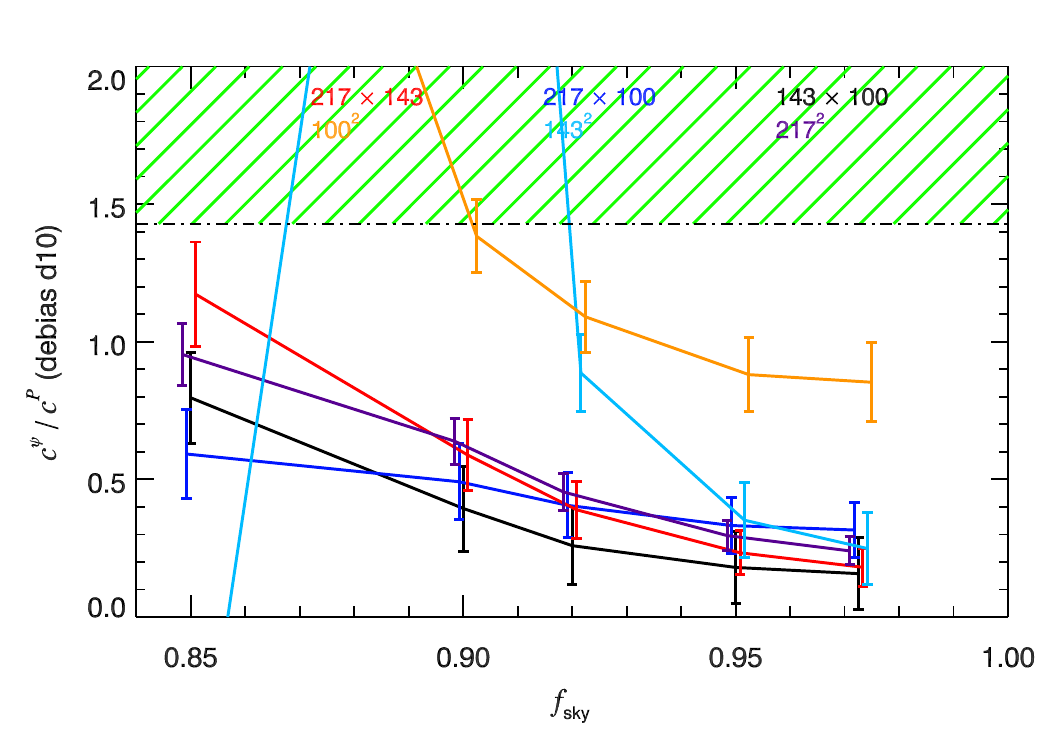}
\includegraphics[width=\half]
{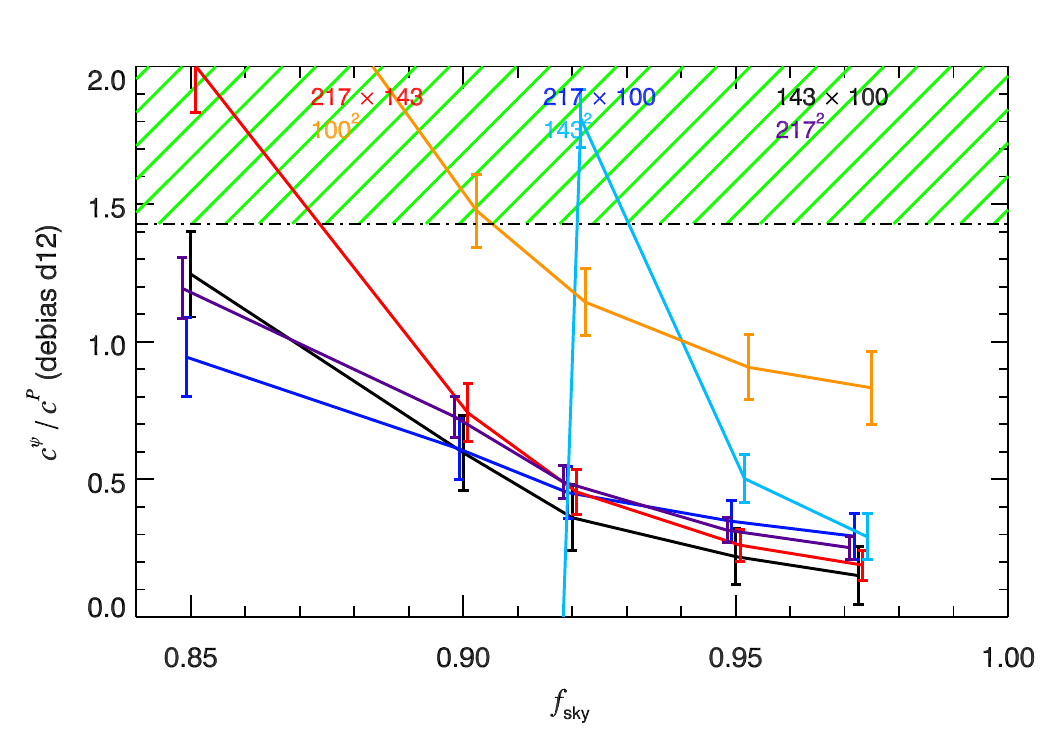}
\caption{
Same as Fig.~\ref{fig:checkratio} for the {\tt d10} \llab\ and {\tt d12} \rlab\ \pysmthree\ dust models.
The hashed region represents forbidden values according to \ref{eq:P5}. 
}
\label{fig:othermodels-checkratio}
\end{figure*}

Figure~\ref{fig:othermodels-checkratio}, like Fig.~\ref{fig:checkratio}, presents how the ratio $\varpsi/\varP$ vary with $\fsky$ when using dust models {\tt d10} \llab\ and {\tt d12} \rlab\ to debias covariances from CMB, noise and systematics. Our conclusions remain the same.

\begin{figure*}
\includegraphics[width=\half]{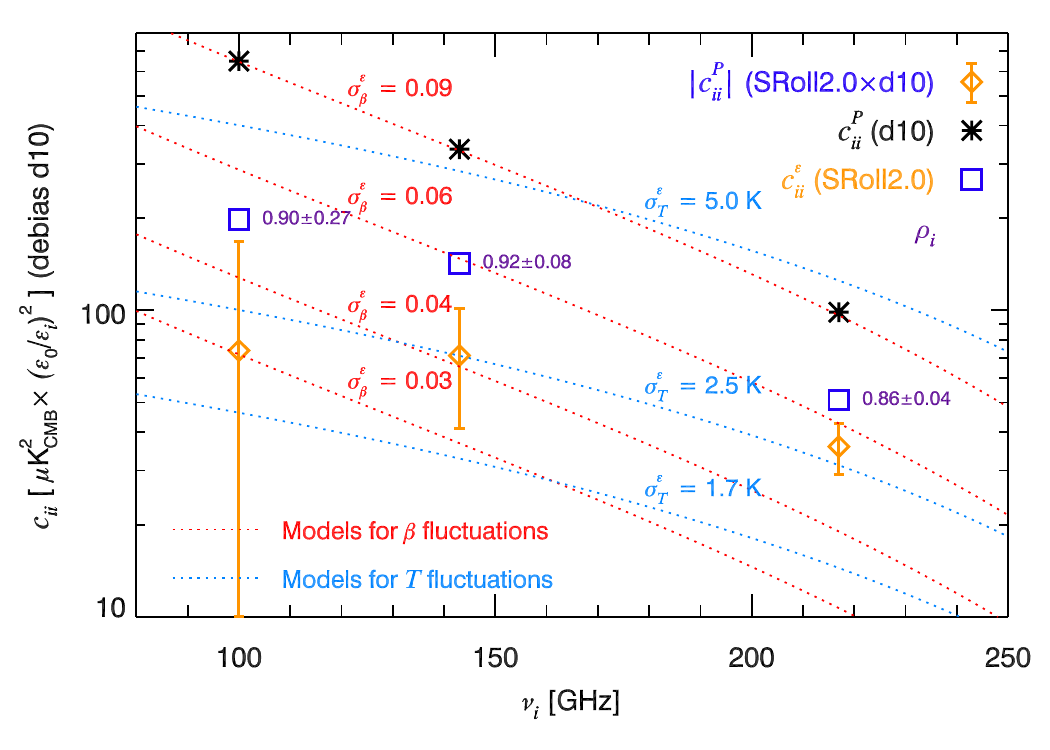}
\includegraphics[width=\half]{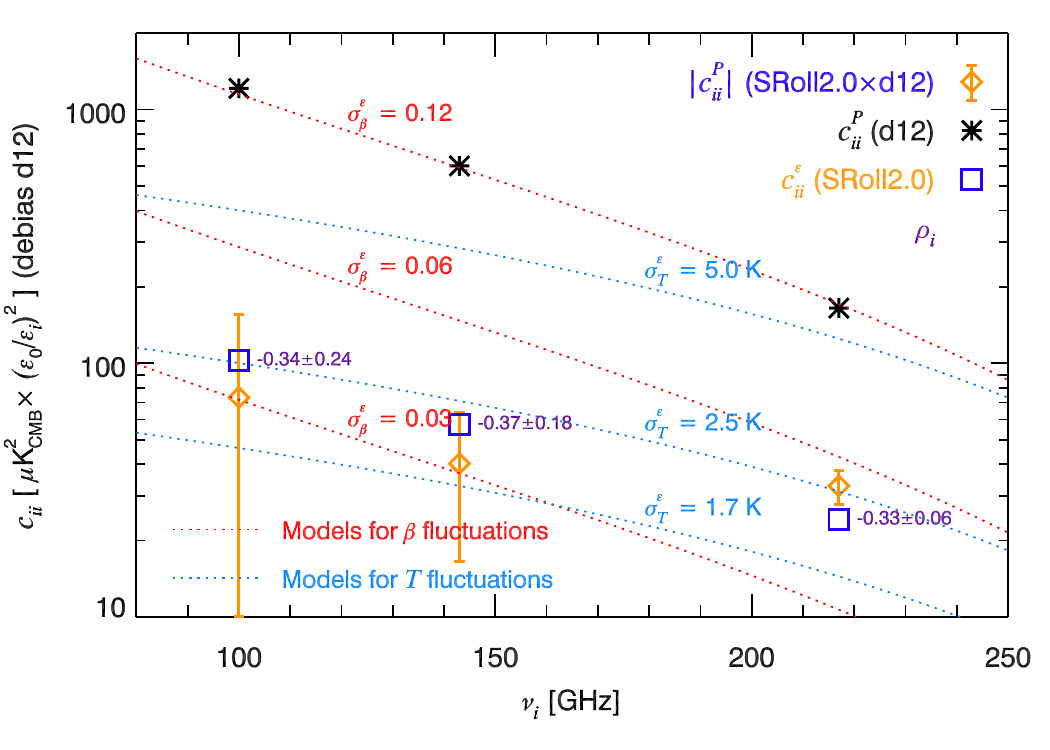}
\caption{Same as Fig.~\ref{fig:PvsI}, but for {\tt d10} \llab\ and {\tt d12} \rlab\ \pysmthree\ dust models. Both models are dominated by fluctuations of $\beta$ and have too much variance with respect to polarization data (by a factor $\sim 4$ and $\sim 7$, respectively), for our 97.3\% mask dominated by the Galactic plane.
}\label{fig:othermodels-PvsI}
\end{figure*}

Figure \ref{fig:othermodels-PvsI} presents the correlation of \Planck\ \srolltwo\ polarization maps residuals with dust models {\tt d10} \llab\ and {\tt d12} \rlab. Variances are for {\tt d10} (based on GNILC maps) around 4 times the one found our reference model \dmod\ (based on \COMMANDER\ maps). For {\tt d12}, this is a factor 7. Note that in our analysis the values of variances are dominated by fluctuations in the Galactic plane owing to the weight $P_0^2$ for the calculation of covariances (see Eq.~\eqref{eq:defCov}). Nevertheless, the correlation of polarization data with {\tt d10} in residual maps is similar to the one obtained in Fig.~\ref{fig:PvsI} for the \dmod\ dust model. SED fluctuations present in the {\tt d12} dust model anticorrelates with polarization data. The reason for this behavior is unknown.

\begin{figure*}
\includegraphics[width=\half]{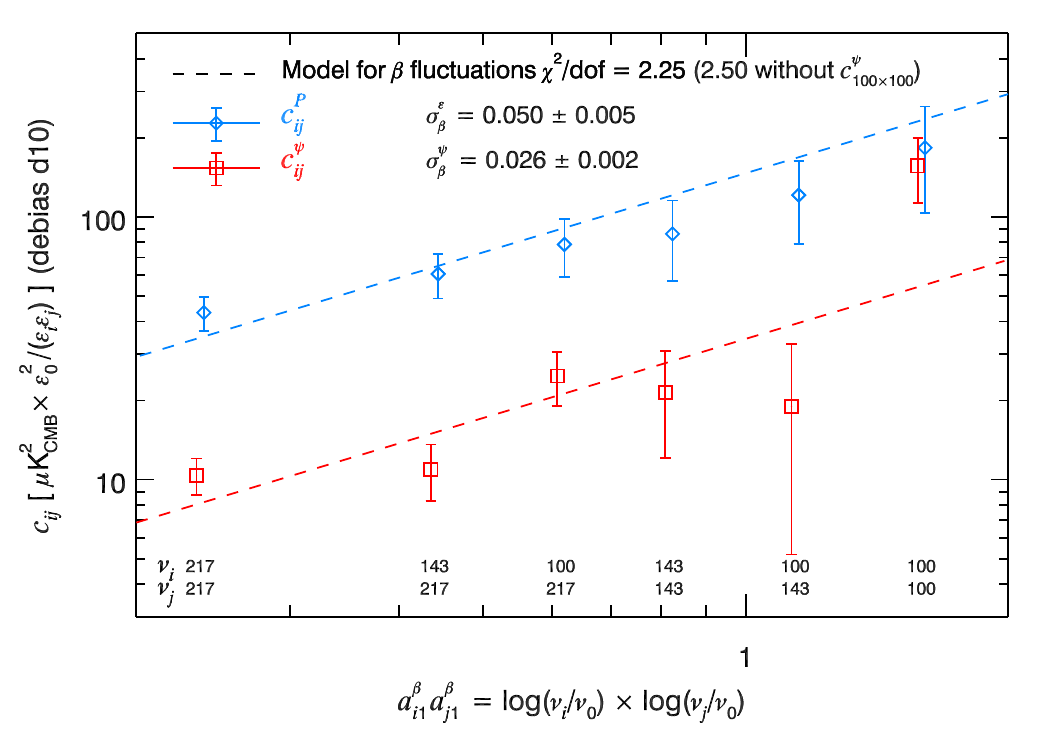}
\includegraphics[width=\half]{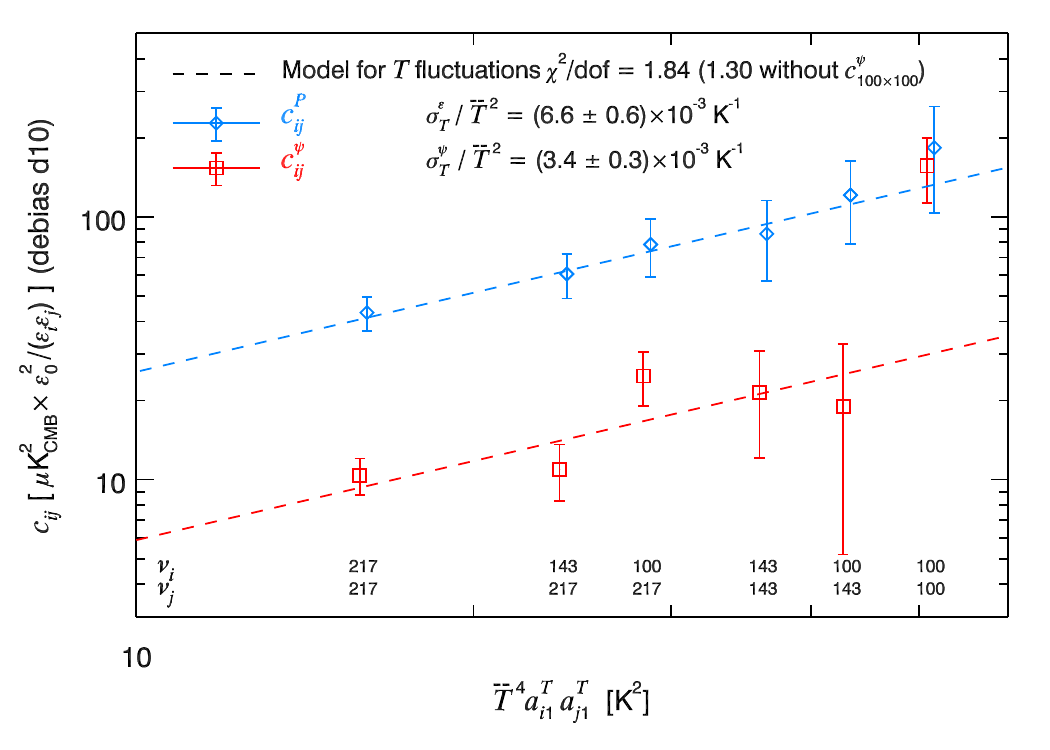}
\includegraphics[width=\half]{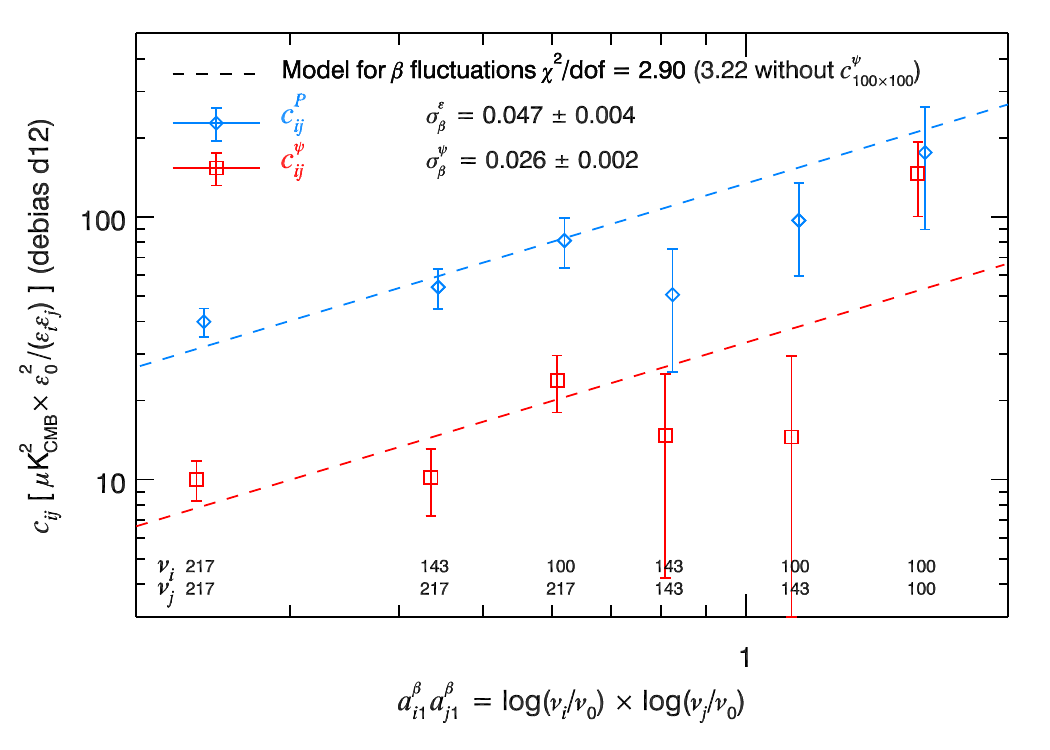}
\includegraphics[width=\half]{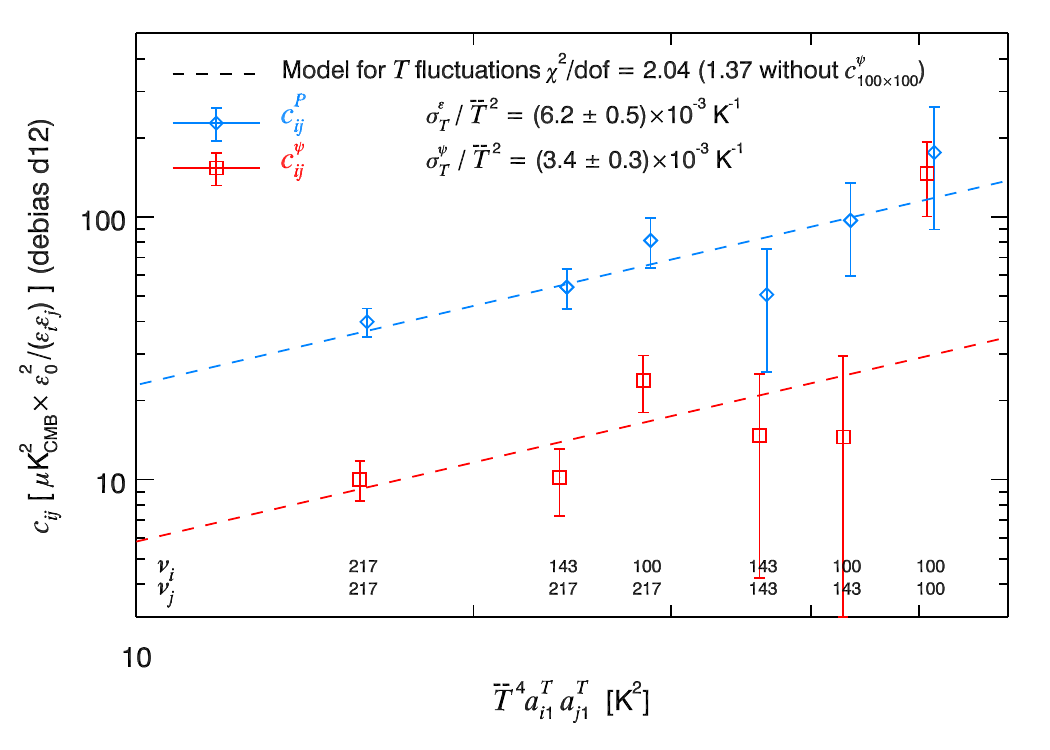}
\caption{Comparison of \srolltwo~data with the predictions of dust model {\tt d10} \tlab\ and {\tt d12} \blab, for $\varP$ (blue) and $\varpsi$ (red), for pure $\beta$ fluctuations \llab\ and pure $T$ fluctuations \rlab. 
} 
\label{fig:othermodels-varbT}
\end{figure*}

Figure~\ref{fig:othermodels-varbT}, like Fig.~\ref{fig:varbT}, presents the spectral dependence of covariances with the product of first-order spectral gradients coefficients $\aio\ajo$, for $\beta$ \llab\ and $T$ \rlab, using the {\tt d10} \tlab\ and {\tt d12} \blab\ dust models for debiasing.

\end{document}